\documentstyle[12pt]{article}
\input amssymb.sty

\ifnum\ht0<\ht2\fi
\let\bls\baselineskip \let\nt\noindent
\let\vp\vphantom 
  
\def\vsk#1>{\vskip#1\bls} \def\vv#1>{\vadjust{\vsk#1>}}
\def\,{\relax\ifmmode\mskip\thinmuskip\relax\else\kern.16667em\fi}
\def\;{\relax\ifmmode\mskip\thickmuskip\relax\else\kern.27777em\fi}
\def\!{\relax\ifmmode\mskip-\thinmuskip\relax\else\kern-.16667em\fi}
\def\&{.\kern.1em} \def\itl#1{{\it #1\/}} 
\def\ftext#1{{\let\thefootnote\relax\footnotetext{\vsk-.8>\noindent #1}}}

\let\ge\geqslant

\let\le\leqslant

\def\bea{\begin{eqnarray}}
\def\ena{\end{eqnarray}}
\def\beq{\begin{equation}}
\def\eeq{\end{equation}}
\def\no{\nonumber}
\def\qed{\quad$\square$\newline}

\def\proof{\noindent{\it Proof.}\quad}
\def\rem{\noindent{\it Remark.}\quad}
\newtheorem{thm}{Theorem}[section]
\newtheorem{prop}[thm]{Proposition}
\newtheorem{lem}[thm]{Lemma}
\newtheorem{cor}[thm]{Corollary}

\makeatletter
\newbox\p@b@ld
\def\poorbold#1{\setbox\p@b@ld\hbox{#1}\kern-.01em\copy\p@b@ld\kern-\wd\p@b@ld
 \kern.02em\copy\p@b@ld\kern-\wd\p@b@ld\kern-.012em\raise.02em\box\p@b@ld}

\@addtoreset{thm}{section}
\@addtoreset{equation}{section}
\makeatother

\begin{document}

\begin{center}
\vp1
{\Large \bf 
Form factors of $SU(N)$ invariant Thirring model
}
\vsk2>
{Yoshihiro Takeyama$^{\,\diamond}$}
\ftext{$^{\diamond\,}$Research Fellow of the Japan Society for
the Promotion of Science. \\
e-mail:ninihuni@kurims.kyoto-u.ac.jp
}
\vsk1.5>
{\it Research Institute for Mathematical Sciences,
Kyoto University, Kyoto 6068502, Japan}
\end{center}
\vsk1.75>

{\narrower\nt
{\bf Abstract.}\enspace
We obtain a new integral formula for solutions of the rational quantum 
Knizhnik-Zamolodchikov equation associated with Lie algebra $sl_{N}$ at level zero. 
Our formula contains the integral representation of form factors 
of $SU(N)$ invariant Thirring model constructed by F. Smirnov. 
We write down recurrence relations 
arising from the construction of the form factors. 
We check that the recurrence relations hold for 
the form factors of the energy momentum tensor. 
\vsk1.4>}
\vsk0>
\thispagestyle{empty}

\tableofcontents 

\section{Introduction}
In this paper we study solutions of the quantum Knizhnik-Zamolodchikov (qKZ) equation 
satisfied by form factors of the $SU(N)$ invariant Thirring model ($SU(N)$ ITM), 
and give recurrence relations for the solutions to be form factors 
of $SU(N)$ ITM. 

In the study of integrable quantum field theories 
it is an important problem to determine all local operators in the theory. 
To study this problem the form factor bootstrap approach is an appropriate method.
The form factors of a local operator ${\cal O}$ 
are functions $\{f^{\cal O}(\beta_{1}, \cdots , \beta_{n})\}_{n}$ satisfying 
certain axioms written by certain difference equations and recurrence relations. 
Thus the problem of determining all local operators is reduced to 
giving all the solutions to the equations. 

In this paper we consider the $SU(N)$ ITM. 
Form factors of some local operators in this model constructed by Smirnov \cite{smirbook} 
are not sufficient in order to determine all local operators. 
To construct a large family of form factors 
one of suitable methods is to use the hypergeometric solutions 
of the qKZ equation \cite{TV1,TV2,NPT}. 

Now let us recall some results in \cite{NT} 
for form factors in the $N=2$ case. 
In \cite{NT} sufficiently many form factors have been constructed 
for the $SU(2)$ ITM using the hypergeometric solutions as follows. 
Form factors $\{f(\beta_{1}, \cdots , \beta_{n}) \}$ 
of $SU(2)$ ITM are functions 
taking values in the tensor product of the $n$-copies of the vector representation $V$ of $SU(2)$, 
and they satisfy the following axioms: 
\bea 
&& 
({\rm I})\, 
P_{j, j+1}S_{j, j+1}(\beta_{j}-\beta_{j+1})f( \cdots , \beta_{j}, \beta_{j+1}, \cdots ) 
=f(\cdots , \beta_{j+1}, \beta_{j}, \cdots ), \no \\
&& 
({\rm II})\,  P_{n-1, n} \cdots P_{1, 2}f(\beta_{1}-2\pi i, \beta_{2}, \cdots , \beta_{n})= 
(-1)^{-\frac{n}{2}}f(\beta_{2}, \cdots , \beta_{n}, \beta_{1}), \no \\
&& 
({\rm III})\, 2\pi i \, {\rm res}_{\beta_{n}=\beta_{n-1}+\pi i}
f(\beta_{1}, \cdots , \beta_{n}) \no \\
&& \quad \! {}=
\left( I-(-1)^{\frac{n}{2}-1}S_{n-1, n-2}(\beta_{n-1}-\beta_{n-2}) \cdots 
         S_{n-1, 1}(\beta_{n-1}-\beta_{1}) \right) 
(f(\beta_{1}, \cdots , \beta_{n-2}) \otimes {\bf e}_{0}), \no 
\ena 
where $S(\beta)$ is the $S$-matrix of the model, 
$P_{i, j}$ is the permutation of the $i$-th and $j$-th components 
and ${\bf e}_{0}$ is the suitably normalized $sl_{2}$ singlet vector in $V^{\otimes 2}$. 
A family of solutions to $({\rm I}), ({\rm II})$ and $({\rm III})$ is constructed 
in the following way. 
First we note that $({\rm I})$ and $({\rm II})$ imply 
the following system of difference equations: 
\bea 
&& 
f(\beta_{1}, \cdots , \beta_{j}-2\pi i, \cdots , \beta_{n})= 
(-1)^{-\frac{n}{2}}
S_{j, j-1}(\beta_{j}-\beta_{j-1}-2\pi i) \cdots S_{j, 1}(\beta_{j}-\beta_{1}-2\pi i) \no \\
&& \qquad \qquad \qquad \qquad \qquad \qquad {}\times 
S_{j, n}(\beta_{j}-\beta_{n}) \cdots S_{j, j+1}(\beta_{j}-\beta_{j+1})
f(\beta_{1}, \cdots , \beta_{j}, \cdots , \beta_{n}). \no 
\ena 
This is nothing but the qKZ equation associated with $sl_{2}$ at level zero. 
%Hence, to construct form factors, it is a first step to construct solutions to the qKZ equation. 
In \cite{NPT}, the integral formulae for solutions of the qKZ equation were given. 
These solutions take values in the space of singular vectors of $sl_{2}$. 
Moreover these solutions span the subspace of singular vectors over the field of 
appropriate periodic functions \cite{T}. 
Any solution is obtained by applying $sl_{2}$ successively to these solutions. 
%it is proved that these solutions span the subspace of singlet vectors \cite{T}. 
Thus we have the complete description of 
the solutions of the $sl_{2}$ qKZ equation at level zero. 
Next let us consider the axiom $({\rm III})$. 
In the hypergeometric description a solution of $({\rm I})$ and $({\rm II})$ is specified 
by a certain function $P_{n}$ called deformed cycle \cite{abelian} 
(or ``p-function'' in \cite{BK}). 
Then the axiom $({\rm III})$ is derived from a recurrence relation for $P_{n}$ and $P_{n-2}$. 
%This is a polynomial recursion equation. 
%Thus the problem to solve $({\rm I}), ({\rm II})$ and $({\rm III})$ is 
%reduced to solving this recursion equation. 
Each local operator corresponds to a sequence of deformed cycles $\{P_{n}\}_{n}$ 
satisfying this recurrence relation. 
A large family of solutions to the recurrence relation has been constructed 
extending Smirnov's construction of the chargeless (or weight zero) local operators 
for the sine-Gordon model \cite{counting}.
This construction was extended to charged local operators and 
a new abelian symmetry was found \cite{NT}. 
%Similar approach is studied for form factors of 
%certain massive models with diagonal $S$-matrix \cite{K,P}. 

In this paper we consider form factors in the $SU(N)$ ITM. 
%In the case of $N>2$, the situation is more complicated than the case of $N=2$ 
%because form factors take values in  not only the vector representation 
%but the other fundamental representations of $SU(N)$. 
The form factors of a local operator ${\cal O}$ are functions 
$\{f^{{\cal O}, (l_{1}, \cdots , l_{n})}(\beta_{1}, \cdots , \beta_{n})\}$ such that 
\bea 
f^{{\cal O}, (l_{1}, \cdots , l_{n})}(\beta_{1}, \cdots , \beta_{n}) \in 
V^{(l_{1})} \otimes \cdots \otimes V^{(l_{n})}, \no
\ena 
where $V^{(l)}$ is the $l$-th fundamental representation of $SU(N)$. 
Now let us recall the axioms satisfied by 
form factors of $SU(N)$ ITM \cite{smirbook}: 
%As in the case of $N=2$, the following equalities hold: 
\bea 
&&
\!\!\!\!\!\!\!\!({\rm I}) \, 
P_{j, j+1}S^{(l_{j}, l_{j+1})}_{j, j+1}(\beta_{j}-\beta_{j+1})
f^{(\cdots , l_{j}, l_{j+1}, \cdots)}( \cdots , \beta_{j}, \beta_{j+1}, \cdots )=
f^{(\cdots , l_{j+1}, l_{j}, \cdots)}( \cdots , \beta_{j+1}, \beta_{j}, \cdots ), \no \\
&&
\!\!\!\!\!\!\!\!({\rm II}) \, 
P_{n-1, n}\cdots P_{1, 2}
f^{(l_{1}, l_{2}, \cdots , l_{n})}(\beta_{1}-2\pi i, \beta_{2}, \cdots , \beta_{n})=
e^{-\frac{N-1}{N}\pi i \sum_{j=1}^{n}l_{j}}
f^{(l_{2}, \cdots , l_{n}, l_{1})}(\beta_{2}, \cdots , \beta_{n}, \beta_{1}), \no \\ 
&&
\!\!\!\!\!\!\!\!({\rm III}) \, 
2\pi i \, {\rm res}_{\beta_{n}=\beta_{n-1}+\pi i} 
f^{(l_{1}, \cdots , l_{n})}(\beta_{1}, \cdots , \beta_{n}) \no \\
&& {}=
\delta_{l_{n-1}+l_{n}, N} 
\left(
I+e^{-\frac{2\pi i}{N}+\frac{N-1}{N}\pi i\sum_{j=1}^{n-2}l_{j}}
 S^{(l_{n-1}, l_{n-2})}_{n-1, n-2}(\beta_{n-1}-\beta_{n-2})\cdots 
S^{(l_{n-1}, l_{1})}_{n-1, 1}(\beta_{n-1}-\beta_{1}) 
\right) \no \\
&& \qquad \qquad {}\times 
f^{(l_{1}, \cdots , l_{n-2})}(\beta_{1}, \cdots , \beta_{n-2})\otimes {\bf e}_{0}, \no 
\ena
where $S^{(l, l')}(\beta)$ is the $S$-matrix  
acting on $V^{(l)} \otimes V^{(l')}$ 
%constructed by fusion procedure \cite{KLS}, 
and ${\bf e}_{0}$ is the suitably normalized singlet vector 
in $V^{(l_{n-1})} \otimes V^{(l_{n})}$. 
Moreover, they satisfy a number of formulae for residues corresponding to bound states. 
The most fundamental one is the following. 
If $l_{n-1}+l_{n}<N$ the residue of 
$f^{(l_{1}, \cdots , l_{n-1}, l_{n})}(\beta_{1}, \cdots , \beta_{n})$ at 
$\beta_{n}=\beta_{n-1}+\frac{l_{n-1}+l_{n}}{N}\pi i$ is given by 
\bea  
({\rm IV}) \, 
2\pi i \, {\rm res}
f^{(l_{1}, \cdots , l_{n-1}, l_{n})}(\beta_{1}, \cdots , \beta_{n})=
a_{l_{n-1}, l_{n}}
f^{(l_{1}, \cdots , l_{n-2}, l_{n-1}+l_{n})}
(\beta_{1}, \cdots , \beta_{n-2}, \beta_{n-1}+\frac{l_{n}\pi i}{N}), \no 
\ena 
where $a_{l, l'}$ is a certain constant. 
In \cite{smirbook} the form factors of some local operators are constructed. 

We study the problem to give the solutions of 
$({\rm I})$-$({\rm IV})$ 
in a similar approach to the case of $N=2$. 
Again the first step is to solve the qKZ equation derived from 
$({\rm I})$ and $({\rm II})$: 
\bea 
&& \!\!\!\!\!\!\!\!\!
f^{(l_{1}, \cdots , l_{n})}(\beta_{1}, \cdots , \beta_{j}-2\pi i, \cdots , \beta_{n})= 
e^{-\frac{N-1}{N}\pi i\sum_{j}l_{j}}
S_{j, j-1}(\beta_{j}-\beta_{j-1}-2\pi i) \cdots S_{j, 1}(\beta_{j}-\beta_{1}-2\pi i) \no \\
&& \qquad \qquad \qquad \qquad \qquad \quad {}\times 
S_{j, n}(\beta_{j}-\beta_{n}) \cdots S_{j, j+1}(\beta_{j}-\beta_{j+1})
f^{(l_{1}, \cdots , l_{n})}(\beta_{1}, \cdots , \beta_{j}, \cdots , \beta_{n}). \no 
\ena
However, it is difficult to construct solutions of the qKZ equation above 
for general $l_{1}, \cdots , l_{n}$. 
Some representations of solutions 
were constructed in \cite{TV, BKZ} in terms of Jackson integrals, that is, formal infinite sums. 
It seems difficult to prove the convergence of these sums. 
In this paper we give a new integral formula for solutions of the qKZ equation 
taking values in the product of the vector representations, 
that is , $l_{1}= \cdots =l_{n}=1$, 
and consider the form factors of type  
\bea 
f^{(1, \cdots , 1, k)}(\beta_{1}, \cdots , \beta_{n-k}, \beta_{n-k+1}) 
\in (V^{(1)})^{\otimes (n-k)} \otimes V^{(k)}
\no 
\ena 
associated with chargeless local operators. 
The conditions $({\rm I})$-$({\rm IV})$ are closed conditions among these functions. 
In fact we suppose that 
$f^{(1, \cdots , 1)}(\beta_{1}, \cdots , \beta_{n})$ 
is a form factor. 
Then, by taking the residue as in the axiom $({\rm IV})$ successively, 
we obtain form factors 
$f^{(1, \cdots , 1, k)}(\beta_{1}, \cdots , \beta_{n-k+1}),$ \\ 
$(k=2, \cdots , N-1)$. 
At last, we obtain a form factor 
$f^{(1, \cdots , 1)}(\beta_{1}, \cdots , \beta_{n-N})$ from $f^{(1, \cdots , 1, N-1)}$ 
by the axiom $({\rm III})$. 
In this way the form factor on $(V^{(1)})^{\otimes (n-N)}$ 
is given by one on $(V^{(1)})^{\otimes n}$.
Moreover we consider form factors of chargeless local operators. 
Then form factors are of weight zero, 
and hence the number of the components of the tensor product 
that $f^{(1, \cdots , 1)}(\beta_{1}, \cdots , \beta_{n})$ takes values in 
is a multiple of $N$, say $n=mN$. 
Then the axioms imply some relation between 
$f_{m}:=f^{(1, \cdots , 1)}(\beta_{1}, \cdots , \beta_{n})$ 
and $f_{m-1}:=f^{(1, \cdots , 1)}(\beta_{1}, \cdots , \beta_{n-N})$. 
We write down this relation by using our integral formula for solutions of the qKZ equation 
taking values in the product of the vector representations. 
As a result we get recurrence relations as in the case of $N=2$. 

First we start from a certain integral formula for solutions on 
the tensor product of the vector representations, that is, $(V^{(1)})^{\otimes n}$. 
This integral formula is obtained as the limit $q \to 1$ 
of the hypergeometric solutions of the trigonometric qKZ equation 
associated with the quantum affine algebra $U_{q}(sl_{N})$ at $|q|=1$, 
which was constructed in \cite{MTT}. 
In \cite{MTT} it is proved that, 
if the parameters in the qKZ equation are generic, 
the set of the solutions become a basis of the weight subspace 
that the solutions take values in. 
Nevertheless, in the case of $N>2$, 
it is not easy to calculate the residues of solutions 
in the axioms $({\rm III})$ and $({\rm IV})$ from this integral formula. 
One reason for this is that 
this integral formula contains much more integrations than 
the integral representation of form factors of $SU(N)$ ITM constructed by Smirnov. 
In order to avoid this difficulty we simplify the integral formula in the following way. 
%Let us consider the solutions of weight zero. 
The integral formula for solutions of the $sl_{N}$ qKZ equation at level zero 
contains as the integrand  
solutions of the $sl_{N-1}$ qKZ equation at level one. 
Substituting the $sl_{N-1}$ part with a special solution, 
we get a simplified integral formula for solutions 
of the $sl_{N}$ qKZ equation at level zero.
The method above of rewriting the integral formula was used by A. Nakayashiki 
in the case of the differential KZ equation \cite{Nnote}. 

The special solution mentioned above of the $sl_{N-1}$ qKZ equation 
is obtained as the limit $q \to 1$ of a solution of the trigonometric qKZ equation 
associated with $U_{q}(\widehat{sl}_{N-1})$ at $|q|<1$. 
The highest-to-highest matrix element of the product 
of intertwining operators 
\bea 
\langle \Lambda_{i} | \Phi(z_{1}) \cdots \Phi(z_{\ell}) | \Lambda_{i} \rangle 
=\sum_{\epsilon_{1}, \cdots , \epsilon_{\ell}} 
\langle \Lambda_{i} | 
\Phi_{\epsilon_{1}}(z_{1}) \cdots \Phi_{\epsilon_{\ell}}(z_{\ell}) | \Lambda_{i} \rangle 
v_{\epsilon_{1}} \otimes \cdots \otimes v_{\epsilon_{\ell}} 
\no 
\ena 
satisfies the qKZ equation \cite{FR}. 
%Note that the matrix element is zero unless it is of weight zero. 
In the case that representations are at level one, 
we can calculate this matrix element 
by using the bosonization of intertwining operators \cite{Ko}. 
The coefficients in this matrix element are given by some integral formulae. 
However, the coefficients in the rhs above 
are determined from functional relations
arising from the commutation relation of intertwining operators \cite{DO} 
and the coefficient of the extremal component 
calculated explicitly in \cite{nakayashiki}. 
We choose the limit $q \to 1$ of this solution 
as the special solution of the rational $sl_{N-1}$ qKZ equation at level one. 

Note that 
we can get a certain integral formula for solutions of the qKZ equation 
by generalizing suitably 
the integral representation of form factors constructed by Smirnov. 
We show that this integral formula is obtained from our simplified integral formula 
in the following way. 
%{}From the simplified integral formula, we can get Smirnov's formula as follows. 
The simplified integral formula still contains one more integration than Smirnov's formula. 
However, we can carry out one-time integration of the simplified integral formula 
in a similar way to the case of $sl_{2}$ \cite{NPT}. 
After this integration, Smirnov's formula is obtained. 

Now let us return to the construction of form factors $\{f_{m}\}$. 
The solutions of the qKZ equation given by the simplified integral formula 
are parameterized by functions called deformed cycles 
as in the case of $N=2$. 
Fix $m$ and let $P_{m}$ be the deformed cycle associated with $f_{m}$. 
By calculating the dimension of the space of deformed cycles, 
we can find that the space spanned by these solutions is quite smaller than 
the weight subspace of weight zero
(See \cite{thetaformula} for a similar argument in the case of the differential KZ equation). 
Hence, even if $f_{m}$ is given in terms of the simplified integral formula, 
the form factor $f_{m-1}$, which is obtained by calculating residues of $f_{m}$ successively, 
may not be represented by the simplified integral formula. 
However, under some conditions for the deformed cycle $P_{m}$, 
the form factor $f_{m-1}$ is also given in terms of the simplified integral formula. 
Then we obtain recurrence relations for $P_{m}$ and $P_{m-1}$ (see Proposition \ref{rescond}), 
where $P_{m-1}$ is the deformed cycle associated with $f_{m-1}$. 
We check that the recurrence relations hold for 
the form factors of the energy momentum tensor presented by Smirnov \cite{smirbook}. 
It is still an open problem 
to construct solutions of the recurrence relations different from the deformed cycles 
associated with the form factors constructed by Smirnov. 

The plan of this paper is as follows. 
In Section 2 we give the qKZ equation studied in this paper. 
The integral formula obtained by taking the limit $q \to 1$ of the hypergeometric solutions 
of the $U_{q}(sl_{N})$ qKZ equation at $|q|=1$ is given in Section 3. 
In Section 4 we construct a special solution of the $sl_{N-1}$ qKZ equation at level one. 
By using this special solution 
we rewrite the integral formula obtained in Section 3 
and get the simplified integral formula for the $sl_{N}$ qKZ equation in Section 5. 
In Section 6 we see that the formula in Section 5 contains Smirnov's formula. 
We study form factors of $SU(N)$ ITM in Section 7 
by using the simplified integral formula and write down recurrence relations for deformed cycles. 
We check that the deformed cycle associated with the energy momentum tensor 
satisfies the recurrence relations. 
In Section 8 we give some supplements about the special solution in Section 4 
and proofs of lemmas and propositions in the previous sections.

\section*{Acknowledgements} 
The author is deeply grateful to Atsushi Nakayashiki 
for providing a note \cite{Nnote} and advising to study the subject of this paper. 
The author also thanks his thesis advisor Professor Tetsuji Miwa 
for reading the manuscript and for kind encouragement, 
and Professor Masaki Kashiwara for valuable remarks.

\section{The qKZ equation}

Let $V_{N}:=\oplus_{j=0}^{N-1}{\Bbb C}\,\, v_{j}$ be the vector representation of $sl_{N}$ 
with the highest weight vector $v_{0}$. 
We denote by $R(\beta)$ the rational $R$-matrix given by
\bea
R(\beta):=\frac{\beta+\hbar P}{\beta+\hbar} \in {\rm End}((V_{N})^{\otimes 2}).
\label{defR}
\ena
Here $\hbar$ is a nonzero complex number 
and $P$ is the permutation operator: $P(u \otimes v):=v \otimes u$.
 
Fix a nonzero complex number $p$. 
We consider the (rational) qKZ equation: 
\bea
\psi(\beta_{1}, \cdots, \beta_{j}+p, \cdots, \beta_{n})=
K_{j}(\beta_{1}, \cdots , \beta_{n})\psi(\beta_{1}, \cdots , \beta_{j}, \cdots, \beta_{n}), 
\quad (j=1, \cdots , n), 
\label{qKZ} 
\ena
where
\bea
K_{j}(\beta_{1}, \cdots , \beta_{n})&:=&
R_{j, j-1}(\beta_{j}-\beta_{j-1}+p) \cdots R_{j, 1}(\beta_{j}-\beta_{1}+p) \no \\
&& \qquad {}\times
R_{j, n}(\beta_{j}-\beta_{n}) \cdots R_{j, j+1}(\beta_{j}-\beta_{j+1}).
\ena
Here $\psi$ is a $(V_{N})^{\otimes n}$-valued unknown function 
and $R_{i, j}(\beta)$ is the operator acting on 
the tensor product of $i$-th and $j$-th components as $R(\beta)$. 
The number $-N+p/\hbar$ is called the {\it level} of this qKZ equation. 

Let $e_{k}$ be the generator of $sl_{N}$ associated with the simple root $\alpha_{k}$. 
The action of $e_{k}$ on $V_{N}$ is given by $e_{k}v_{j}=\delta_{k, j}v_{j-1}$. 

In the following, we consider the qKZ equation at level zero, that is, the case of
\bea
p=N\hbar
\label{levelzerocond} 
\ena
and solutions of (\ref{qKZ}) satisfying the highest weight condition:
\bea
E_{k}\psi(\beta_{1}, \cdots , \beta_{n})=0, \quad
{\rm where} \quad
E_{k}:=\sum_{j=1}^{n}1\otimes \cdots \otimes \stackrel{j-{\rm th}}{e_{k}} 
        \otimes \cdots \otimes 1, 
\quad (k=1, \cdots , N-1).
\label{hw}
\ena
Hereafter we assume that ${\rm Im}\hbar <0$.

\section{General solution at level zero}

Let us write down an integral formula for solutions of (\ref{qKZ}).
We can obtain this formula by taking the limit $q \to 1$ of solutions to 
the qKZ equation associated with $U_{q}(sl_{N})$ at $|q|=1$ \cite{MTT}. 

First we introduce some notations.
For non-negative integers $\nu_{1}, \cdots , \nu_{N-1}$ satisfying
\bea
\nu_{0}:=n \ge \nu_{1} \ge \cdots \ge \nu_{N-1} \ge \nu_{N}:=0, 
\ena
we denote by ${\cal Z}_{\nu_{1}, \cdots , \nu_{N-1}}$ the set of all $n$-tuples 
$J=(J_{1}, \cdots , J_{n}) \in ({\Bbb Z}_{\ge 0})^{n}$ such that 
\bea 
\# \{ r; J_{r} \ge j \}=\nu_{j}. 
\ena 
For $J=(J_{1}, \cdots , J_{n}) \in {\cal Z}_{\nu_{1}, \cdots , \nu_{N-1}}$, we set 
\bea 
v_{J}:=v_{J_{1}} \otimes \cdots \otimes v_{J_{n}} \in (V_{N})^{\otimes n}. 
\ena 
We set 
\bea 
{\cal N}_{j}^{J}:= \{ r ; J_{r}\ge j \} 
\ena 
and define integers $r_{j, m}^{J}, (0 \le j \le N-1, 1 \le m \le \nu_{j})$ by 
\bea 
{\cal N}_{j}^{J}=\{ r_{j, 1}^{J}, \cdots , r_{j, \nu_{j}}^{J} \}, \quad 
r_{j, 1}^{J}< \cdots < r_{j, \nu_{j}}^{J}. 
\ena 
Note that $r_{0, m}^{J}=m$. 
For example, for $J=(1, 2, 0, 1, 0, 2) \in {\cal Z}_{4, 2}$, 
we have ${\cal N}_{1}^{J}=\{1, 2, 4, 6\}$ and ${\cal N}_{2}^{J}=\{2, 6\}$. 

For $J \in {\cal Z}_{\nu_{1}, \cdots , \nu_{N-1}}$, 
we define sets $M_{k}^{J}, (k=1, \cdots N-1)$ as follows. 
%\bea 
%{\Bbb M}(J):=(M_{1}^{J}, \cdots , M_{N-1}^{J})
%\ena 
%as follows. 
The set $M_{k}^{J}$ satisfies 
%we denote by ${\cal I}_{\nu_{1}, \cdots , \nu_{N-1}}$ the set of all $(N-1)$-tuples 
%$M=(M_{1}, \cdots , M_{N-1})$ such that
\bea
M_{k}^{J} \subset \{1, 2, \cdots , \nu_{k-1}\}, \quad \# M_{k}^{J}=\nu_{k}. 
\ena
The elements of $M_{k}^{J}:=\{m^{J}_{k, 1}, \cdots , m^{J}_{k, \nu_{k}}\}, \, 
(m^{J}_{k, 1} < \cdots < m^{J}_{k, \nu_{k}})$ are defined by the following rule: 
%Let us define a vector $v_{M} \in (V_{N})^{\otimes n}$ in the following manner.
%Define a set ${\cal N}^{M}_{k}=\{ r^{M}_{k, 1}, \cdots , r^{M}_{k, \nu_{k}}\}$ 
%inductively by
\bea
r^{J}_{k, j}=r^{J}_{k-1, m^{J}_{k, j}}.
\label{defMJ}
\ena
For example, for $J=(1, 2, 0, 1, 0, 2) \in {\cal Z}_{4, 2}$, 
we have $M^{J}_{1}=\{1, 2, 4, 6\}$ and $M^{J}_{2}=\{2, 4\}$. 

%Then a $n$-tuple $J^{M}:=(J^{M}_{1}, \cdots , J^{M}_{n}) \in ({\Bbb Z}_{\ge 0})^{n}$ 
%is uniquely determined by the following condition:
%\bea
%\{m | J^{M}_{m} \ge k \}={\cal N}^{M}_{k}.
%\label{defJ}
%\ena
%Now we set 
%\bea
%v_{M}:=v_{J^{M}_{1}} \otimes \cdots \otimes v_{J^{M}_{n}}.
%\ena

Let us introduce some functions. 
For a subset 
$K=\{k_{1}, \cdots , k_{l}\} \subset \{1, \cdots , m \}, (k_{1}< \cdots < k_{l})$, 
we define the rational function $g_{K}$ and the trigonometric function $P_{K}$ by
\bea
&&
g_{K}(t_{1}, \cdots , t_{l} | z_{1}, \cdots , z_{m}):=
\prod_{a=1}^{l}\left(
\frac{1}{t_{a}-z_{k_{a}}} \prod_{j=1}^{k_{a}-1} \frac{t_{a}-z_{j}-\hbar}{t_{a}-z_{j}} \right)
\prod_{1 \le a<b \le l}(t_{a}-t_{b}-\hbar), \\
&&
P_{K}(e^{\frac{2\pi i}{p}t_{1}}, \cdots , e^{\frac{2\pi i}{p}t_{l}} | 
      e^{\frac{2\pi i}{p}z_{1}}, \cdots , e^{\frac{2\pi i}{p}z_{m}})
:=
\prod_{a=1}^{l} \left(
\prod_{j=1}^{k_{a}-1}(1-e^{\frac{2\pi i}{p}(t_{a}-z_{j}-\hbar)}) \!\!\!
\prod_{j=k_{a}+1}^{m} \!\!\! (1-e^{\frac{2\pi i}{p}(t_{a}-z_{j})}) \right). \no \\
&&
\ena
Introduce a set of variables $\{\gamma_{j, m}\}, \, (1 \le j \le N-1, \, 1 \le m \le \nu_{j})$. 
For $J \in {\cal Z}_{\nu_{1}, \cdots , \nu_{N-1}}$, we set 
\bea
w_{J}^{(N)}(\{\gamma_{j, m}\}| \beta_{1}, \cdots , \beta_{n}):=
{\rm Skew}_{N-1}\circ \cdots \circ {\rm Skew}_{1} 
\left( \prod_{k=1}^{N-1}g_{M_{k}^{J}}(\{\gamma_{k, m}\} | \{\gamma_{k-1, m'}\}) \right),
\ena
where $\gamma_{0, m}:=\beta_{m}$ and the operator ${\rm Skew}_{k}$ is the skew-symmetrization 
with respect to the variables $\gamma_{k, m}, \, (1 \le m \le \nu_{k})$:
\bea
{\rm Skew}_{k}X(\gamma_{k, 1}, \cdots , \gamma_{k, \nu_{k}}):=
\sum_{\sigma \in S_{\nu_{k}}}({\rm sgn}\sigma)
X(\gamma_{k, \sigma(1)}, \cdots , \gamma_{k, \sigma(\nu_{k})}).
\label{asym}
\ena
Next we set
\bea
P_{J}(\{e^{\frac{2\pi i}{p}\gamma_{j, m}}\}| 
      e^{\frac{2\pi i}{p}\beta_{1}}, \cdots , e^{\frac{2\pi i}{p}\beta_{n}}):=
\prod_{k=1}^{N-1}
P_{M_{k}^{J}}(\{e^{\frac{2\pi i}{p}\gamma_{k, m}}\} | \{e^{\frac{2\pi i}{p}\gamma_{k-1, m'}}\}),
\ena
and define the space ${\cal P}_{\nu_{1}, \cdots , \nu_{N-1}}$ by
\bea
{\cal P}_{\nu_{1}, \cdots , \nu_{N-1}}:=
\sum_{J \in {\cal Z}_{\nu_{1}, \cdots , \nu_{N-1}}} {\Bbb C} \, P_{J}.
\ena 

For $J \in {\cal Z}_{\nu_{1}, \cdots , \nu_{N-1}}$ and 
$P \in {\cal P}_{\nu_{1}, \cdots , \nu_{N-1}}$, 
we define a function $I_{J}[P]=I_{J}[P](\beta_{1}, \cdots , \beta_{n})$ by
\bea
I_{J}[P]&:=&\!\!\!\!
\left( \prod_{j=1}^{N-1}\prod_{m=1}^{\nu_{j}} \int_{C_{j}}d\gamma_{j, m} \right)
\prod_{k=1}^{N-1}
\Big( \phi(\{\gamma_{k, m}\}| \{\gamma_{k-1, m'}\}) \, \varphi(\{\gamma_{k, m}\}) \Big) 
\label{gensol} \\
&&{}\times
w_{J}^{(N)}(\{\gamma_{j, m}\}| \{\beta_{m} \}) 
\prod_{j=1}^{N-1}
\frac{\prod_{1 \le a<b \le \nu_{j}}{\rm sh}\frac{\pi i}{p}(\gamma_{j, a}-\gamma_{j, b}-\hbar)}
     {\prod_{m=1}^{\nu_{j}}\prod_{m'=1}^{\nu_{j-1}}
        (1-e^{\frac{2\pi i}{p}(\gamma_{j, m}-\gamma_{j-1, m'})})}
P(\{e^{\frac{2\pi i}{p}\gamma_{j, m}}\}),
\no
\ena
where
\bea
&&
\phi(t_{1}, \cdots , t_{l} | z_{1}, \cdots , z_{m}):=
\prod_{a=1}^{l}\prod_{j=1}^{m}
\frac{\Gamma\left( \frac{t_{a}-z_{j}-\hbar}{p} \right)}
     {\Gamma\left( \frac{t_{a}-z_{j}}{p} \right)}, \label{defphi} \\
&&
\varphi(t_{1}, \cdots , t_{l}):= \!\!\!\! \prod_{1 \le a<b \le l} \!\!\!\!
\textstyle 
\Gamma\left( \frac{t_{a}-t_{b}+\hbar}{p} \right)
\Gamma\left( \frac{t_{b}-t_{a}+\hbar}{p} \right). 
\ena
The contour $C_{j}$ for $\gamma_{j, m}, \, (1 \le m \le \nu_{j})$ 
is a deformation of the real axis $(-\infty, \infty)$ such that the poles at
\bea
\gamma_{j-1, m'}+\hbar-p{\Bbb Z}_{\ge 0}, \, (1 \le m' \le \nu_{j-1}), 
\quad \gamma_{j, a}-\hbar-p{\Bbb Z}_{\ge 0}, \, (a \not= m) 
\label{pole1}
\ena
are above $C_{j}$ and the poles at 
\bea
\gamma_{j-1, m'}+p{\Bbb Z}_{\ge 0}, \, (1 \le m' \le \nu_{j-1}), 
\quad \gamma_{j, a}+\hbar+p{\Bbb Z}_{\ge 0}, \, (a \not= m) 
\label{pole2}
\ena
are below $C_{j}$. 
These conditions are not compatible if all the poles really exist. 
However, we can define $I_{J}[P_{J'}]$ for each 
$P_{J'} \in {\cal P}_{\nu_{1}, \cdots , \nu_{N-1}}$ 
because $P_{J'}$ has zeroes at some points of (\ref{pole1}) and (\ref{pole2}), 
and we can deform the real axis such that the conditions above are satisfied 
for the actual poles of the integrand of (\ref{gensol}).
Then we define $I_{J}[P]$ for 
$P \in {\cal P}_{\nu_{1}, \cdots , \nu_{N-1}}=
 \sum_{J' \in {\cal Z}_{\nu_{1}, \cdots , \nu_{N-1}}}{\Bbb C} \, \, P_{J'}$ 
as a linear combination of $I_{J}[P_{J'}]$ 
(See \cite{MTT} for details). 

Set
\bea
\psi_{P}(\beta_{1}, \cdots , \beta_{n}):=
\sum_{J \in {\cal Z}_{\nu_{1}, \cdots , \nu_{N-1}}}
I_{J}[P](\beta_{1}, \cdots , \beta_{n})v_{J}
\label{defgensol}
\ena

\begin{thm} \label{gensolth}
If $\nu_{1}, \cdots , \nu_{N-1}$ satisfy
\bea
\nu_{j-1}+\nu_{j+1} \ge 2\nu_{j}, \quad {\rm for \,\,  all} \,\,  j=1, \cdots , N-1, 
\label{condnu}
\ena
then the integral (\ref{gensol}) converges and 
$\psi_{P}$ is a solution of the qKZ equation (\ref{qKZ}) 
satisfying the highest weight condition (\ref{hw}).
\end{thm}

\rem
In the case of $N=2$, (\ref{defgensol}) is nothing but the integral formula 
for solutions of the $sl_{2}$ qKZ equation at level zero constructed in \cite{NPT}.
\newline
 
\proof
The convergence of the integral (\ref{gensol}) under the condition (\ref{condnu}) 
can be proved in a similar way to the proof of Proposition 2 in \cite {MT}. 

Set
\bea
R(\beta)v_{\epsilon_{1}}\otimes v_{\epsilon_{2}}=
\sum_{\epsilon_{1}', \epsilon_{2}'}
R(\beta)_{\epsilon_{1}', \epsilon_{2}'}^{\epsilon_{1}, \epsilon_{2}}
v_{\epsilon_{1}'} \otimes v_{\epsilon_{2}'}. 
\ena
For $J \in {\cal Z}_{\nu_{1}, \cdots , \nu_{N-1}}$, 
we abbreviate $w_{J_{1}, \cdots , J_{n}}^{(N)}(\{\gamma_{j, m}\}|\beta_{1}, \cdots , \beta_{n})$ to
$w_{J_{1}, \cdots , J_{n}}(\beta_{1}, \cdots , \beta_{n})$, 
and we write down dependence on $\beta_{1}, \cdots , \beta_{n}$ 
of the integrand $w_{J}^{(N)}$ and $P$ in $I_{J}[P]$ as follows:
\bea
I_{J}[P]=
I(w_{J_{1}, \cdots , J_{n}}(\beta_{1}, \cdots , \beta_{n}), 
  P(\beta_{1}, \cdots , \beta_{n})).
\ena
Then we can show the following formulae in the same way as 
the proof of Lemma 1 and Lemma 3 in \cite{MT}:
\bea
&&
w_{J_{1}, \cdots , J_{k+1}, J_{k}, \cdots , J_{n}}
(\beta_{1}, \cdots , \beta_{k+1}, \beta_{k}, \cdots , \beta_{n}) \no \\
&& {}=
\sum_{J_{k}', J_{k+1}'}R(\beta_{k}-\beta_{k+1})_{J_{k}, J_{k+1}}^{J_{k}', J_{k+1}'}
w_{J_{1}, \cdots , J_{k}', J_{k+1}', \cdots , J_{n}}
(\beta_{1}, \cdots , \beta_{k}, \beta_{k+1}, \cdots , \beta_{n}), \label{wrel1} \\
&&
I(w_{J_{n}, J_{1}, \cdots , J_{n-1}}(\beta_{n}, \beta_{1}, \cdots , \beta_{n-1}), 
 P(\beta_{1}, \cdots , \beta_{n}))|_{\beta_{n} \to \beta_{n}+p} \no \\
&& {}=
I(w_{J_{1}, \cdots , J_{n}}(\beta_{1}, \cdots , \beta_{n-1}, \beta_{n}), 
 P(\beta_{1}, \cdots , \beta_{n})). 
\label{wrel2}
\ena
It is easy to see that $\psi_{P}$ is a solution of the qKZ equation 
{}from (\ref{wrel1}) and (\ref{wrel2}). 

Let us prove that $\psi_{P}$ satisfies the highest weight condition. 
%For $M'=(M_{1}', \cdots , M_{N-1}') \in 
%{\cal I}_{\nu_{1}, \cdots , \nu_{k}-1, \cdots , \nu_{N-1}}$ 
%and $1 \le a \le \nu_{k-1}, (a \not\in M_{k}')$, 
%we define $M'+\{a\}_{k} \in {\cal I}_{\nu_{1}, \cdots , \nu_{k}, \cdots , \nu_{N-1}}$ by 
%the following condition: 
%\bea
%J^{M'+\{a\}_{k}}=(J_{1}^{M'}, \cdots , J_{r}^{M'}+1, \cdots , J_{n}^{M'}), \quad 
%{\rm where} \quad r=r_{k-1, a}^{M'}.  
%\ena
Note that
\bea
E_{k}\psi_{P}=
\sum_{J' \in {\cal Z}_{\nu_{1}, \cdots , \nu_{k}-1, \cdots , \nu_{N-1}}}
\left( \sum_{j=1 \atop J'_{j}=k-1}^{n}
 I_{J'_{1}, \cdots , J'_{j}+1, \cdots , J'_{n}}[P] \right) v_{J'}.  
\label{Eattacked}
\ena
Hence it suffices to prove that
\bea
\sum_{j=1 \atop J'_{j}=k-1}^{n}I_{J'_{1}, \cdots , J'_{j}+1, \cdots , J'_{n}}[P]=0
%\sum_{1 \le a \le \nu_{k-1} \atop a \not\in M_{k}'} I_{M'+\{a\}_{k}}[P]=0
\label{hwcond}
\ena
for $J' \in {\cal Z}_{\nu_{1}, \cdots , \nu_{k}-1, \cdots , \nu_{N-1}}$. 

First we prove (\ref{hwcond}) in the case of $N=3$. 
%The proof for the case of $N>3$ is similar. 
The proof for the highest weight condition with $k=2$, that is, $E_{2}\psi_{P}=0$, 
is similar to the proof for the case of $sl_{2}$ in \cite{NPT}. 
Let us prove the case of $N=3$ and $k=1$. 

In this proof we set $\gamma_{1, a}=:\alpha_{a}$ and $\gamma_{2, m}=:\gamma_{m}$ and
\bea
\Phi(\{\gamma_{m}\}|\{\alpha_{a}\}|\{\beta_{j}\}):=
\prod_{k=1}^{2}\left(
\phi(\{\gamma_{k, m}\}|\{\gamma_{k-1, m}\})\varphi(\{\gamma_{k, m}\}) \!\!\!\!\!\!\!\!\!\!
\prod_{1 \le m<m' \le \nu_{k}} \!\!\!\!\!\!\!
{\rm sh}\frac{\pi i}{p}(\gamma_{k, m}-\gamma_{k, m'}-\hbar)
\right)
\ena
for simplicity's sake. 
Then the following equality holds. 

\begin{lem}\label{hw1}
\bea
&&
\hbar\sum_{j=1 \atop J'_{j}=0}^{n}
 w^{(3)}_{J'_{1}, \cdots , J'_{j}+1, \cdots , J'_{n}}=
{\rm Skew}\Bigl( 
g_{M_{1}^{J'}}(\{\alpha_{a}\}_{2 \le a \le \nu_{1}}|\{\beta_{j}\})
g_{M_{2}^{J'}}(\{\gamma_{m}\}| \{\alpha_{a}\}_{2 \le a \le \nu_{1}})  \label{hw11} \\
&& \qquad \qquad {}\times 
\left(\prod_{a=2}^{\nu_{1}}(\alpha_{1}-\alpha_{a}-\hbar)
\prod_{m=1}^{\nu_{2}}\frac{\gamma_{m}-\alpha_{1}-\hbar}{\gamma_{m}-\alpha_{1}}-
\prod_{j=1}^{n}\frac{\alpha_{1}-\beta_{j}-\hbar}{\alpha_{1}-\beta_{j}}
\prod_{a=2}^{\nu_{1}}(\alpha_{1}-\alpha_{a}+\hbar)\right) \Bigr).  \no 
%&&
%\hbar\sum_{1 \le a \le \nu_{1} \atop a \not\in M_{2}}w^{(3)}_{M+\{a\}_{2}} \no \\
%&& {}=
%{\rm Skew}\Big( 
%g_{M_{1}}(\{\gamma_{m}\}|\{\beta_{j}\})
%g_{M_{2}}(\{\alpha_{m}\}_{2 \le m \le \nu_{2}}| \{\gamma_{m}\}) \label{hw12} \\
%&& \qquad \quad {}\times 
%\left( 
%\prod_{p=2}^{\nu_{2}}(\alpha_{1}-\alpha_{p}-\hbar)-
%\prod_{m=1}^{\nu_{1}}\frac{\alpha_{1}-\gamma_{m}-\hbar}{\alpha_{1}-\gamma_{m}}
%\prod_{p=2}^{\nu_{2}}(\alpha_{1}-\alpha_{p}+\hbar) \right) \Bigr). \no
\ena
Here $J'=(J'_{1}, \cdots , J'_{n}) \in {\cal Z}_{\nu_{1}-1, \nu_{2}}$ and
%$M \in {\cal I}_{\nu_{1}, \nu_{2}-1}$ in (\ref{hw12}). 
${\rm Skew}$ is the skew-symmetrizations 
with respect to $\gamma_{1}, \cdots , \gamma_{\nu_{1}}$ 
and $\alpha_{1}, \cdots , \alpha_{\nu_{2}}$.
\end{lem}

This lemma is proved in Section \ref{app2}.

{}From Lemma \ref{hw1}, we have 
\bea 
&& 
\hbar \sum_{j=1 \atop J'_{j}=0}^{n} 
I_{J'_{1}, \cdots , J'_{j}+1, \cdots , J'_{n}}[P]= 
\left( \prod_{a=1}^{\nu_{1}} \int_{C_{1}} d\alpha_{a} 
       \prod_{m=1}^{\nu_{2}} \int_{C_{2}} d\gamma_{m} \right) 
\Phi(\{\gamma_{m}\} | \{\alpha_{a}\} | \{\beta_{j}\}) \label{hwpf11} \\
&& \qquad \qquad \qquad {}\times 
({\rm the \, rhs \, of} \, (\ref{hw11}))
\frac{P(\{e^{\frac{2\pi i}{p}\gamma_{m}}, e^{\frac{2\pi i}{p}\alpha_{a}}\})}
     {\prod_{a=1}^{\nu_{1}}
      \left( \prod_{m=1}^{\nu_{2}}(1-e^{\frac{2\pi i}{p}(\gamma_{m}-\alpha_{a})})
      \prod_{j=1}^{n}(1-e^{\frac{2\pi i}{p}(\alpha_{a}-\beta_{j})}) \right)}. 
\no
\ena 
Note that 
\bea
\frac{\Phi |_{\alpha_{1} \to \alpha_{1}+p}}{\Phi }=
\prod_{j=1}^{n}\frac{\alpha_{1}-\beta_{j}-\hbar}{\alpha_{1}-\beta_{j}}
\prod_{m=1}^{\nu_{2}}\frac{\gamma_{m}-\alpha_{1}-p}{\gamma_{m}-\alpha_{1}-\hbar-p}
\prod_{a=2}^{\nu_{1}}\frac{\alpha_{1}-\alpha_{a}+\hbar}{\alpha_{1}-\alpha_{a}-\hbar+p}.
\ena
Hence the integration in (\ref{hwpf11}) with respect to $\alpha_{1}$ 
is given by
\bea
&&
(\int_{C_{1}}-\int_{C_{1}+p})d\alpha_{1}\Phi(\{\gamma_{m}\}|\alpha_{1}|\{\beta_{j}\})
\prod_{a=2}^{\nu_{1}}(\alpha_{1}-\alpha_{a}-\hbar)
\prod_{m=1}^{\nu_{2}}\frac{\gamma_{m}-\alpha_{1}-\hbar}{\gamma_{m}-\alpha_{1}} \no \\
&& \qquad {}\times 
\frac{P(\{e^{\frac{2\pi i}{p}\gamma_{j, m}}\})}
     {\prod_{m=1}^{\nu_{2}}(1-e^{\frac{2\pi i}{p}(\gamma_{m}-\alpha_{1})})
      \prod_{j=1}^{n}(1-e^{\frac{2\pi i}{p}(\alpha_{1}-\beta_{j})})}.
\label{hwdeform}
\ena
It is easy to see that 
the contour $C_{1}$ can be deformed to $C_{1}+p$ 
without crossing the poles of the integrand in (\ref{hwdeform}).
Hence (\ref{hwdeform}) equals zero, 
and this completes the proof for the case of $N=3$ and $k=1$.

The proof for the cases $N>3$ is similar. 
The case of $k=N-1$ can be proved in a similar manner to 
the proof for the $sl_{2}$ case in \cite{NPT}, 
and the other case can be proved from Lemma \ref{hw1} and the calculation (\ref{hwdeform}) 
for $\beta_{j}=\gamma_{k-1, j}, \alpha_{a}=\gamma_{k, a}$ and $\gamma_{m}=\gamma_{k+1, m}$. 
\qed

Now let us see that the formula (\ref{gensol}) contains as the integrand 
the integral representation of solutions to the $sl_{N-1}$ qKZ equation at level one. 
%investigate the structure of the integral formula (\ref{gensol}). 
Set $\alpha_{a}:=\gamma_{1, a}$ and $\ell :=\nu_{1}$.
Let us consider $I_{J}[P]$, 
where $P \in {\cal P}_{\nu_{1}, \cdots , \nu_{N-1}}$ is in the following form:
\bea
P(\{e^{\frac{2\pi i}{p}\alpha_{a}}\}, \{e^{\frac{2\pi i}{p}\gamma_{k, m}}\}_{k \ge 2})=
P_{1}(\{e^{\frac{2\pi i}{p}\alpha_{a}}\} | \{e^{\frac{2\pi i}{p}\beta_{j}}\} )
\bar{P}(\{e^{\frac{2\pi i}{p}\gamma_{k, m}}\}_{k \ge 2}| \{e^{\frac{2\pi i}{p}\alpha_{a}}\}).
\ena
We write down the skew-symmetrization in $w_{J}^{(N)}$ with respect to $\alpha_{a}$'s. 
Then we get
\bea
&& \!\!\!\!
I_{J}[P]=\left( \prod_{a=1}^{\ell} \int_{C_{1}} d\alpha_{a} \right)
\phi(\{\alpha_{a}\}|\{\beta_{j}\})\varphi(\{\alpha_{a}\}) \label{source} \\
&& \!\!\!\!\!\!\!\!\!\! {}\times
\sum_{\sigma \in S_{\ell}} ({\rm sgn}\sigma) 
g_{M_{1}^{J}}(\{\alpha_{\sigma(a)}\}| \{\beta_{j}\})
I_{\bar{J}}^{(\sigma)}[\bar{P}](\{ \alpha_{a} \})
\frac{\prod_{1 \le a<b \le \ell}{\rm sh}\frac{\pi i}{p}(\alpha_{a}-\alpha_{b}-\hbar)}
     {\prod_{a=1}^{\ell}\prod_{j=1}^{n}(1-e^{\frac{2\pi i}{p}(\alpha_{a}-\beta_{j})})}
P_{1}(\{e^{\frac{2\pi i}{p}\alpha_{a}}\}| \{e^{\frac{2\pi i}{p}\beta_{j}} \}). \no 
\ena
The notation in the above formula is as follows. 
For $J=(J_{1}, \cdots , J_{n}) \in {\cal Z}_{\ell, \nu_{2}, \cdots , \nu_{N-1}}$, 
we define 
$\bar{J}:=(\bar{J}_{1}, \cdots ,\bar{J}_{\ell}) \in {\cal Z}_{\nu_{2}, \cdots , \nu_{N-1}}$ 
by 
\bea 
\bar{J}_{a}:=J_{r_{1, a}^{J}}-1. 
\label{defJbar} 
\ena 
In other words, $\bar{J}$ is obtained by picking up non-zero components 
of $J=(J_{1}, \cdots , J_{n})$ and adding $(-1)$ to each component. 
For example, for $J=(1, 2, 0, 1, 0, 2) \in {\cal Z}_{4, 2}$, 
we have $\bar{J}=(0, 1, 0, 1) \in {\cal Z}_{2}$. 
{}For $\bar{J}=(\bar{J}_{1}, \cdots , \bar{J}_{\ell})
 \in {\cal Z}_{\nu_{2}, \cdots , \nu_{N-1}}$, 
we define sets $M_{k}^{\bar{J}}, (k=1, \cdots , N-2)$ in the same way as (\ref{defMJ}) 
and the function $w_{\bar{J}}^{(N-1)}$ from $M_{k}^{\bar{J}}$'s. 
Then we set 
\bea
&& \!\!\!\!
I_{\bar{J}}^{(\sigma)}[\bar{P}](\alpha_{1}, \cdots , \alpha_{\ell}):=
\left( \prod_{k=2}^{N-1}\prod_{m=1}^{\nu_{k}} \int_{C_{k}}d\gamma_{k, m} \right)
\prod_{k=2}^{N-1}
\Big( \phi(\{\gamma_{k, m}\}| \{\gamma_{k-1, m'}\}) \, \varphi(\{\gamma_{k, m}\}) \Big)  \\
&&\!\!\!\!\!\!\!\!\!\! {}\times
w_{\bar{J}}^{(N-1)}(\{\gamma_{k, m}\}_{k \ge 2} | \{ \alpha_{\sigma(a)}\}) 
\prod_{k=2}^{N-1}
\frac{\prod_{1 \le a<b \le \nu_{k}}{\rm sh}\frac{\pi i}{p}(\gamma_{k, a}-\gamma_{k, b}-\hbar)}
     {\prod_{m=1}^{\nu_{k}}\prod_{m'=1}^{\nu_{k-1}}
      (1-e^{\frac{2\pi i}{p}(\gamma_{k, m}-\gamma_{k-1, m'})})}
\bar{P}(\{e^{\frac{2\pi i}{p}\gamma_{k, m}}\}_{k \ge 2}| \{e^{\frac{2\pi i}{p}\alpha_{a}}\}). \no
\ena

%For $\bar{M} \in {\cal I}_{\nu_{2}, \cdots , \nu_{N-1}}$, 
%we can define a $\ell$-tuple $(J_{1}^{\bar{M}}, \cdots , J_{\ell}^{\bar{M}})$ 
%in the same way as before by (\ref{defJ}). 
%Then we define a vector $v_{\bar{M}}$ by
Set 
\bea
v_{\bar{J}}:=v_{\bar{J}_{1}} \otimes \cdots \otimes v_{\bar{J}_{\ell}}
\in (V_{N-1})^{\otimes \ell}.
\ena
and 
\bea
\bar{\psi}_{\bar{P}}(\alpha_{1}, \cdots , \alpha_{\ell}):=
\sum_{\bar{J} \in {\cal Z}_{\nu_{2}, \cdots , \nu_{N-1}}}
I_{\bar{J}}^{({\rm id})}[\bar{P}](\alpha_{1}, \cdots , \alpha_{\ell})
v_{\bar{J}}.
\ena
%{}From the proof of Theorem \ref{gensolth}, 
Recall that $p=N\hbar$. 
Then we see that $\bar{\psi}_{\bar{P}}$ is a solution of the qKZ equation 
associated with $sl_{N-1}$ at level $-(N-1)+p/\hbar=1$ satisfying the highest weight condition. 
In the next section we construct a special solution to this $sl_{n-1}$ qKZ eqaution 
at level one without any integration.

\section{Special solution at level one}

In the following we fix a positive integer $m$ and assume that
\bea
%\hbar=-\frac{2\pi i}{N}, \quad p=-2\pi i, 
%\quad {\rm and} \quad 
\nu_{j}=(N-j)m, \,\, (0 \le j \le N),
\ena
that is, we consider singlet solutions in $(V_{N})^{\otimes Nm}$ at level zero. 
%Then $\bar{\psi}_{\bar{P}}$ is a singlet solution of the $sl_{N-1}$ qKZ equation at level one. 

Let us construct a special solution of the qKZ equation 
associated with $sl_{N-1}$ at level one. 
Note that $\ell =\nu_{1}=(N-1)m$. 

\begin{lem}
There exists a set of rational functions 
$\{H_{\epsilon_{1}, \cdots , \epsilon_{\ell}}(\alpha_{1}, \cdots , \alpha_{\ell})\}_{
 (\epsilon_{1}, \cdots , \epsilon_{\ell}) \in {\cal Z}_{(N-2)m, \cdots 2m, m}} $ 
uniquely determined by the following conditions:

%The indicies $\epsilon_{a}$ run satisfying
%\bea
%0 \le \epsilon_{a} \le N-2, \quad \# \{a|\epsilon_{a}=k\}=m \quad (k=0, \cdots , N-2), 
%\ena
%and the following relations hold:
\bea
H_{\cdots, \epsilon_{p+1}, \epsilon_{p}, \cdots }
(\cdots , \alpha_{p+1}, \alpha_{p}, \cdots )\!\!\!&=&\!\!\!
\frac{\alpha_{p}-\alpha_{p+1}}{\alpha_{p}-\alpha_{p+1}+\hbar}
H_{\cdots, \epsilon_{p}, \epsilon_{p+1}, \cdots }
(\cdots, \alpha_{p}, \alpha_{p+1}, \cdots) \no  \\
&& {}+
\frac{\hbar}{\alpha_{p}-\alpha_{p+1}+\hbar}
H_{\cdots, \epsilon_{p+1}, \epsilon_{p}, \cdots }
(\cdots, \alpha_{p}, \alpha_{p+1}, \cdots ),  \label{rel1} \\
\!\!\! H_{\epsilon_{1}, \cdots , \epsilon_{\ell}}
(\alpha_{1}, \cdots , \alpha_{\ell-1}, \alpha_{\ell}-N\hbar)\!\!\!&=&\!\!\!
\prod_{a=1}^{\ell -1}
\frac{\alpha_{a}-\alpha_{\ell}+\hbar}{\alpha_{a}-\alpha_{\ell}+(N-1)\hbar}
H_{\epsilon_{\ell}, \epsilon_{1}, \cdots , \epsilon_{\ell -1}}
(\alpha_{\ell}, \alpha_{1}, \cdots , \alpha_{\ell -1}).
\label{rel2}
\ena
Moreover, 
\bea
H_{\scriptsize{\underbrace{0 \cdots 0}_{m}\underbrace{1 \cdots 1}_{m} \cdots 
   \underbrace{(N-2) \cdots (N-2)}_{m}}}(\alpha_{1}, \cdots , \alpha_{\ell})=
\prod_{a, b \atop{(\epsilon_{a}<\epsilon_{b})}}
\frac{1}{\alpha_{a}-\alpha_{b}-\hbar}.
\label{rel3}
\ena
\end{lem}

\rem
{}From (\ref{rel1}) and (\ref{rel3}), it is easy to see that 
\bea
\prod_{1 \le a<b \le \ell} \!\!\! (\alpha_{a}-\alpha_{b}-\hbar) \,
H_{\epsilon_{1}, \cdots , \epsilon_{\ell}}(\alpha_{1}, \cdots , \alpha_{\ell}) 
\label{Hpol}
\ena
is a polynomial in $\alpha_{1}, \cdots , \alpha_{\ell}$ 
of order less than or equal to $m-1$ in each variable.
\newline

In the following we construct $\{H_{\epsilon_{1}, \cdots , \epsilon_{\ell}}\}$ 
{}from a solution of the qKZ equation associated with $U_{q}(\widehat{sl}_{N-1})$ with $0<q<1$ 
at level one. 

Let $\Lambda_{i}, \, (0 \le i \le N-2)$ be the fundamental weights of $\widehat{sl}_{N-1}$ 
and $V(\Lambda_{i})$ the level one irreducible highest weight $U_{q}(\widehat{sl}_{N-1})$ module 
with the highest weight $\Lambda_{i}$ and the highest weight vector $| \Lambda_{i} \rangle$. 
Then there exist the type I intertwining operators $\tilde{\Phi}^{(i)}(z)$ \cite{DO}:
\bea
\tilde{\Phi}^{(i)}(z) : V(\Lambda_{i+1}) \longrightarrow V(\Lambda_{i}) \otimes V_{z}, \qquad
\tilde{\Phi}^{(i)}(z)| \Lambda_{i+1} \rangle =
| \Lambda_{i} \rangle \otimes v_{i}+ \cdots , 
\ena
where $V_{z}$ is the homogeneous evaluation module 
associated with the vector representation $V_{N-1}$. 

Set 
\bea
\Phi^{(i)}(z):=z^{\Delta_{i}-\Delta_{i+1}}\tilde{\Phi}^{(i)}(z), \qquad
\Delta_{i}:=\frac{(\Lambda_{i}+2\rho | \Lambda_{i})}{2N}, 
\ena
and
\bea
G(z_{1}, \cdots , z_{\ell}):=
\langle \Lambda_{i} | \Phi(z_{1}) \cdots \Phi(z_{\ell}) | \Lambda_{i} \rangle  
\in V_{z_{1}} \otimes \cdots \otimes V_{z_{\ell}}. 
\ena
Then $G$ satisfies the (trigonometric) qKZ equation at level one \cite{FR}:
\bea
&&
G(z_{1}, \cdots , q^{2N}z_{j}, \cdots , z_{\ell})=
R^{q}_{j, j-1}(q^{2N}z_{j}/z_{j-1}) \cdots R^{q}_{j, 1}(q^{2N}z_{j}/z_{1}) 
\cdot (q^{-2\Lambda_{i}-2\rho})_{j} \no \\
&& \qquad \qquad {}\times 
\left( R^{q}_{\ell, j}(z_{\ell}/z_{j}) \right)^{-1} \cdots 
\left( R^{q}_{j+1, j}(z_{j+1}/z_{j}) \right)^{-1} 
G(z_{1}, \cdots , z_{j}, \cdots , z_{\ell}). 
\label{tqKZ}
\ena
Here $R^{q}(z)$ is the trigonometric $R$-matrix given by
\bea
R^{q}(z)=q^{\frac{1}{N-1}-1}
\frac{(q^{2}z ; q^{2(N-1)})_{\infty} (q^{2N-4}z ; q^{2(N-1)})_{\infty}}
     {(z ; q^{2(N-1)})_{\infty} (q^{2N-2}z ; q^{2(N-1)})_{\infty}}
\bar{R}^{q}(z), \quad \bar{R}^{q}(z)(v_{0} \otimes v_{0})=v_{0} \otimes v_{0}. 
\ena

Now we set
\bea
H^{q}(z_{1}, \cdots , z_{\ell}):=
\prod_{a=1}^{\ell}z^{(1-\frac{1}{N-1})a} 
\prod_{1 \le a<b \le \ell}
\frac{(q^{2(N-1)}z_{b}/z_{a}; q^{2(N-1)})_{\infty}}{(q^{2}z_{b}/z_{a}; q^{2(N-1)})_{\infty}}
G(z_{1}, \cdots , z_{\ell}). 
\ena
{}From the commutation relation 
\bea
&&
\Phi(z_{2}) \Phi(z_{1}) \\
&& {}=
\left( \frac{z_{1}}{z_{2}} \right)^{\frac{1}{N-1}-1} \!\!
\frac{(q^{2(N-1)}z_{2}/z_{1}; q^{2(N-1)})_{\infty} (q^{2}z_{1}/z_{2}; q^{2(N-1)})_{\infty}}
     {(q^{2(N-1)}z_{1}/z_{2}; q^{2(N-1)})_{\infty} (q^{2}z_{2}/z_{1}; q^{2(N-1)})_{\infty}}
\bar{R}^{q}(z_{1}/z_{2}) \Phi(z_{1}) \Phi(z_{2}), \no
\ena
we find 
\bea
P_{p, p+1}\bar{R}^{q}(z_{p}/z_{p+1})H^{q}( \cdots , z_{p}, z_{p+1}, \cdots )=
H^{q}( \cdots , z_{p+1}, z_{p}, \cdots ). 
\label{trel1}
\ena

Expand $H^{q}$ as
\bea
H^{q}(z_{1}, \cdots , z_{\ell})=
\sum_{\epsilon_{1}, \cdots , \epsilon_{\ell}}
H^{q}_{\epsilon_{1}, \cdots , \epsilon_{\ell}}(z_{1}, \cdots , z_{\ell})
v_{\epsilon_{1}} \otimes \cdots \otimes v_{\epsilon_{\ell}}. 
\ena
{}From (\ref{tqKZ}) and (\ref{trel1}), we have
\bea
H^{q}_{\epsilon_{1}, \cdots , \epsilon_{\ell}}(z_{1}, \cdots , z_{\ell-1}, q^{-2N}z_{\ell})=
q^{c_{N, \ell}}
\prod_{a=1}^{\ell -1}
\frac{z_{\ell}-q^{2}z_{a}}{z_{\ell}-q^{2(N-1)}z_{a}}
H^{q}_{\epsilon_{\ell}, \epsilon_{1}, \cdots , \epsilon_{\ell -1}}
(z_{\ell}, z_{1}, \cdots , z_{\ell -1}),
\label{trel2}
\ena
where $c_{N, \ell}$ is a certain constant. 

The extremal component is calculated in \cite{nakayashiki}:
\bea
H^{q}_{\scriptsize{\underbrace{0 \cdots 0}_{m}\underbrace{1 \cdots 1}_{m} \cdots 
       \underbrace{(N-2) \cdots (N-2)}_{m}}}\!\!\!&=&\!\!\!
\prod_{a=1}^{\ell}z_{a}^{(1-\frac{1}{N-1})(\ell+\frac{1}{2})+\frac{i}{N-1}}
\prod_{a=1}^{(i+1)m}\!\!\! z_{a}^{-1} \!\!\! 
\prod_{a<b \atop (\epsilon_{a}<\epsilon_{b})}
\frac{1}{z_{a}-q^{2}z_{b}}.
\label{trel3}
\ena

Now we consider the scaling limit of $H^{q}$ as
\bea
q=e^{\lambda \frac{\hbar}{2}}, \quad z_{a}=e^{\lambda \alpha_{a}}, \qquad \lambda \to \infty.
\ena
Then $\bar{R}(z)$ goes to the rational $R$-matrix (\ref{defR}). 

Set
\bea
H_{\epsilon_{1}, \cdots , \epsilon_{\ell}}(\alpha_{1}, \cdots , \alpha_{\ell}):=
\lim_{\lambda \to \infty}\lambda^{\frac{(\ell-m)\ell}{2}}
H^{q}_{\epsilon_{1}, \cdots , \epsilon_{\ell}}(z_{1}, \cdots , z_{\ell}). 
\ena
It is easy to see that $\{H_{\epsilon_{1}, \cdots , \epsilon_{\ell}}\}$ satisfies
(\ref{rel1}), (\ref{rel2}) and (\ref{rel3}) from 
(\ref{trel1}), (\ref{trel2}) and (\ref{trel3}), respectively. 
\newline

\rem 
We have a more explicit formula for the scaling limit of $H^{q}$.
See Proposition \ref{Hanotherformula}. 
\newline

By using $\{H_{\epsilon_{1}, \cdots , \epsilon_{\ell}}\}$, 
we have a special solution of the qKZ equation at level one. 
\begin{prop}
Set
\bea
&&
\bar{\psi}(\alpha_{1}, \cdots , \alpha_{\ell})   \label{ssol} \\
&& {}:=
\prod_{1 \le a<b \le \ell} \left( 
\frac{\Gamma( \frac{\alpha_{a}-\alpha_{b}-\hbar}{p})}
     {\Gamma( \frac{\alpha_{a}-\alpha_{b}+\hbar}{p})}
(\alpha_{a}-\alpha_{b}-\hbar) \right)
\sum_{\epsilon_{1}, \cdots , \epsilon_{\ell}}
H_{\epsilon_{1}, \cdots , \epsilon_{\ell}}(\alpha_{1}, \cdots , \alpha_{\ell})
v_{\epsilon_{1}} \otimes \cdots \otimes v_{\epsilon_{\ell}}. \no
\ena
Then $\bar{\psi}$ is a solution of the qKZ equation (\ref{qKZ}) 
associated with $sl_{N-1}$ at level one 
satisfying the highest weight condition. 
\end{prop}

\proof
It is easy to see that $\bar{\psi}$ is a solution of the qKZ equation 
{}from (\ref{rel1}) and (\ref{rel2}). 
The highest weight condition is proved in Section \ref{app1}. 
\qed

\section{Simplified integral formula at level zero}

By using the special solution (\ref{ssol}), 
we can find a simpler integral formula in the case of level zero as follows. 

Suppose that there exists $\bar{P} \in {\cal P}_{(N-2)m, \cdots , 2m, m}$ 
such that
\bea
I_{\bar{J}}^{({\rm id})}[\bar{P}](\alpha_{1}, \cdots , \alpha_{\ell})=
\prod_{1 \le a<b \le \ell} \left( 
\frac{\Gamma( \frac{\alpha_{a}-\alpha_{b}-\hbar}{p})}
     {\Gamma( \frac{\alpha_{a}-\alpha_{b}+\hbar}{p})}
(\alpha_{a}-\alpha_{b}-\hbar) \right)
H_{\bar{J}}(\alpha_{1}, \cdots , \alpha_{\ell})
\label{monoassume}
\ena
for all $\bar{J} \in {\cal Z}_{(N-2)m, \cdots , 2m, m}$. 
%Here we set $J^{\bar{M}}:=(J_{1}^{\bar{M}}, \cdots , J_{\ell}^{\bar{M}})$. 
In the following we omit $\bar{P}$ and 
abbreviate $I_{\bar{M}}^{(\sigma)}[\bar{P}]$ to $I_{\bar{M}}^{(\sigma)}$.
In order to rewrite (\ref{source}) using $H_{J^{\bar{M}}}$, 
we need a formula for $I_{\bar{M}}^{(\sigma)}$ for any $\sigma \in S_{\ell}$. 
This formula is given by 
\bea
I_{\bar{J}}^{(\sigma)}=
\prod_{1 \le a<b \le \ell} \left( 
\frac{\Gamma( \frac{\alpha_{a}-\alpha_{b}-\hbar}{p})}
     {\Gamma( \frac{\alpha_{a}-\alpha_{b}+\hbar}{p})}
(\alpha_{a}-\alpha_{b}-\hbar) \right)
T_{\sigma}(H_{\bar{J}}), \quad {\rm for \quad all} \quad \sigma \in S_{\ell},
\label{mono}
\ena
where $T_{\sigma}$ is the permutation of variables defined by 
\bea
T_{\sigma}(X)(\alpha_{1}, \cdots , \alpha_{\ell}):=
X(\alpha_{\sigma(1)}, \cdots , \alpha_{\sigma(\ell)})
\ena
for a function $X$.

%In this proof, 
%we abbreviate $I_{\bar{M}}^{\sigma}$ and $H_{\bar{J}^{\bar{M}}}$ 
%to $I_{\bar{J}_{1}, \cdots , \bar{J}_{\ell}}^{\sigma}$ and 
%$H_{\bar{J}_{1}, \cdots , \bar{J}_{\ell}}$ respectively, 
%where $(\bar{J}_{1}, \cdots , \bar{J}_{\ell})=\bar{J}^{\bar{M}}$.  
%
We can see (\ref{mono}) by induction on the length of $\sigma$. 
In the case of $\sigma={\rm id}$, (\ref{mono}) is nothing but (\ref{monoassume}). 
Assume that (\ref{mono}) holds for $\sigma \in S_{\ell}$.
Set $\tau :=(k, k+1) \in S_{\ell}$. 
{}From (\ref{wrel1}), we find
\bea
\frac{\alpha_{\sigma(k)}-\alpha_{\sigma(k+1)}+\hbar}{\alpha_{\sigma(k)}-\alpha_{\sigma(k+1)}}
I_{\cdots , \bar{J}_{k}, \bar{J}_{k+1}, \cdots }^{(\sigma\tau)}=
I_{\cdots , \bar{J}_{k+1}, \bar{J}_{k}, \cdots }^{(\sigma)}+
\frac{\hbar}{\alpha_{\sigma(k)}-\alpha_{\sigma(k+1)}}
I_{\cdots , \bar{J}_{k}, \bar{J}_{k+1}, \cdots }^{(\sigma)}.
\ena
Then we have
\bea
&&
\left\{ \prod_{1 \le a<b \le \ell} 
\left( 
\frac{\Gamma( \frac{\alpha_{a}-\alpha_{b}-\hbar}{p})}
     {\Gamma( \frac{\alpha_{a}-\alpha_{b}+\hbar}{p})}
(\alpha_{a}-\alpha_{b}-\hbar) \right) \right\}^{-1} 
\left( 
I_{\cdots , \bar{J}_{k+1}, \bar{J}_{k}, \cdots }^{(\sigma)}+
\frac{\hbar}{\alpha_{\sigma(k)}-\alpha_{\sigma(k+1)}}
I_{\cdots , \bar{J}_{k}, \bar{J}_{k+1}, \cdots }^{(\sigma)} 
\right) \no \\
&& {}=
T_{\sigma}\left( 
H_{\cdots , \bar{J}_{k+1}, \bar{J}_{k}, \cdots }+
\frac{\hbar}{\alpha_{k}-\alpha_{k+1}}
H_{\cdots , \bar{J}_{k}, \bar{J}_{k+1}, \cdots } \right) \quad {\rm from} \quad 
(\ref{mono}) \no \\
&& {}= 
%\sigma \left[\left[
%\frac{\alpha_{k}-\alpha_{k+1}+\hbar}{\alpha_{k}-\alpha_{k+1}}
%H_{\cdots , \bar{J}_{k}, \bar{J}_{k+1}, \cdots }(\cdots , \alpha_{k+1}, \alpha_{k}, \cdots )
%\right]\right] \no \\
%&& {}=
T_{\sigma} \left(
\frac{\alpha_{k}-\alpha_{k+1}+\hbar}{\alpha_{k}-\alpha_{k+1}}
T_{\tau}( H_{\cdots , \bar{J}_{k}, \bar{J}_{k+1}, \cdots})
\right) \quad {\rm from} \quad (\ref{rel1}) \no \\
&& {}= 
\frac{\alpha_{\sigma(k)}-\alpha_{\sigma(k+1)}+\hbar}{\alpha_{\sigma(k)}-\alpha_{\sigma(k+1)}}
T_{\sigma\tau}\left( H_{\cdots , \bar{J}_{k}, \bar{J}_{k+1}, \cdots} \right).
\ena
Hence we have (\ref{mono}) for $\sigma\tau \in S_{\ell}$.

By substituting (\ref{source}) with (\ref{mono}), 
we get
\bea
&&
I_{J}[P]=\left( -p \pi i \right)^{\frac{\ell(\ell-1)}{2}}
\left( \prod_{a=1}^{\ell} \int_{C_{1}} d\alpha_{a} \right)
\phi(\{\alpha_{a}\}|\{\beta_{j}\})
\frac{P_{1}(\{e^{\frac{2\pi i}{p}\alpha_{a}}\}| \{e^{\frac{2\pi i}{p}\beta_{j}} \})}
     {\prod_{a=1}^{\ell}\prod_{j=1}^{n}(1-e^{\frac{2\pi i}{p}(\alpha_{a}-\beta_{j})})} \no \\
&& \qquad \qquad {}\times
\sum_{\sigma \in S_{\ell}} ({\rm sgn}\sigma)
T_{\sigma}(g_{M_{1}^{J}}H_{J^{\bar{M}}})
(\alpha_{1}, \cdots , \alpha_{\ell}).
\label{simplerep1} 
\ena
Here we used
\bea
\Gamma( \frac{x-\hbar}{p} )
\Gamma( \frac{-x+\hbar}{p} ) 
(x-\hbar){\rm sh}\frac{\pi i}{p}(x-\hbar)=-p\pi i 
\ena
and note that the function $\varphi(\{\alpha_{a}\})$ is canceled out. 
In this way we find a simplified representation (\ref{simplerep1}) 
of the integral (\ref{gensol}). 

Now let us prove that the formula (\ref{simplerep1}) gives really 
an integral formula for solutions of the $sl_{N}$ qKZ equation at level zero. 
First we set
\bea
w_{J}(\{\alpha_{a}\}| \{\beta_{j}\})\!\!\!&:=&\!\!\!
\sum_{\sigma \in S_{\ell}} ({\rm sgn}\sigma) 
T_{\sigma}(g_{M_{1}^{J}}H_{\bar{J}})(\alpha_{1}, \cdots , \alpha_{\ell}) \no \\
\!\!\!&=&\!\!\!
{\rm Skew}( g_{M_{1}^{J}}(\{\alpha_{a}\}| \{\beta_{j}\}) H_{\bar{J}}(\{\alpha_{a}\})),
\label{defw} 
\ena
where ${\rm Skew}$ is the skew-symmetrization 
with respect to $\alpha_{1}, \cdots , \alpha_{\ell}$.
Note that $w_{J}$ is a rational function of $\alpha_{1}, \cdots , \alpha_{\ell}$ 
with at most simple poles at points $\beta_{1}, \cdots , \beta_{n}$ from Remark (\ref{Hpol}).

Next we consider the part of $P_{1}$ in (\ref{simplerep1}). 
Let us define the space of ``deformed cycles'' as follows. 
Let ${\cal C}_{n}$ be the space of $p$-periodic entire functions 
of $\beta_{1}, \cdots , \beta_{n}$. 
We denote by $\widehat{\cal P}_{n}^{\otimes \ell}$ 
the space of polynomials in 
$e^{\frac{2\pi i}{p}\alpha_{1}}, \cdots , e^{\frac{2\pi i}{p}\alpha_{\ell}}$ of 
order less than or equal to $n$ in each vartiable $e^{\frac{2\pi i}{p}\alpha_{a}}$ 
with the coefficients in ${\cal C}_{n}$. 
Then we set
\bea
\widehat{\cal F}_{q}^{\otimes \ell}:=
\left\{ \frac{ P(\{e^{\frac{2\pi i}{p}\alpha_{a}}\})}
             { \prod_{a=1}^{\ell}\prod_{j=1}^{n}
               (1-e^{\frac{2\pi i}{p}(\alpha_{a}-\beta_{j})})} ; 
        P \in \widehat{\cal P}_{n}^{\otimes \ell} \right\}.
\ena
We call the elements of $\widehat{\cal F}_{q}^{\otimes \ell}$ deformed cycles. 

For a deformed cycle $W \in \widehat{\cal F}_{q}^{\otimes \ell}$, we set
\bea
F_{J}[W]:=
\left( \prod_{a=1}^{\ell} \int_{C} d\alpha_{a} \right)
\phi(\{\alpha_{a}\}|\{\beta_{j}\})
w_{J}(\{\alpha_{a}\}|\{\beta_{j}\})
W(\{e^{\frac{2\pi i}{p}\alpha_{a}}\}| \{\beta_{j} \})
\label{simpleint}
\ena
where the contour $C$ is a deformation of the real axis $(-\infty, \infty)$ 
such that the poles at $\beta_{j}+\hbar-p{\Bbb Z}_{\ge 0}$ are above $C$ and
the poles at $\beta_{j}+p{\Bbb Z}_{\ge 0}$ are below $C$. 
Note that, unlike the integral (\ref{gensol}), 
the integrand in (\ref{simpleint}) does not have poles at the points 
$\alpha_{a}=\alpha_{b}\pm \hbar \pm p {\Bbb Z}_{\ge 0}, ( a \not= b)$ 
because the function $\varphi(\{\alpha_{a}\})$ was canceled out. 

The function $F_{J}[W]$ gives the simplified integral formula 
of solutions to the $sl_{N}$ qKZ equation as follows. 
First we have that 

\begin{lem}
The integral (\ref{simpleint}) converges for any deformed cycle $W$.
\end{lem}

\proof
{}From the Stirling formula, we have
\bea
\phi(\alpha | \beta_{1}, \cdots , \beta_{n})=
(\alpha/p)^{-n\hbar/p}(1+o(1)), \quad {\rm as} \quad \alpha \to \pm \infty.
\label{stir}
\ena
Note that $n\hbar/p=m$. 

Recall Remark (\ref{Hpol}). 
By using (\ref{stir}), 
we see that there exist two constants $C$ and $M>0$ such that
\bea
\phi(\{\alpha_{a}\})w_{J}(\{\alpha_{a}\})W(\{e^{\frac{2\pi i}{p}\alpha_{a}}\})
\le C\prod_{a=1}^{\ell}|\alpha_{a}|^{-2} \quad 
{\rm for} \quad |\alpha_{a}|>M \,\, (a=1, \cdots , \ell). 
\ena
This completes the proof. 
\qed

\begin{thm}
For $W \in \widehat{\cal F}_{q}^{\otimes \ell}$, set
\bea
\psi_{W}(\beta_{1}, \cdots , \beta_{n}):=\!\!\!
\sum_{J \in {\cal I}_{(N-1)m, \cdots , 2m, m}}\!\!\!
F_{J}[W](\beta_{1}, \cdots , \beta_{n})v_{J}.
\label{levelzerosol}
\ena
Then $\psi_{W}$ is a solution of the qKZ equation (\ref{qKZ}) 
satisfying the highest weight condition. 
\end{thm}

\rem 
In the case of $N=2$ we have $H_{\bar{J}}=1$ for all $J$. 
Then (\ref{levelzerosol}) is the integral formula constructed in \cite{NPT}. 

{}From the definition (\ref{simpleint}), 
we have that $\psi_{{\rm Skew}W}=\ell ! \psi_{W}$. 
Hence the dimension of the space spanned by the solutions (\ref{levelzerosol}) 
is at most that of $\bigwedge^{\ell}\widehat{\cal F}_{q}$, 
where $\bigwedge^{\ell}\widehat{\cal F}_{q}$ is the subspace of deformed cycles 
skew-symmetric with respect to $\alpha_{1}, \cdots , \alpha_{\ell}$. 
In the case of $N>2$ and $m>1$, 
the dimension of $\bigwedge^{\ell}\widehat{\cal F}_{q}$ is much less than 
that of the subspace of singlet vectors in $(V_{N})^{\otimes n}$ 
(This can be shown by a similar argument to Discussions in \cite{thetaformula}). 
Therefore the space of solutions given by the simplified integral formula is 
quite smaller than the space of singlet vectors. 
\newline

\proof 
We abbreviate 
$w_{J_{1}, \cdots , J_{n}}(\alpha_{1}, \cdots , \alpha_{\ell} |\beta_{1}, \cdots , \beta_{n})$ to 
$w_{J_{1}, \cdots , J_{n}}(\beta_{1}, \cdots , \beta_{n})$.
%and 
%write down dependence on $\beta_{1}, \cdots , \beta_{n}$ in $F_{M}[W]$ as follows:
%\bea
%F_{M}[W]=F(w_{J_{1}^{M}, \cdots , J_{n}^{M}}(\beta_{1}, \cdots , \beta_{n})|
%           W(\beta_{1}, \cdots , \beta_{n})).
%\ena
In order to see that $\psi_{W}$ is a solution, 
it suffices to prove (\ref{wrel1}) and (\ref{wrel2}) for $w_{J}$ and $F_{J}[W]$
as in the proof of Theorem \ref{gensolth}.

We can prove (\ref{wrel2}) for $F_{J}[W]$ 
in a similar way to the proof of Lemma 3 in \cite{MT} 
by using (\ref{rel2}). 
Here let us prove (\ref{wrel1}). 
If $J_{k}=0$ or $J_{k+1}=0$, it is easy to see (\ref{wrel1}) 
in the same way as the proof in the case of $sl_{2}$ (see \cite{NPT}). 
Here we consider the case of $J_{k}>0$ and $J_{k+1}>0$. 

Let $\alpha_{a}$ be  
the integral variable attached to 
the $k$-th component of $(V_{N})^{\otimes n}$, that is, $r_{1, a}^{J}=k$. 
For two functions $f_{1}$ and $f_{2}$ we write $f_{1} \sim f_{2}$ 
if $f_{1}-f_{2}$ is symmetric with respect to $\alpha_{a}$ and $\alpha_{a+1}$. 
We use the following abbreviation:
\bea
H_{\bar{J}_{1}, \cdots , \bar{J}_{a}, \bar{J}_{a+1}, \cdots , \bar{J}_{\ell}}
 (\alpha_{1}, \cdots , \alpha_{a}, \alpha_{a+1}, \cdots , \alpha_{\ell})=
H_{\bar{J}_{a}, \bar{J}_{a+1}}(\alpha_{a}, \alpha_{a+1}).
\ena

The rhs of (\ref{wrel1}) for $w_{J}$ in (\ref{simpleint}) is the skew-symmetrization of 
\bea
&&
Q(\alpha_{a}, \alpha_{a+1})
\frac{1}{\alpha_{a}-\beta_{k}}\frac{1}{\alpha_{a+1}-\beta_{k+1}}
\frac{\alpha_{a+1}-\beta_{k}-\hbar}{\alpha_{a+1}-\beta_{k}}
(\alpha_{a}-\alpha_{a+1}-\hbar) \label{q1} \\
&& \!\!\! {}\times
\left\{
\frac{\beta_{k}-\beta_{k+1}}{\beta_{k}-\beta_{k+1}+\hbar}
 H_{\bar{J}_{a}, \bar{J}_{a+1}}(\alpha_{a}, \alpha_{a+1})+
\frac{\hbar}{\beta_{k}-\beta_{k+1}+\hbar}
 H_{\bar{J}_{a+1}, \bar{J}_{a}}(\alpha_{a}, \alpha_{a+1}) 
\right\}, \no
\ena
where $Q(\alpha_{a}, \alpha_{a+1})$ is a certain symmetric function 
with respect to $\alpha_{a}$ and $\alpha_{a+1}$. 
{}From (\ref{rel1}), we have
\bea
&&
(\ref{q1})=
Q(\alpha_{a}, \alpha_{a+1})
\frac{1}{\alpha_{a}-\beta_{k}}\frac{1}{\alpha_{a+1}-\beta_{k+1}}
\frac{\alpha_{a+1}-\beta_{k}-\hbar}{\alpha_{a+1}-\beta_{k}}
(\alpha_{a}-\alpha_{a+1}-\hbar) \no \\
&& \quad {}\times \Bigl\{
\frac{\beta_{k}-\beta_{k+1}}{\beta_{k}-\beta_{k+1}+\hbar}
 H_{\bar{J}_{a}, \bar{J}_{a+1}}(\alpha_{a}, \alpha_{a+1}) \no \\
&& \quad {}+
\frac{\hbar}{\beta_{k}-\beta_{k+1}+\hbar}
\left(
\frac{\alpha_{a}-\alpha_{a+1}+\hbar}{\alpha_{a}-\alpha_{a+1}}
 H_{\bar{J}_{a}, \bar{J}_{a+1}}(\alpha_{a+1}, \alpha_{a})-
\frac{\hbar}{\alpha_{a}-\alpha_{a+1}}
 H_{\bar{J}_{a}, \bar{J}_{a+1}}(\alpha_{a}, \alpha_{a+1}) 
\right) \Bigr\} \no \\
&& \!\!\!{}\sim Q(\alpha_{a}, \alpha_{a+1})
H_{\bar{J}_{a}, \bar{J}_{a+1}}(\alpha_{a}, \alpha_{a+1}) \no \\
&& {}\times
\Bigl\{
\frac{1}{\alpha_{a}-\beta_{k}}\frac{1}{\alpha_{a+1}-\beta_{k+1}}
\frac{\alpha_{a+1}-\beta_{k}-\hbar}{\alpha_{a+1}-\beta_{k}}
(\alpha_{a}-\alpha_{a+1}-\hbar)
\frac{\beta_{k}-\beta_{k+1}}{\beta_{k}-\beta_{k+1}+\hbar} \no \\
&& \quad {}-
\frac{\alpha_{a+1}-\alpha_{a}+\hbar}{\alpha_{a+1}-\alpha_{a}}
\frac{1}{\alpha_{a+1}-\beta_{k}}\frac{1}{\alpha_{a}-\beta_{k+1}}
\frac{\alpha_{a}-\beta_{k}-\hbar}{\alpha_{a}-\beta_{k}}
(\alpha_{a+1}-\alpha_{a}-\hbar)
\frac{\hbar}{\beta_{k}-\beta_{k+1}+\hbar} \no \\
&& \quad {}-
\frac{\hbar}{\alpha_{a}-\alpha_{a+1}}
\frac{1}{\alpha_{a}-\beta_{k}}\frac{1}{\alpha_{a+1}-\beta_{k+1}}
\frac{\alpha_{a+1}-\beta_{k}-\hbar}{\alpha_{a+1}-\beta_{k}}
(\alpha_{a}-\alpha_{a+1}-\hbar)
\frac{\hbar}{\beta_{k}-\beta_{k+1}+\hbar} \Bigr\} \no \\
&& \!\!\!{}=
Q(\alpha_{a}, \alpha_{a+1})
H_{\bar{J}_{a}, \bar{J}_{a+1}}(\alpha_{a}, \alpha_{a+1}) \no \\
&& {}\times
\frac{
(\alpha_{a}-\alpha_{a+1}-\hbar)
\left\{(\alpha_{a}-\beta_{k+1})(\alpha_{a+1}-\beta_{k})
  -\hbar(\alpha_{a+1}-\beta_{k+1})+\hbar^{2}\right\}}
{(\alpha_{a}-\beta_{k})(\alpha_{a}-\beta_{k+1})
 (\alpha_{a+1}-\beta_{k})(\alpha_{a+1}-\beta_{k+1})}.
\label{qlast}
\ena

On the other hand, the lhs of (\ref{wrel1}) is the skew-symmetrization of
\bea
&&
Q(\alpha_{a}, \alpha_{a+1})H_{\bar{J}_{a+1}, \bar{J}_{a}}(\alpha_{a+1}, \alpha_{a}) \no \\
&& {}\times 
\frac{1}{\alpha_{a}-\beta_{k+1}}\frac{1}{\alpha_{a+1}-\beta_{k}}
\frac{\alpha_{a+1}-\beta_{k+1}-\hbar}{\alpha_{a+1}-\beta_{k+1}}
(\alpha_{a}-\alpha_{a+1}-\hbar), 
\label{q2}
\ena
where $Q(\alpha_{a}, \alpha_{a+1})$ is the same function as 
$Q(\alpha_{a}, \alpha_{a+1})$ in (\ref{q1}). 
{}From (\ref{rel1}), we have 
\bea
&& 
(\ref{q2})=
Q(\alpha_{a}, \alpha_{a+1})
\frac{1}{\alpha_{a}-\beta_{k+1}}\frac{1}{\alpha_{a+1}-\beta_{k}}
\frac{\alpha_{a+1}-\beta_{k+1}-\hbar}{\alpha_{a+1}-\beta_{k+1}}
(\alpha_{a}-\alpha_{a+1}-\hbar) \no \\
&& {}\times
\left\{
\frac{\alpha_{a}-\alpha_{a+1}+\hbar}{\alpha_{a}-\alpha_{a+1}}
H_{\bar{J}_{a}, \bar{J}_{a+1}}(\alpha_{a+1}, \alpha_{a})-
\frac{\hbar}{\alpha_{a}-\alpha_{a+1}}
H_{\bar{J}_{a}, \bar{J}_{a+1}}(\alpha_{a}, \alpha_{a+1}) \right\} \no \\
&& \!\!\!{}\sim
Q(\alpha_{a}, \alpha_{a+1})
H_{\bar{J}_{a}, \bar{J}_{a+1}}(\alpha_{a}, \alpha_{a+1}) \no \\
&& {}\times
\Bigl\{
-\frac{1}{\alpha_{a+1}-\beta_{k+1}}\frac{1}{\alpha_{a}-\beta_{k}}
\frac{\alpha_{a}-\beta_{k+1}-\hbar}{\alpha_{a}-\beta_{k+1}}
(\alpha_{a+1}-\alpha_{a}-\hbar)
\frac{\alpha_{a}-\alpha_{a+1}+\hbar}{\alpha_{a}-\alpha_{a+1}} \no \\
&& \quad {}-
\frac{1}{\alpha_{a}-\beta_{k+1}}\frac{1}{\alpha_{a+1}-\beta_{k}}
\frac{\alpha_{a+1}-\beta_{k+1}-\hbar}{\alpha_{a+1}-\beta_{k+1}}
(\alpha_{a}-\alpha_{a+1}-\hbar)
\frac{\hbar}{\alpha_{a}-\alpha_{a+1}} \Bigr\} \no \\
&& \!\!\! {}=(\ref{qlast}).
\ena
Hence (\ref{wrel1}) holds.

Let us prove the highest weight condition. 
In the same way as (\ref{Eattacked}), it suffices to prove that
\bea
\sum_{a=1 \atop J'_{a}=k-1}^{n} F_{J'_{1}, \cdots , J'_{a}+1, \cdots , J'_{n}}[W]=0, \,\, 
(k=1, \cdots , N-1),
\label{FMhwcond} 
\ena
where 
$(J'_{1}, \cdots , J'_{n}) \in {\cal Z}_{\nu_{1}, \cdots , \nu_{k}-1, \cdots , \nu_{N}}$.

First consider the case of $k>1$. 
{}From the highest weight condition for $\bar{\psi}$, we can see that 
\bea
\sum_{a=1 \atop J'_{a}=k-1}^{n} w_{J'_{1}, \cdots , J'_{a}+1, \cdots , J'_{n}}[W]=0.
\ena
Hence (\ref{FMhwcond}) holds in the case of $k>1$. 

Now we consider the case of $k=1$. 

\begin{lem}\label{simplehw}
For $J'=(J'_{1}, \cdots , J'_{N}) \in {\cal Z}_{\ell-1, \nu_{2}, \cdots , \nu_{N}}$, 
the following equality holds:
\bea
&&
\hbar \sum_{a=1 \atop J'_{a}=0}^{n} 
w_{J'_{1}, \cdots , J'_{a}+1, \cdots , J'_{n}}={\rm Skew}\Bigl(
g_{M_{1}^{J'}}(\{\alpha_{a}\}_{2 \le a \le \ell}|\{\beta_{j}\}) \label{FMhweq} \\
&& \quad {}\times
\left\{ 
\prod_{a=2}^{\ell}(\alpha_{1}-\alpha_{a}-\hbar)
H(\alpha_{1})-
\prod_{j=1}^{n}\frac{\alpha_{1}-\beta_{j}-\hbar}{\alpha_{1}-\beta_{j}}
\prod_{a=2}^{\ell}(\alpha_{1}-\alpha_{a}+(N-1)\hbar)
H(\alpha_{1}+N\hbar)
\right\} \Bigr), \no 
\ena
Here we used the following abbreviation: 
\bea
H(\alpha_{1})=
H_{0, \bar{J}'_{1}, \cdots , \bar{J}'_{\ell -1}}
 (\alpha_{1}, \alpha_{2}, \cdots , \alpha_{\ell}). 
\ena
\end{lem}

This lemma is proved in Section \ref{app2}. 

{}From (\ref{FMhweq}), we can get (\ref{FMhwcond}) for $k=1$ 
by the same calculation as (\ref{hwdeform}). 
\qed

\section{Modification of the integral formula}
\subsection{One-time integration}\label{sectiononetime}

%We denote by ${\cal Z}_{\scriptsize{\underbrace{m, \cdots , m}_{N}}}$ 
%the set of all $n$-tuples 
%$(\epsilon_{1}, \cdots , \epsilon_{n}) \in {\Bbb Z}_{\ge 0}$ such that 
%\bea
%\# \{j| \epsilon_{j}=k \} =m, \qquad {\rm for \,\, all} \quad k=0, \cdots , N-1.
%\ena
Recall that $n=Nm$.
%and $J^{M} \in {\cal Z}_{m, \cdots , m}$ for 
%$M \in {\cal I}_{(N-1)m, \cdots , 2m, m}$. 
Let 
$\{\omega_{\epsilon_{1}, \cdots , \epsilon_{n}}(\beta_{1}, \cdots , \beta_{n})\}_{
(\epsilon_{1}, \cdots , \epsilon_{n}) \in {\cal Z}_{(N-1)m, \cdots , 2m, m}}$ 
be the set of vectors in $(V_{N})^{\otimes Nm}$ 
uniquely defined by the following conditions:
\bea
\omega_{\cdots , \epsilon_{j+1}, \epsilon_{j}, \cdots}
(\cdots , \beta_{j+1}, \beta_{j}, \cdots )=
P_{j, j+1}R_{j, j+1}(\beta_{j}-\beta_{j+1})
\omega_{\cdots , \epsilon_{j}, \epsilon_{j+1}, \cdots }
(\cdots , \beta_{j}, \beta_{j+1}, \cdots), 
\label{omegarel}
\ena
and
\bea
&&
\omega_{\scriptsize{\underbrace{0 \cdots 0}_{m}\underbrace{1 \cdots 1}_{m} \cdots 
       \underbrace{(N-1) \cdots (N-1)}_{m}}}(\beta_{1}, \cdots , \beta_{n}) \no \\
&& \qquad {}=
\underbrace{v_{0} \otimes \cdots \otimes v_{0}}_{m} \otimes
\underbrace{v_{1} \otimes \cdots \otimes v_{1}}_{m} \otimes \cdots \otimes
\underbrace{v_{N-1} \otimes \cdots \otimes v_{N-1}}_{m}. 
\ena
Here $P_{j, j+1}$ is the permutation operator acting on the tensor product 
of $j$-th and $(j+1)$-th components. 

For $J \in {\cal Z}_{(N-1)m, \cdots , 2m, m}$, we set
\bea
K_{r}^{J}:={\cal N}_{r}^{J} \setminus {\cal N}_{r+1}^{J}
=:\{ k_{r, 1}^{J}, \cdots , k_{r, m}^{J} \}, 
\quad k_{r, 1}^{J}< \cdots < k_{r, m}^{J}.
\label{defK}
\ena
Note that $\{1, \cdots , n \}=\sqcup_{r=0}^{N-1}K_{r}^{J}$. 

We define rational functions $\mu_{J}^{(a)}$ and $\tilde{w}_{J}$ by
\bea
&&
\mu_{J}^{(a)}(\alpha | \{ \beta_{j} \}):=
\frac{1}{\alpha-\beta_{a}}
\prod_{k \in K_{r}^{J} \atop k \not=a} \frac{\alpha-\beta_{k}-\hbar}{\beta_{a}-\beta_{k}-\hbar}
\prod_{j \in K_{p}^{J} \atop p>r} \frac{\alpha-\beta_{j}-\hbar}{\alpha-\beta_{j}}, 
\quad (a \in K_{r}^{J}), \\
&&
\tilde{w}_{J}(\{ \alpha_{a} \}| \{ \beta_{j} \}):=
{\rm Skew} \left(
\prod_{s=1}^{N-1}\prod_{j=1}^{m} \mu_{J}^{(k_{N-s, j}^{J})}(\alpha_{j+(s-1)m}| \{\beta_{j}\})
\right),
\ena
where ${\rm Skew}$ is the skew-symmetrization 
with respect to $\alpha_{1}, \cdots , \alpha_{\ell}$. 
Note that $\ell=(N-1)m$.

\begin{prop}\label{basechange}
\bea
&&
\sum_{J \in {\cal Z}_{(N-1)m, \cdots , 2m, m}} \!\!\!\!\!\!\!\!\!\! w_{J}v_{J} 
\label{basechangeeq} \\
&& {}=
(-1)^{\frac{\ell(\ell-1)}{2}} \!\!\!\!\!\!\!\!\!\!
\sum_{J \in {\cal Z}_{(N-1)m, \cdots , 2m, m}} \!\!\!\!\!\!\!\!\!\! 
\tilde{w}_{J}\omega_{J}
\prod_{r=1}^{N-1}\prod_{a, b \in K_{r}^{J} \atop a<b}
 \frac{(\beta_{b}-\beta_{a}-\hbar)(\beta_{a}-\beta_{b}-\hbar)}{\beta_{a}-\beta_{b}}
\prod_{a \in K_{s}^{J}, b \in K_{r}^{J} \atop 0 \le r<s \le N-1}
 \frac{\beta_{a}-\beta_{b}-\hbar}{\beta_{a}-\beta_{b}}. \no
\ena
\end{prop}

This proposition is proved in Section \ref{app1}.

By using Proposition \ref{basechange}, 
we rewrite $\psi_{W}$ in terms of $\tilde{w}_{J}$ and $\omega_{J}$. 
Then we can carry out the integration once as follows. 

Recall the definition of $\phi$ (\ref{defphi}):
\bea
\phi(\alpha | \beta_{1}, \cdots , \beta_{n}):=
\prod_{j=1}^{n}
\frac{\Gamma (\frac{\alpha-\beta_{j}-\hbar}{p})}
     {\Gamma (\frac{\alpha-\beta_{j}}{p} )}.
\ena

For a function $f(\alpha)$, we define a function $Df$ by
\bea
(Df)(\alpha):=f(\alpha)-f(\alpha+p)
\frac{\phi(\alpha+p)}{\phi(\alpha)}=
f(\alpha)-f(\alpha+p)
\prod_{j=1}^{n}\frac{\alpha-\beta_{j}-\hbar}{\alpha-\beta_{j}}. 
\label{defD}
\ena

Set
\bea
L_{J}^{(0)}(\alpha):=\prod_{k \in k_{0}^{J}}(\alpha-\beta_{k}-N\hbar).
\ena

\begin{prop}\label{putD}
\bea
(DL_{J}^{(0)})(\alpha)=
\hbar \sum_{r=1}^{N-1} \sum_{k \in K_{r}^{J}}
\prod_{j \in K_{0}^{J}}(\beta_{k}-\beta_{j}-r\hbar)
\prod_{j \in K_{r}^{J} \atop j \not= k}\frac{\beta_{k}-\beta_{j}-\hbar}{\beta_{k}-\beta_{j}}
\mu_{J}^{(k)}(\alpha).
\label{DLformula} 
\ena
\end{prop}
 
This proposition is proved in Section \ref{app2}. 

{}From Proposition \ref{putD}, we have
\bea
\tilde{w}_{J}(\{\alpha_{a}\})&=&
\frac{\hbar}
{\prod_{j \in K_{0}^{J}}(\beta_{k_{1, m}^{J}}-\beta_{j}-\hbar)}
\prod_{j \in K_{1}^{J} \atop j \not= k_{1, m}^{J}}
 \frac{\beta_{k_{1, m}^{J}}-\beta_{j}}{\beta_{k_{1, m}^{J}}-\beta_{j}-\hbar} \no \\
&& {}\times 
{\rm Skew}\left( 
\mu_{J}^{(k_{N-1, 1}^{J})}(\alpha_{1}) \cdots \mu_{J}^{(k_{1, m-1}^{J})}(\alpha_{\ell -1})
(D L_{J}^{(0)})(\alpha_{\ell}) \right) 
\label{puttedD}
\ena

Take the deformed cycle of the following form: 
\bea
W(\{e^{\frac{2\pi i}{p}\alpha_{a}}\})=\prod_{a=1}^{\ell}
\frac{P_{a}(e^{\frac{2\pi i}{p}\alpha_{a}})}{\prod_{j=1}^{n}
            (1-e^{\frac{2\pi i}{p}(\alpha_{a}-\beta_{j})})}
     \in \widehat{\cal F}_{q}^{\otimes \ell}. 
\label{Wassume}
\ena 
Let us consider the following integral:
\bea
\left( \prod_{a=1}^{\ell} \int_{C} d\alpha_{a} \right)
\phi(\{\alpha_{a}\}| \{\beta_{j}\}) \tilde{w}_{J}(\{\alpha_{a}\}) 
W(\{e^{\frac{2\pi i}{p}\alpha_{a}}\})
\label{1}
\ena
{}Using (\ref{puttedD}), 
we can carry out the integration once in (\ref{1}) by using the following formula:
\bea
&& \quad 
\int_{C}d\alpha \phi(\alpha ) (D L_{J}^{(0)})(\alpha) 
\frac{P_{a}(e^{\frac{2\pi i}{p}\alpha})}{\prod_{j=1}^{n}
      (1-e^{\frac{2\pi i}{p}(\alpha-\beta_{j})})} \no \\
&& {}=
\left( \int_{C}-\int_{C+p} \right)d\alpha \phi(\alpha) L_{J}^{(0)}(\alpha)
\frac{P_{a}(e^{\frac{2\pi i}{p}\alpha})}
     {\prod_{j=1}^{n}(1-e^{\frac{2\pi i}{p}(\alpha-\beta_{j})})}=
p^{m}(P_{a}^{-\infty}-P_{a}^{+\infty}), 
\label{integration}
\ena
where
\bea
P_{a}^{\pm \infty}:=\lim_{\alpha \to \pm \infty}
\frac{P_{a}(e^{\frac{2\pi i}{p}\alpha})}
     {\prod_{j=1}^{n}(1-e^{\frac{2\pi i}{p}(\alpha-\beta_{j})})}. 
\ena
The formula (\ref{integration}) can be obtained from (\ref{stir}). 

Especially, if $P_{a}^{\pm \infty}=0, \, (1 \le a \le \ell-1)$, then we have
\bea
(\ref{1})&=&
p^{m}(P_{\ell}^{-\infty}-P_{\ell}^{+\infty})\frac{\hbar}
{\prod_{j \in K_{0}^{J}}(\beta_{k_{1, m}^{J}}-\beta_{j}-\hbar)}
\prod_{j \in K_{1}^{J} \atop j \not= k_{1, m}^{J}}
 \frac{\beta_{k_{1, m}^{J}}-\beta_{j}}{\beta_{k_{1, m}^{J}}-\beta_{j}-\hbar} \no \\
&\times& \left( \prod_{a=1}^{\ell -1} \int_{C} d\alpha_{a} \right)
\phi(\{\alpha_{a}\}_{1 \le a \le \ell -1}| \{\beta_{j}\})
\frac{\prod_{a=1}^{\ell -1}P_{a}(e^{\frac{2\pi i}{p}\alpha_{a}})}
     {\prod_{a=1}^{\ell -1}\prod_{j=1}^{n}(1-e^{\frac{2\pi i}{p}(\alpha_{a}-\beta_{j})})} \no \\
&& {}\times 
{\rm Skew}\left( 
\mu_{J}^{(k_{N-1, 1}^{J})}(\alpha_{1}) \cdots \mu_{J}^{(k_{1, m-1}^{J})}(\alpha_{\ell -1})\right).
\label{onetime}
\ena

\subsection{Smirnov's formula}
{}From (\ref{onetime}), we can get the integral formula for solutions of the qKZ equation 
constructed by Smirnov \cite{smirbook} as follows. 

For a rational function $f(\alpha)$, 
let $[f(\alpha)]_{+} \in {\Bbb C}\, [\alpha]$ be its polynomial part: 
\bea
f(\alpha)=[f(\alpha)]_{+}+o(1), \quad {\rm as} \quad \alpha \to \infty. 
\ena
Denote by $T_{\hbar}$ the difference operator defined by 
$T_{\hbar}f(\alpha):=f(\alpha)-f(\alpha+\hbar)$. 

Set
\bea
L_{J}^{(r)}(\alpha):=\prod_{j \in K_{r}^{J}}(\alpha -\beta_{j}-N\hbar), \quad (r=0, \cdots , N-1)
\ena
and
\bea
Q_{J}^{(k)}(\alpha):=
\sum_{r=0}^{N-1}L_{J}^{(r)}(\alpha+r\hbar)
T_{\hbar}\left( \left[
\frac{\prod_{s=o}^{r-1}L_{J}^{(s)}(\alpha+(r-1)\hbar)\prod_{s=r+1}^{N-1}L_{J}^{(s)}(\alpha+r\hbar)}
     {(\alpha+r\hbar)^{k}} \right]_{+} \right)
\ena
for $k=1, \cdots , \ell$. 
Here we note that $\ell=(N-1)m$ and hence $Q_{J}^{(\ell)}=0$.

\begin{prop}\label{toQ}
For $a \in K_{r}^{J}, (r>0)$, the following equality holds: 
\bea
&&
\hbar^{-1}\left(
D\Bigl( \prod_{j=1 \atop j \not=a}^{n}(\alpha-\beta_{j}-N\hbar) \Bigr)
-\sum_{k=1}^{\ell -1}(\beta_{a}+N\hbar)^{k-1}Q_{J}^{(k)}(\alpha) 
\right) \label{Q} \\
&& {}=
\prod_{j \in K_{t}^{J} \atop t<r}(\beta_{a}-\beta_{j}-\hbar)
\prod_{j \in K_{r}^{J} \atop j \not=a}(\beta_{a}-\beta_{j}-\hbar)
\prod_{j \in K_{t}^{J} \atop t>r}(\beta_{a}-\beta_{j}) \mu_{J}^{(a)}(\alpha) \no \\
&& \quad {}+
\sum_{u=1}^{N-r-1}\sum_{b \in K_{r+u}^{J}}
\prod_{j \in K_{r+u}^{J} \atop j \not=b}
\frac{\beta_{b}-\beta_{j}-\hbar}{\beta_{b}-\beta_{j}} \no \\
&& \qquad {}\times
\sum_{s=0}^{u-1}\prod_{j \in K_{r+s}^{J}}(\beta_{b}-\beta_{j}-(u-s)\hbar)
\frac{
\prod_{j \in K_{t}^{J} \atop t<r+s}(\beta_{a}-\beta_{j}-\hbar)
\prod_{j \in K_{t}^{J} \atop t>r+s}(\beta_{a}-\beta_{j})}
{(\beta_{b}-\beta_{a}-(u-s-1)\hbar)(\beta_{b}-\beta_{a}-(u-s)\hbar)}
\mu_{J}^{(b)}(\alpha). \no
\ena
\end{prop}

This proposition is proved in Section \ref{app2}.

{}From the same calculation as (\ref{integration}), 
it is easy to see that, 
if $P=P(e^{\frac{2\pi i}{p}\alpha})$ satisfies 
\bea 
\lim_{\alpha \to \pm\infty}
\frac{P(e^{\frac{2\pi i}{p}\alpha})}
{\prod_{j=1}^{n}(1-e^{\frac{2\pi i}{p}(\alpha-\beta_{j})})}=0,
\ena
then we have 
\bea
\int_{C} d\alpha \phi(\alpha | \{ \beta_{j} \}) 
D\Bigl( \prod_{j=1 \atop j \not=a}^{n}(\alpha -\beta_{j}-N\hbar) \Bigr)
\frac{P(e^{\frac{2\pi i}{p}\alpha})}
     {\prod_{j=1}^{n}(1-e^{\frac{2\pi i}{p}(\alpha-\beta_{j})})}=0.  
\label{zerocohomology} 
\ena 

{}From Proposition \ref{toQ} 
we see that, for $a \in K_{r}^{J}, (r>0)$, by adding the linear sum of 
$\mu_{J}^{(b)}$'s $(b \in K_{s}^{J}, s>r)$ to $\mu_{J}^{(a)}$ as in the rhs of (\ref{Q}), 
we get the lhs of (\ref{Q}). 
Moreover, the first term in the lhs of (\ref{Q}) vanishes after the integral over $C$ 
{}from (\ref{zerocohomology}). 
Therefore, in (\ref{onetime}), we can replace
\bea
{\rm Skew}\left( 
\mu_{J}^{(k_{N-1, 1}^{J})}(\alpha_{1}) \cdots \mu_{J}^{(k_{1, m-1}^{J})}(\alpha_{\ell -1})\right)
\ena
by
\bea
\det[(\beta_{a}+N\hbar)^{k-1}]_{a \not\in K_{0}^{J}, \, a \not= k_{1, m}^{J} 
                                \atop{ 1 \le k \le \ell-1}}
\det[ Q_{J}^{(k)}(\alpha_{b})]_{1 \le k, b \le \ell-1}
\ena
multiplied by a certain rational function of $\beta_{1}, \cdots , \beta_{n}$ 
determined from (\ref{Q}).

Finally we get the following formula for solutions:

\begin{thm}\label{smirnov}
Suppose that $W$ is a deformed cycle of the form (\ref{Wassume}) 
with $P_{a}^{\pm\infty}=0, \, (1 \le a \le \ell-1)$. 
Then 
\bea
\psi_{W}&=&(-1)^{\frac{N}{2}m(m+1)+m^2}(P_{\ell}^{-\infty}-P_{\ell}^{+\infty}) \label{smirform}\\
&&  {}\times \!\!\!\!\!
\sum_{J \in {\cal Z}_{(N-1)m, \cdots , 2m, m}} \!\!\!\!\! \omega_{J}
\prod_{a \in K_{r}^{J}, b \in K_{s}^{J} \atop 0 \le r<s \le N-1}
 \frac{1}{\beta_{a}-\beta_{b}} \no \\
&& \quad {}\times
\left( \prod_{a=1}^{\ell -1} \int_{C} d\alpha_{a} \right)
\phi(\{\alpha_{a}\} | \{\beta_{j}\})
\det[Q_{J}^{(b)}(\alpha_{a})]_{a, b=1}^{\ell-1}
\frac{\prod_{a=1}^{\ell -1}P_{a}(e^{\frac{2\pi i}{p}\alpha_{a}})}
     {\prod_{a=1}^{\ell -1}\prod_{j=1}^{n}(1-e^{\frac{2\pi i}{p}(\alpha_{a}-\beta_{j})})}. \no
\ena
\end{thm}

\begin{cor}\label{zerocycle} 
Suppose that $W$ is  a deformed cycle satisfying the assumption in Theorem \ref{smirnov}. 
If it also holds that $P_{\ell}^{\pm\infty}=0$, 
then $\psi_{W}=0$.
\end{cor}

\rem
The formula (\ref{smirform}) is nothing but the integral formula 
constructed by Smirnov \cite{smirbook}. 
Note that indices for basis of the vector representation in \cite{smirbook} 
are reverse to that of $V_{N}$, that is, $e_{j}$ in \cite{smirbook} is equal to $v_{N-j}$.  
Let $A_{k}(\alpha | B^{(1)} | \cdots | B^{(N)})$ be the polynomial 
defined in \cite{smirbook}, page 185. 
Then 
\bea
Q_{J}^{(k)}|_{\hbar=-\frac{2\pi i}{N}, p=-2\pi i}=
A_{\ell -k}\left( \alpha-\pi i \Big| 
\{\beta_{j}-2\pi i-\frac{\pi i}{N}\}_{j \in K_{N-1}^{J}} \Big| \cdots \Big| 
\{\beta_{j}-2\pi i-\frac{\pi i}{N}\}_{j \in K_{0}^{J}} \right).
\ena

\section{Form factors of $SU(N)$ invariant Thirring model}
\subsection{Axioms for form factors}

In the following we assume that
\bea
\hbar=-\frac{2\pi i}{N}, \qquad p=N\hbar=-2\pi i.
\ena

Consider the $l$-th fundamental representation of $SU(N)$:
\bea
V^{(l)} \simeq \wedge^{l}V_{N}.
\ena
This space is realized as the subspace of $(V_{N})^{\otimes l}$ spanned by the following vectors
\bea
(V_{N})^{\otimes l} \ni v_{[\epsilon_{1}, \cdots , \epsilon_{l}]}:=
\sum_{\sigma \in S_{l}}({\rm sgn}\sigma)
v_{\epsilon_{\sigma(1)}} \otimes \cdots \otimes v_{\epsilon_{\sigma(l)}}, 
\quad (0 \le \epsilon_{1} < \cdots < \epsilon_{l} \le N-1).
\ena
In the following we denote by $V^{(l)}$ this subspace.

Fix a positive integer $m$ and assume that $n=Nm$. 
As mentioned in Introduction we consider form factors of type 
\bea 
f^{(1, \cdots , 1, k)}(\beta_{1}, \cdots , \beta_{n-k}, \beta_{n-k+1}) 
\in (V^{(1)})^{\otimes (n-k)} \otimes V^{(k)}. 
\ena 
In the following we abbreviate $f^{(1, \cdots , 1, k)}$ to $f^{(k)}$ for $k=2, \cdots , N-1$. 

The form factor associated with $n$ rank-1 particles 
$f_{m}(\beta_{1}, \cdots , \beta_{n}):=f^{(1, \cdots , 1)}(\beta_{1}, \cdots , \beta_{n})$ 
takes values in $(V^{(1)})^{\otimes n}$ 
and satisfies the following conditions:
\bea
&&
P_{j, j+1}S_{j, j+1}(\beta_{j}-\beta_{j+1})f_{m}( \cdots , \beta_{j}, \beta_{j+1}, \cdots )=
f_{m}( \cdots , \beta_{j+1}, \beta_{j}, \cdots ), \label{ax1} \\
&&
P_{n-1, n}\cdots P_{1, 2}f_{m}(\beta_{1}-2\pi i, \beta_{2}, \cdots , \beta_{n})=
e^{-\frac{(N-1)n}{N}\pi i}f_{m}(\beta_{2}, \cdots , \beta_{n-1}, \beta_{1}),
\label{ax2}
\ena
where $S(\beta)$ is the $S$-matrix defined by
\bea
S(\beta):=S_{0}(\beta)R(\beta), \qquad 
S_{0}(\beta):=\frac{\Gamma(\frac{N-1}{N}+\frac{\beta}{2\pi i})
                   \Gamma(-\frac{\beta}{2\pi i})}
                  {\Gamma(\frac{N-1}{N}-\frac{\beta}{2\pi i})
                   \Gamma(\frac{\beta}{2\pi i})}. 
\ena
The function $f_{m}$ has a simple pole at the point $\beta_{n}=\beta_{n-1}-\hbar$ 
with the following residue:
\bea
2\pi i{\rm res}_{\beta_{n}=\beta_{n-1}-\hbar}f_{m}(\beta_{1}, \cdots , \beta_{n})=
f^{(2)} \left( \beta_{1}, \cdots , \beta_{n-2}, \beta_{n-1}-\frac{\hbar}{2} \right). 
\label{res0}
\ena
where $f^{(2)}$ is the form factor 
associated with $(n-2)$ rank-1 particles and one rank-2 particle, that is, 
a vector in $(V^{(1)})^{\otimes (n-2)}\otimes V^{(2)} \subset (V^{(1)})^{\otimes n}$. 
Generally, for $2 \le k \le N-2$, the form factor 
\bea
f^{(k)}(\beta_{1}, \cdots , \beta_{n-k+1}) \in (V^{(1)})^{\otimes (n-k)}\otimes V^{(k)}
\ena
has a simple pole at the point $\beta_{n-k+1}=\beta_{n-k}-\frac{k+1}{2}\hbar$ with the residue
\bea
&& 
2\pi i{\rm res}f^{(k)}(\beta_{1}, \cdots , \beta_{n-k+1}) \no \\
&& {}=
f^{(k+1)}\left(\beta_{1}, \cdots , \beta_{n-k-1}, \beta_{n-k}-\frac{k}{2}\hbar\right)
\in (V^{(1)})^{\otimes (n-k-1)}\otimes V^{(k+1)}. 
\label{res1}
\ena
In the case of $k=N-1$, the residue at 
$\beta_{n-N+2}=\beta_{n-N+1}-\frac{N}{2}\hbar=\beta_{n-N+1}+\pi i$ is given by
\bea
&&
2\pi i{\rm res}f^{(N-1)}(\beta_{1}, \cdots , \beta_{n-N+2}) \no \\
&& {}=
\left(
I+e^{-\frac{2\pi i}{N}+\frac{N-1}{N}(n-N)\pi i}
 S_{n-N+1, n-N}(\beta_{n-N+1}-\beta_{n-N})\cdots S_{n-N+1, 1}(\beta_{n-N+1}-\beta_{1}) 
\right) \no \\
&& \quad {}\times
f_{m-1}(\beta_{1}, \cdots , \beta_{n-N})\otimes v_{[0, 1, \cdots , N-1]},
\label{res2}
\ena
where 
$f_{m-1}(\beta_{1}, \cdots , \beta_{n-N}):=f^{(1, \cdots , 1)}(\beta_{1}, \cdots , \beta_{n-N})$ 
is the form factor 
associated with $(n-N)=(m-1)N$ rank-1 particles 
satisfying (\ref{ax1}) and (\ref{ax2}).

\subsection{Recurrence relations for deformed cycles}

Define a function $\zeta(\beta)$ by
\bea
\zeta(\beta):=
\frac{\Gamma_{2}(-i\beta+\frac{2(2N-1)}{N}\pi )\Gamma_{2}(i\beta+\frac{2(N-1)}{N}\pi)}
     {\Gamma_{2}(-i\beta+2\pi)\Gamma_{2}(i\beta)}, 
\quad \Gamma_{2}(x)=\Gamma_{2}(x | 2\pi, 2\pi).
\ena
Here $\Gamma_{2}(x | \omega_{1}, \omega_{2})$ is the double gamma function satisfying
\bea
\frac{\Gamma_{2}(x+\omega_{1}| \omega_{1}, \omega_{2})}
     {\Gamma_{2}(x | \omega_{1}, \omega_{2})}
=\frac{1}{\Gamma_{1}(x | \omega_{2})}, 
\label{gamma2}
\ena
where
\bea
\Gamma_{1}(x | \omega ):=\frac{\omega^{\frac{x}{\omega}-\frac{1}{2}}}{\sqrt{2\pi}}
\Gamma( \frac{x}{\omega}).
\ena
We refer the reader to \cite{JM} for other properties of the double gamma function. 
{}From the definiton of $\zeta(\beta)$, we can see that
\bea
\zeta(\beta-2\pi i)=\zeta(-\beta) \quad {\rm and} \quad 
\frac{\zeta(-\beta)}{\zeta(\beta)}=S_{0}(\beta).
\label{zetarel1}
\ena

For $P \in \widehat{\cal P}_{n}^{\otimes \ell}$, we set
\bea
f_{P}:=e^{\frac{(N-1)n}{2N} \sum_{j=1}^{n}\beta_{j}}
\prod_{1 \le j<j' \le n}\zeta(\beta_{j}-\beta_{j'}) \Psi_{P}.
\label{defff}
\ena
Here $\Psi_{P}$ is the solution (\ref{levelzerosol}) of the qKZ equation given by 
\bea
\Psi_{P}:=\psi_{W}, \quad {\rm where} \quad 
W(\{e^{-\alpha_{a}}\}):=
\frac{P(\{e^{-\alpha_{a}}\})}
     {\prod_{a=1}^{\ell}\prod_{j=1}^{n}(1-e^{-(\alpha_{a}-\beta_{j})})}
\in \widehat{\cal F}_{q}^{\otimes \ell}.
\label{WtoP} 
\ena

It is easy to see the following proposition 
{}from (\ref{wrel1}) and (\ref{wrel2}) for $F_{J}[W]$. 

\begin{prop} 
If $P$ is symmetric with respect to $\beta_{1}, \cdots , \beta_{n}$, 
then $f_{P}$ satisfies (\ref{ax1}) and (\ref{ax2}).
\end{prop}

Suppose that the form factor $f_{m} \in (V^{(1)})^{\otimes n}$ 
is parametrized by $P_{m} \in \widehat{\cal P}^{\otimes \ell}$ 
as (\ref{defff}): $f_{m}=f_{P_{m}}$. 
Similarly, suppose that $f_{m-1}=f_{P_{m-1}}$ for 
$P_{m-1} \in \widehat{\cal P}^{\otimes (\ell-N+1)}$.
Now we give a sufficient condition for $P_{m}$ and $P_{m-1}$ 
to satisfy (\ref{res0}), (\ref{res1}) and (\ref{res2}) 
for certain functions 
$f^{(k)} \in (V^{(1)})^{\otimes (n-k)}\otimes V^{(k)}, \,(k=2, \cdots , N-1)$.

%In the following we set
%\bea
%A_{a}:=e^{-\alpha_{a}}, B_{j}:=e^{\beta_{j}}.
%\ena
For two polynomials $P_{1}$ and $P_{2}$ of $e^{-\alpha_{1}}, \cdots , e^{-\alpha_{r}}$,  
we write 
\bea
P_{1} \sim P_{2} \quad {\rm if} \quad {\rm Skew}(P_{1}-P_{2})=0,
\ena
where ${\rm Skew}$ is the skew-symmetrization 
with respect to $\alpha_{1}, \cdots , \alpha_{r}$. 

\begin{prop}\label{rescond}
For $P_{m} \in \widehat{\cal P}_{n}^{\otimes \ell}$ and 
$P_{m-1} \in \widehat{\cal P}_{n-N}^{\otimes (\ell-N+1)}$, suppose that 
there exists a set of polynomials of $e^{-\alpha_{a}}$'s 
\bea
&&
\widehat{P}^{(k)}(\alpha_{1}, \cdots , \alpha_{\ell-k+1} | 
                  \beta_{1}, \cdots , \beta_{n-k-1} | \beta_{n-k}) 
\in \widehat{\cal P}_{n-k}^{\otimes (\ell-k+1)}, 
\quad {\rm and} \\
&&
P^{(k)}(\alpha_{1}, \cdots , \alpha_{\ell-k+1} | \beta_{1}, \cdots , \beta_{n-k} | \beta_{n-k+1})
\in \widehat{\cal P}_{n-k+1}^{\otimes (\ell-k+1)}, \quad (k=1, \cdots , N-1)
\ena
satisfying the following conditions:
\bea
&&
P^{(1)}=P_{m}(\alpha_{1}, \cdots , \alpha_{\ell} | \beta_{1}, \cdots , \beta_{n}),  
\label{cond1} \\
&&
P^{(k)}|_{\beta_{n-k+1}=\beta_{n-k}-\frac{k+1}{2}\hbar} \sim
\prod_{a=1}^{\ell -k}(1-e^{-(\alpha_{a}-\beta_{n-k})})
\widehat{P}^{(k)}, \quad (k=1, \cdots , N-2), \label{cond2} \\
&&
P^{(k+1)}=
\widehat{P}^{(k)}
(\alpha_{1}, \cdots , \alpha_{\ell-k}, \beta_{n-k}+\frac{k}{2}\hbar | 
 \beta_{1}, \cdots , \beta_{n-k-1} | \beta_{n-k}+\frac{k}{2}\hbar ), 
\quad (k=1, \cdots , N-2), \no \\
&& \label{cond3} \\
&& 
P^{(N-1)}|_{\beta_{n-N+2}=\beta_{n-N+1}-\frac{N}{2}\hbar} \sim
\prod_{a=1}^{\ell -N+1}(1-e^{-(\alpha_{a}-\beta_{n-N+1})})(1-e^{-(\alpha_{a}-\beta_{n-N+1}-\hbar)})
\widehat{P}^{(N-1)}, \label{cond3.5} \\
&&
\widehat{P}^{(N-1)}|_{\alpha_{\ell-N+2}=\beta_{n-N+1}-\delta(N-1)\hbar} \no \\
&& \quad {}=
d_{m}^{-1}e^{-\frac{(N-1)(2n-N)}{2}(\beta_{n-N+1}-\delta(N-1)\hbar)}
P_{m-1}(\alpha_{1}, \cdots , \alpha_{\ell-N+1}| \beta_{1}, \cdots , \beta_{n-N}),
\label{cond4}
\ena
where $\delta=0, 1$ and $d_{m}$ is a constant defined by (\ref{defdm}).

Then there exists a set of functions 
\bea
f^{(k)}(\beta_{1}, \cdots , \beta_{n-k+1}) \in (V^{(1)})^{\otimes(n-k)}\otimes V^{(k)}, 
\quad (k=2, \cdots , N-1)
\ena
satisfying (\ref{res0}), (\ref{res1}) and (\ref{res2}). 
\end{prop}

In the rest of this subsection, we prove this proposition. 

Recall that $\ell=(N-1)m$. 
We denote by ${\cal Z}_{N-1}^{(m)}$ the additive group 
freely generated by the elements  
$(\epsilon_{1}, \cdots , \epsilon_{\ell}) \in 
 {\cal Z}_{(N-2)m, \cdots , 2m, m}$. 
We set 
\bea
&& 
(\epsilon_{1}, \cdots , \epsilon_{\ell-a}, [\epsilon_{\ell-a+1}, \cdots , \epsilon_{\ell-b}], 
\epsilon_{\ell-b+1}, \cdots , \epsilon_{\ell}) \no \\
&& {}:= 
\sum_{\sigma \in S_{a-b}}({\rm sgn}\sigma) 
(\epsilon_{1}, \cdots , \epsilon_{\ell-a}, 
 \epsilon_{\ell-a+\sigma(1)}, \cdots , \epsilon_{\ell-a+\sigma(a-b)}, 
 \epsilon_{\ell-b}, \cdots , \epsilon_{\ell}) \in {\cal Z}_{N-1}^{(m)}. 
\ena

Set 
\bea
G_{\epsilon_{1}, \cdots , \epsilon_{\ell}}(\alpha_{1}, \cdots , \alpha_{\ell}):= 
\prod_{1 \le a<b \le \ell}(\alpha_{a}-\alpha_{b}-\hbar) 
H_{\epsilon_{1}, \cdots , \epsilon_{\ell}}(\alpha_{1}, \cdots , \alpha_{\ell}). 
\label{Gdef} 
\ena 
Note that $G_{\epsilon_{1}, \cdots , \epsilon_{\ell}}$ is a polynomial 
of $\alpha_{1}, \cdots , \alpha_{\ell}$ from Remark (\ref{Hpol}). 
For $\epsilon \in {\cal Z}_{N-1}^{(m)}$, we define $G_{\epsilon}$ by (\ref{Gdef}) and 
$G_{\epsilon+\epsilon'}:=G_{\epsilon}+G_{\epsilon'}$. 

Here we note that the function $w_{J}$ defined in (\ref{defw}) is given by 
\bea 
w_{J}={\rm Skew}\Bigl( 
\prod_{a=1}^{\ell}\left( \frac{1}{\alpha_{a}-\beta_{m_{a}}} 
\prod_{j=1}^{m_{a}-1}\frac{\alpha_{a}-\beta_{j}-\hbar}{\alpha_{a}-\beta_{j}} \right) 
G_{\bar{J}_{1}, \cdots , \bar{J}_{\ell}}(\alpha_{1}, \cdots , \alpha_{\ell}) \Bigr), 
\ena 
where $\{m_{1}, \cdots , m_{\ell}\}:=M_{1}^{J}, (m_{1}< \cdots < m_{\ell})$ and 
$(\bar{J}_{1}, \cdots , \bar{J}_{\ell})$ is defined in (\ref{defJbar}). 
Here we recall that the set 
\bea 
M_{1}^{J}=\{r ; J_{r} \ge 1\}=\{r; J_{r} \not=0 \} 
\ena 
parametrizes the position of non-zero components in $J=(J_{1}, \cdots , J_{n})$, 
and the values on these componentes are determined from 
$\bar{J}=(\bar{J}_{1}, \cdots , \bar{J}_{\ell})$ by (\ref{defJbar}). 

{}From (\ref{rel1}), (\ref{rel2}) and (\ref{rel3}), we see that 
\bea
G_{\cdots , \epsilon_{k+1}, \epsilon_{k}, \cdots }
 (\cdots , \alpha_{k+1}, \alpha_{k}, \cdots )\!\!\!&=&\!\!\!
-\frac{\alpha_{k}-\alpha_{k+1}}{\alpha_{k}-\alpha_{k+1}-\hbar}
G_{\cdots , \epsilon_{k}, \epsilon_{k+1}, \cdots }
 (\cdots , \alpha_{k}, \alpha_{k+1}, \cdots ) 
\label{Grel1} \\
&& {}- 
\frac{\hbar}{\alpha_{k}-\alpha_{k+1}-\hbar} 
G_{\cdots , \epsilon_{k+1}, \epsilon_{k}, \cdots }
 (\cdots , \alpha_{k}, \alpha_{k+1}, \cdots ), \no 
\ena
\bea
&& 
G_{\epsilon_{1}, \cdots , \epsilon_{\ell}}
 (\alpha_{1}, \cdots , \alpha_{\ell -1}, \alpha_{\ell}-N\hbar)=
(-1)^{\ell -1}G_{\epsilon_{\ell}, \epsilon_{1}, \cdots , \epsilon_{\ell -1}}
 (\alpha_{\ell}, \alpha_{1}, \cdots , \alpha_{\ell -1}), 
\label{Grel2} \\
&& 
G_{0, \cdots , 0, 1, \cdots , 1, \cdots , N-2, \cdots , N-2}
(\alpha_{1}, \cdots , \alpha_{\ell})=
\prod_{s=0}^{N-2}\prod_{a<b \atop (\epsilon_{a}=s=\epsilon_{b})}(\alpha_{a}-\alpha_{b}-\hbar). 
\label{Grel3}
\ena

\begin{lem}
The following formulae hold: 
\bea 
&& 
G_{\cdots , \epsilon_{k}, \epsilon_{k+1}, \cdots }
 (\cdots , \alpha, \alpha -\hbar, \cdots )=
-G_{\cdots, \epsilon_{k+1}, \epsilon_{k}, \cdots }
 (\cdots , \alpha, \alpha -\hbar, \cdots ), \label{Geval1} \\
&&
G_{\cdots , [\epsilon_{k}, \epsilon_{k+1}], \cdots }
 (\cdots , \alpha, \alpha-\hbar, \cdots )=
G_{\cdots , [\epsilon_{k}, \epsilon_{k+1}], \cdots }
 (\cdots , \alpha-\hbar, \alpha, \cdots ), \label{Geval2} \\
&& 
G_{\epsilon_{1}, \cdots , \epsilon_{\ell-N+1}, 0, 1, \cdots , N-2}
 (\alpha_{1}, \cdots , \alpha_{\ell-N+1}, \beta, \beta-\hbar, \cdots , \beta-(N-2)\hbar) \no \\
&& {}=
(-1)^{\frac{(N-1)(N-2)}{2}m}\prod_{a=1}^{\ell-N+1}(\alpha_{a}-\beta-\hbar)
G_{\epsilon_{1}, \cdots , \epsilon_{\ell-N+1}} 
 (\alpha_{1}, \cdots , \alpha_{\ell-N+1}). \label{Geval3} 
\ena
\end{lem}

\proof 
It is easy to see (\ref{Geval1}) and (\ref{Geval2}) 
{}from (\ref{Grel1}). 

Let us prove (\ref{Geval3}). 
Note that both sides of (\ref{Geval3}) satisfy (\ref{Grel1}) 
as functions of $\alpha_{1}, \cdots , \alpha_{\ell-N+1}$. 
Hence it sufficies to prove that (\ref{Geval3}) holds for 
\bea
(\epsilon_{1}, \cdots , \epsilon_{\ell-N+1})=
(\underbrace{0, \cdots , 0}_{m-1}, \cdots , \underbrace{N-2, \cdots , N-2}_{m-1}). 
\label{epsiloncond} 
\ena
In the case of $N=2$ this is trivial. 
In the case of $N=3$ we can prove this from (\ref{Geval1}) and the following formula: 
\bea 
&& 
G_{\scriptsize{ \underbrace{0, \cdots , 0}_{m-1}, \underbrace{1, \cdots , 1}_{m}, 0}}
 (\alpha_{1}, \cdots , \alpha_{2m}) \label{N=3} \\
&& {}=
(-1)^{m-1}\prod_{a=1}^{m-1}(\alpha_{a}-\alpha_{2m}-2\hbar)
          \prod_{a=m}^{2m-2}(\alpha_{a}-\alpha_{2m-1}-\hbar) 
G_{\scriptsize{ \underbrace{0, \cdots , 0}_{m-1}, \underbrace{1, \cdots , 1}_{m-1}}}
 (\alpha_{1}, \cdots , \alpha_{2m-2}). 
\no 
\ena 
This formula can be proved easily from (\ref{Grel2}) and (\ref{Grel3}). 

In the case of $N>3$, from (\ref{N=3}), we have 
\bea
&& 
G_{\scriptsize{\underbrace{0, \cdots , 0}_{m-1}, \underbrace{1, \cdots , 1}_{m}}, 0, 
2, \cdots , 2, \cdots , N-2, \cdots , N-2}(\alpha_{1}, \cdots , \alpha_{\ell}) \\
&& {}=
(-1)^{m-1} \!\!\!\! \prod_{a, b \atop (a<b, 1 \le \epsilon_{a}=\epsilon_{b})} \!\!\!\!\!\!
 (\alpha_{a}-\alpha_{b}-\hbar) \,
G_{\scriptsize{\underbrace{0, \cdots , 0}_{m}, \underbrace{1, \cdots , 1}_{m}}}
 (\alpha_{1}, \cdots , \alpha_{m-1}, \alpha_{2m}+\hbar, \alpha_{2m+1}, \cdots , \alpha_{3m}). 
\no 
\ena
Repeating this calculation, we find 
\bea
&& 
G_{\scriptsize{\underbrace{0, \cdots , 0}_{m-1}, \cdots , \underbrace{N-2, \cdots , N-2}_{m-1}}, 
   N-2, N-1, \cdots , 0}(\alpha_{1}, \cdots , \alpha_{\ell}) \no \\
&& {}=
(-1)^{\frac{(N-1)(N-2)}{2}(m-1)} 
\prod_{s=0}^{N-2}\prod_{a \atop (\epsilon_{a}=s, a \le \ell-N+1)} 
 (\alpha_{a}-\alpha_{\ell-s}-(N-1-s)\hbar) \no \\
&& \qquad \qquad \qquad {}\times 
G_{\scriptsize{\underbrace{0, \cdots , 0}_{m-1}, \cdots , \underbrace{N-2, \cdots , N-2}_{m-1}}}
 (\alpha_{1}, \cdots , \alpha_{\ell-N+1}). 
\ena
By setting $\alpha_{\ell-s}=\beta-(N-2-s)\hbar, \, (0 \le s \le N-2)$ and using (\ref{Geval1}), 
we see (\ref{Geval3}) for (\ref{epsiloncond}). 
\qed

Now let us calculate residues of $f_{P_{m}}$ for $P_{m}$ satisfying 
the assumption of Proposition \ref{rescond}. 
It is easy to see that, at each point of taking residues 
(\ref{res0}), (\ref{res1}) and (\ref{res2}), 
the coefficient part $e^{\frac{N-1}{2N}n\pi i\sum_{j}\beta_{j}}\prod \zeta(\beta_{j}-\beta_{j'})$ 
is regular. 
Hence it suffices to consider residues of $\psi_{W}$. 

%In the following we set 
%\bea 
%f_{J}[P]:=F_{J}[W], 
%\ena 
%abbreviate $F_{M}[W]$ to $F_{M}[P]$, 
%where $W$ is given by (\ref{WtoP}).
Set 
\bea
{\rm RES}_{k}(F):=
\left( 2\pi i{\rm res}_{\beta_{n-k+1}=\beta_{n-k}-\frac{k+1}{2}\hbar}F \right)
|_{\beta_{n-k} \to \beta_{n-k}+\frac{k}{2}\hbar}
\ena
for a function $F=F(\beta_{1}, \cdots , \beta_{n-k+1})$ and $k=1, \cdots , N-2$. 
Then we have 
\bea
f^{(k+1)}(\beta_{1}, \cdots , \beta_{n-k})=
{\rm RES}_{k}f^{(k)}(\beta_{1}, \cdots , \beta_{n-k+1})
\ena 
{}from (\ref{res1}). 

\begin{lem}\label{resIIlem}
Let $P_{m}$ be a polynomial satisfying the assumption in Proposition \ref{rescond} 
and $W_{m}$ is the deformed cycle determined from $P_{m}$ by (\ref{WtoP}).  
Suppose that $J \in {\cal Z}_{(N-1)m, \cdots , 2m, m}$ satisfies 
$J_{a} \not= 0, \, (a=n-k, \cdots , n)$ for some $k, \, (1 \le k \le N-2)$.
Then the following formula holds:
\bea
&& 
{\rm RES}_{k}\circ \cdots \circ {\rm RES}_{1}\left( F_{J}[W_{m}] \right)  \label{resII} \\
&=& \!\!\!
\frac{1}{k!}\prod_{j=1}^{k}a_{m, j} 
\prod_{j=1}^{n-k-1} 
\left( e^{\frac{k}{2}(\beta_{n-k}-\beta_{j})+\frac{k}{4}\hbar} 
 \prod_{t=0}^{k-1} \Gamma( \frac{\beta_{n-k}-\beta_{j}+(t-\frac{k}{2})\hbar}{-2\pi i}+1) 
                   \Gamma( \frac{\beta_{n-k}-\beta_{j}+(\frac{k}{2}-t)\hbar}{2\pi i}) 
\right) \no \\
&\times& \!\!\!
\left( \prod_{a=1}^{\ell -k}\int_{C^{(k)}} d\alpha_{a} \right) 
\phi^{(k)}(\{\alpha_{a}\}|\{\beta_{j}\}_{j \le n-k-1}| \beta_{n-k})
w_{J}^{(k)}(\{\alpha_{a}\}|\{\beta_{j}\}_{j \le n-k-1}|\beta_{n-k}) 
W^{(k+1)}(\{e^{-\alpha_{a}}\}). \no
\ena 
In the formula above, the functions $\phi^{(k)}, w_{J}^{(k)}$ and $W^{(k+1)}$ are defined by 
\bea 
&& 
\phi^{(k)}(\{\alpha_{a}\}|\{ \beta_{j}\}_{j \le n-k-1} | \beta_{n-k}):=
\prod_{a=1}^{\ell-k} \left( 
\prod_{j=1}^{n-k-1}
\frac{\Gamma(\frac{\alpha_{a}-\beta_{j}-\hbar}{-2\pi i})}
     {\Gamma(\frac{\alpha_{a}-\beta_{j}}{-2\pi i})} 
\frac{\Gamma(\frac{\alpha_{a}-\beta_{n-k}-(\frac{k}{2}+1)\hbar}{-2\pi i})}
     {\Gamma(\frac{\alpha_{a}-\beta_{n-k}+\frac{k}{2}\hbar}{-2\pi i})} \right), \label{defphik}\\
&& 
w_{J}^{(k)}(\{\alpha_{a}\}|\{ \beta_{j}\}_{j \le n-k-1} | \beta_{n-k})
:={\rm Skew}\left( 
            g_{J}^{(k)}(\{\alpha_{a}\}|\{ \beta_{j}\}_{j \le n-k-1} | \beta_{n-k}) \right) \no  \\
&& 
g_{J}^{(k)}(\{\alpha_{a}\}|\{ \beta_{j}\}_{j \le n-k-1} | \beta_{n-k}):=
g_{M_{1}^{J}\setminus \{n-k+1, \cdots ,n\}}
(\{\alpha_{a}\}|\beta_{1}, \cdots , \beta_{n-k-1}, \beta_{n-k}-\frac{k}{2}\hbar)  \\
&& \quad {}\times 
G_{\bar{J}_{1}, \cdots \bar{J}_{\ell-k-1}, [\bar{J}_{\ell-k}, \cdots , \bar{J}_{\ell}]}
(\alpha_{1}, \cdots , \alpha_{\ell-k}, 
 \beta_{n-k}+\frac{k}{2}\hbar, \beta_{n-k}+\frac{k-2}{2}\hbar, \cdots , 
 \beta_{n-k}-\frac{k-2}{2}\hbar), \no \\
&& W^{(k+1)}(\{ e^{-\alpha_{a}} \}):=
\frac{P^{(k+1)}(\{e^{-\alpha_{a}}\})}
{\prod_{a=1}^{\ell-k}\left( 
 \prod_{j=1}^{n-k-1}(1-e^{-(\alpha_{a}-\beta_{j})}) 
 (1-e^{-(\alpha_{a}-\beta_{n-k}+\frac{k}{2}\hbar)})\right)}. \label{defWk} 
\ena
%where $(\bar{J}_{1}, \cdots , \bar{J}_{\ell}):=\bar{J}^{\bar{M}}$. 
The contour $C^{(k)}$ is a deformation of the real axis $(-\infty, \infty)$ 
such that the poles at 
\bea
\beta_{j}+\hbar+2\pi i{\Bbb Z}_{\ge 0}, \, (1 \le j \le n-k-1), \quad 
\beta_{n-k}+(\frac{k}{2}+1)\hbar+2\pi i{\Bbb Z}_{\ge 0}
\ena
are above $C^{(k)}$ and the poles at 
\bea
\beta_{j}-2\pi i{\Bbb Z}_{\ge 0}, \, (1 \le j \le n-k-1), \quad 
\beta_{n-k}-\frac{k}{2}\hbar-2\pi i{\Bbb Z}_{\ge 0} 
\ena
are below $C^{(k)}$. 
The constant $a_{r}$ is defined by 
\bea
a_{m, r}:=(2\pi i)^{-(Nm-r-2)}e^{-\frac{r(r-3)}{4}\hbar}
\Gamma(-\frac{1}{N})\Gamma(\frac{N-r}{N})
\prod_{j=1}^{r-1}\Gamma(\frac{N-j-1}{N})\Gamma(\frac{j}{N}).
\ena
\end{lem}

\rem 
Note that, under the assumption in Proposition \ref{resIIlem}, 
the residue (\ref{resII}) is skew-symmetric with respect to 
$J_{n-k}, \cdots , J_{n}$ from the definition of $w_{J}^{(k)}$. 
\newline 

\proof 
Let us calculate ${\rm RES}_{1}F_{J}[W_{m}]$. 
It can be shown that the point $\beta_{n}=\beta_{n-1}-\hbar$ is a simple pole 
of $F_{J}[W]$ for any $W \in \widehat{\cal F}_{q}^{\otimes \ell}$ 
in the same way as the proof of Proposition 3 in \cite {NT}. 
Hence, in the calculation of the residue, we can replace $P_{m}$ by 
$P_{m}|_{\beta_{n}=\beta_{n-1}-\hbar}=\prod_{a=1}^{\ell-1}(1-e^{-(\alpha_{a}-\beta_{n-1})}) \widehat{P}^{(1)}$. 

Then we consider the integral 
\bea
\left( \prod_{a=1}^{\ell} \int_{C} d\alpha_{a} \right) 
\phi(\{\alpha_{a}\}) w_{J}(\{\alpha_{a}\}) 
\frac{\prod_{a=1}^{\ell-1}(1-e^{-(\alpha_{a}-\beta_{n-1})})\widehat{P}^{(1)}(\{e^{-\alpha_{a}}\})}
     {\prod_{a=1}^{\ell}\prod_{j=1}^{n}(1-e^{-(\alpha_{a}-\beta_{j})})}.
\label{pfint1} 
\ena
The singularity of this integral at $\beta_{n} \to \beta_{n-1}-\hbar$ 
comes from the pinch of the contour $C$ by the poles of the integrand at 
$\alpha_{a}=\beta_{n-1}$ and $\alpha_{a}=\beta_{n}+\hbar$. 
Note that the integrand of (\ref{pfint1}) is regular 
at $\alpha_{a}=\beta_{n-1}, \, (1 \le a \le \ell-1)$. 
Hence only the contour for $\alpha_{\ell}$ may be pinched. 
The singularity at $\beta_{n}=\beta_{n-1}-\hbar$ comes from the residue at $\alpha_{\ell}=\beta_{n-1}$. 

Let us rewrite the integrand of (\ref{pfint1}) as follows. 
First we have
\bea
&& 
\phi(\alpha_{\ell}| \{\beta_{j}\})
\frac{1}{\prod_{j=1}^{n}(1-e^{-(\alpha_{\ell}-\beta_{j})})} \no \\
&& {}= 
\prod_{j=1}^{n}\left( 
(-2\pi i)^{-1}e^{\frac{1}{2}(\alpha_{\ell}-\beta_{j})}
\Gamma(\frac{\alpha_{\ell}-\beta_{j}-\hbar}{-2\pi i})
\Gamma(\frac{\alpha_{\ell}-\beta_{j}}{2\pi i}+1) \right). 
\label{pfint2}
\ena
When we put $\alpha_{\ell}=\beta_{n-1}$ in the rhs of (\ref{pfint2}), 
the factor $\Gamma(\frac{\beta_{n-1}-\beta_{n}-\hbar}{-2\pi i})$ appears. 
The singularity of (\ref{pfint1}) at $\beta_{n}=\beta_{n-1}-\hbar$ comes from this factor. 

Expand $w_{J}$ in (\ref{pfint1}) as follows: 
\bea
w_{J}=\sum_{\sigma \in S_{\ell}}({\rm sgn}\sigma) 
\prod_{a=1}^{\ell}\left( \frac{1}{\alpha_{\sigma(a)}-\beta_{m_{a}}}
\prod_{j=1}^{m_{a}-1}\frac{\alpha_{\sigma(a)}-\beta_{j}-\hbar}{\alpha_{\sigma(a)}-\beta_{j}} 
\right) 
G_{\bar{J}}(\{\alpha_{\sigma(a)}\}),  
\label{pfint3} 
\ena 
where $M_{1}^{J}=:\{m_{1}, \cdots , m_{\ell}\}, \, m_{1}< \cdots < m_{\ell}$. 
It is easy to see that the pinch of the contour for $\alpha_{\ell}$ occurs 
only when $\sigma(\ell-1)=\ell$ or $\sigma(\ell)=\ell$. 
For such terms, 
we deform the contour by taking the residue at $\alpha_{\ell}=\beta_{n-1}$, 
that is, 
\bea
\int_{C}(*)d\alpha_{\ell}= 
({\rm regular \, term})+(-2\pi i){\rm res}_{\alpha_{\ell}=\beta_{n-1}}(*). 
\label{takeres} 
\ena 
%we take the residue at $\alpha_{\ell}=\beta_{n-1}$ as in (\ref{takeres}). 
Because the function (\ref{pfint2}) is regular at $\alpha_{\ell}=\beta_{n-1}$, 
it suffices to calculate the residue of the rational function (\ref{pfint3}). 

Consider the case of $\sigma(\ell-1)=\ell$. 
Then the residue of (\ref{pfint3}) at $\alpha_{\ell}=\beta_{n-1}$ is given by 
\bea
&& 
(-1)\prod_{j=1}^{n-1}\frac{\beta_{n-1}-\beta_{j}-\hbar}{\beta_{n-1}-\beta_{j}} \label{pfsum1} \\
&& {}\times 
\sum_{\tau \in S_{\ell-1}} ({\rm sgn}\tau) 
\prod_{a=1}^{\ell -2}\left( \frac{1}{\alpha_{\tau(a)}-\beta_{m_{a}}} 
\prod_{j=1}^{m_{a}-1}\frac{\alpha_{\tau(a)}-\beta_{j}-\hbar}{\alpha_{\tau(a)}-\beta_{j}} 
\right) 
\frac{1}{\alpha_{\tau(\ell-1)}-\beta_{n}}
\prod_{j=1}^{n-1}\frac{\alpha_{\tau(\ell-1)}-\beta_{j}-\hbar}{\alpha_{\tau(\ell-1)}-\beta_{j}} 
\no \\ 
&& {}\times 
G_{\bar{J}}
(\alpha_{\tau(1)}, \cdots , \alpha_{\tau(\ell-2)}, \beta_{n-1}, \alpha_{\tau(\ell-1)}). \no
\ena 
Here we set $\tau:=\sigma \cdot (\ell-1, \ell) \in S_{\ell}$. 
Similarly, we find that the residue in the case of $\sigma(\ell)=\ell$ is given by 
\bea
&& 
\prod_{j=1}^{n-1}\frac{\beta_{n-1}-\beta_{j}-\hbar}{\beta_{n-1}-\beta_{j}} 
\sum_{\sigma \in S_{\ell-1}} ({\rm sgn}\sigma) 
\prod_{a=1}^{\ell -1}\left( \frac{1}{\alpha_{\sigma(a)}-\beta_{m_{a}}} 
\prod_{j=1}^{m_{a}-1}\frac{\alpha_{\sigma(a)}-\beta_{j}-\hbar}{\alpha_{\sigma(a)}-\beta_{j}} 
\right) 
\frac{-\hbar}{\beta_{n}-\beta_{n-1}} \no \\
&& \qquad \qquad \qquad \qquad \qquad {}\times 
G_{\bar{J}}
(\alpha_{\sigma(1)}, \cdots , \alpha_{\sigma(\ell-2)}, \alpha_{\sigma(\ell-1)}, \beta_{n-1}). 
\label{pfsum2} 
\ena 

Then the limit as $\beta_{n} \to \beta_{n-1}-\hbar$ 
of the sum of (\ref{pfsum1}) and (\ref{pfsum2}) is given by 
\bea 
&& 
(-1)\prod_{j=1}^{n-1}\frac{\beta_{n-1}-\beta_{j}-\hbar}{\beta_{n-1}-\beta_{j}} \times{} \\
&& {}\times 
\sum_{\sigma \in S_{\ell-1}} ({\rm sgn}\sigma) 
\prod_{a=1}^{\ell -2}\left( \frac{1}{\alpha_{\sigma(a)}-\beta_{m_{a}}} 
\prod_{j=1}^{m_{a}-1}\frac{\alpha_{\sigma(a)}-\beta_{j}-\hbar}{\alpha_{\sigma(a)}-\beta_{j}} 
\right) \no \\
&& \quad {}\times 
\frac{1}{\alpha_{\sigma(\ell-1)}-\beta_{n-1}} 
\prod_{j=1}^{n-2}
\frac{\alpha_{\sigma(\ell-1)}-\beta_{n-1}-\hbar}{\alpha_{\sigma(\ell-1)}-\beta_{n-1}} \no \\
&& \quad {}\times 
\Bigl( 
\frac{\alpha_{\sigma(\ell-1)}-\beta_{n-1}-\hbar}{\alpha_{\sigma(\ell-1)}-\beta_{n-1}+\hbar}
G_{\bar{J}}
(\alpha_{\sigma(1)}, \cdots , \alpha_{\sigma(\ell-2)}, \beta_{n-1}, \alpha_{\sigma(\ell-1)}) \no \\
&& \qquad \qquad \qquad \qquad \qquad \qquad {}+ 
G_{\bar{J}}
(\alpha_{\sigma(1)}, \cdots , \alpha_{\sigma(\ell-2)}, \alpha_{\sigma(\ell-1)}, \beta_{n-1}) 
\Bigr). \no
\ena 
Using (\ref{Grel1}), we have 
\bea 
&&
\frac{\alpha_{\sigma(\ell-1)}-\beta_{n-1}-\hbar}{\alpha_{\sigma(\ell-1)}-\beta_{n-1}+\hbar}
G_{\epsilon_{1}, \cdots , \epsilon_{\ell}}
(\alpha_{\sigma(1)}, \cdots , \alpha_{\sigma(\ell-2)}, \beta_{n-1}, \alpha_{\sigma(\ell-1)}) \no \\
&& {}+ 
G_{\epsilon_{1}, \cdots , \epsilon_{\ell}}
(\alpha_{\sigma(1)}, \cdots , \alpha_{\sigma(\ell-2)}, \alpha_{\sigma(\ell-1)}, \beta_{n-1}) 
\no \\
&& {}=
\frac{\alpha_{\sigma(\ell-1)}-\beta_{n-1}}{\alpha_{\sigma(\ell-1)}-\beta_{n-1}+\hbar} 
G_{\epsilon_{1}, \cdots , \epsilon_{\ell-2}, [\epsilon_{\ell-1}, \cdots , \epsilon_{\ell}]} 
(\alpha_{\sigma(1)}, \cdots , \alpha_{\sigma(\ell-2)}, \alpha_{\sigma(\ell-1)}, \beta_{n-1}). 
\label{keyformula0} 
\ena 
{}From this calculation, it is easy to get the formula (\ref{resII}) in the case of $k=1$. 

We can prove (\ref{resII}) in the case of $k>1$ by a similar calculation. 
Then we use the following formula 
\bea
&&
\frac{\alpha_{\ell-k}-\beta-\hbar}{\alpha_{\ell-k}-\beta+k\hbar}
G_{\epsilon_{1}, \cdots , \epsilon_{\ell-k}, [\epsilon_{\ell-k+1}, \cdots , \epsilon_{\ell}]}
(\alpha_{1}, \cdots , \alpha_{\ell-k-1}, \beta, \alpha_{\ell-k}, 
 \beta-\hbar, \cdots , \beta-(k-1)\hbar) \no \\
&& {}+\frac{1}{k}
G_{\epsilon_{1}, \cdots , \epsilon_{\ell-k}, [\epsilon_{\ell-k+1}, \cdots , \epsilon_{\ell}]}
(\alpha_{1}, \cdots , \alpha_{\ell-k-1}, \alpha_{\ell-k}, 
 \beta, \beta-\hbar, \cdots , \beta-(k-1)\hbar) 
\label{keyformula} \\
&& {}=\frac{1}{k}
\frac{\alpha_{\ell-k}-\beta}{\alpha_{\ell-k}-\beta+k\hbar} 
G_{\epsilon_{1}, \cdots , \epsilon_{\ell-k-1}, [\epsilon_{\ell-k}, \cdots, \epsilon_{\ell}]} 
(\alpha_{1}, \cdots , \alpha_{\ell-k-1}, \alpha_{\ell-k}, 
 \beta, \beta-\hbar, \cdots , \beta-(k-1)\hbar).
\no
\ena 
instead of (\ref{keyformula0}). 
This formula can be proved from (\ref{Grel1}) and (\ref{Geval1}). 
\qed

\begin{lem}\label{resIlem}
Let $P_{m}$ be a polynomial satisfying the assumption in Proposition \ref{rescond}. 
Suppose that $J \in {\cal Z}_{(N-1)m, \cdots , 2m, m}$ satisfies 
$J_{a} \not= 0, \, (a=n-k+1, \cdots , n)$ and $J_{n-k}=0$ 
for some $k, \, (0 \le k \le N-2)$. 
Fix $s$ such that $1 \le s \le N-k-2$. 
Then 
\bea
{\rm RES}_{k+s} \circ \cdots \circ {\rm RES}_{1} \left( F_{J}[W_{m}] \right)=0
\label{reszero} 
\ena
if $J_{a}=0$ for some $a, (n-k-s \le a \le n-k-1)$. 

If $J_{a} \not=0, \, (a=n-k-s, \cdots , n-k-1)$, the following formula holds: 
\bea
&& 
{\rm RES}_{k+s}\circ \cdots \circ {\rm RES}_{1}\left( F_{J}[W_{m}] \right) \no \\
&& {}=
\frac{(-1)^{s}}{k!}\prod_{j=1}^{k+s}a_{m, j} 
\prod_{j=1}^{n-k-s} 
\Bigl( e^{\frac{k+s}{2}(\beta_{n-k-s}-\beta_{j})+\frac{k+s}{4}\hbar} \no \\
&& \qquad \qquad \qquad {}\times 
 \prod_{t=0}^{k+s-1} \Gamma( \frac{\beta_{n-k-s}-\beta_{j}+(t-\frac{k+s}{2})\hbar}{-2\pi i}+1) 
                   \Gamma( \frac{\beta_{n-k-s}-\beta_{j}+(\frac{k+s}{2}-t)\hbar}{2\pi i}) 
\Bigr) \times{} \no \\
&& {}\times  
\left( \prod_{a=1}^{\ell -k-s}\int_{C^{(k+s)}} d\alpha_{a} \right) 
\phi^{(k+s)}(\{\alpha_{a}\}|\{\beta_{j}\}_{j \le n-k-s-1}| \beta_{n-k-s}) \no \\
&& \qquad \qquad \qquad {}\times 
w_{J}^{(k, k+s)}(\{\alpha_{a}\}|\{\beta_{j}\}_{j \le n-k-s-1}|\beta_{n-k-s}) 
W^{(k+s+1)}(\{e^{-\alpha_{a}}\}). \label{resI}
\ena
Here $\psi^{(k+s)}$ and $W^{(k+s+1)}$ are given by (\ref{defphik}) and (\ref{defWk}), 
respectively. 
The function $w_{J}^{(k, k')}$ is defined by 
\bea
&& 
w_{J}^{(k, k')}(\{\alpha_{a}\}|\{ \beta_{j}\}_{j \le n-k'-1} | \beta_{n-k'})
:={\rm Skew} \left( 
g_{J}^{(k, k')}(\{\alpha_{a}\}|\{ \beta_{j}\}_{j \le n-k'-1} | \beta_{n-k'}) \right) \no  \\
&& 
g_{J}^{(k, k')}(\{\alpha_{a}\}|\{ \beta_{j}\}_{j \le n-k'-1} | \beta_{n-k'}):=
g_{M_{1}^{J}\setminus \{n-k', \cdots ,n\}}
(\{\alpha_{a}\}|\beta_{1}, \cdots , \beta_{n-k'-1})  \\
&& \qquad {}\times 
G_{\bar{J}_{1}, \cdots \bar{J}_{\ell-k-1}, [\bar{J}_{\ell-k}, \cdots , \bar{J}_{\ell}]}
(\alpha_{1}, \cdots , \alpha_{\ell-k'}, 
 \beta_{n-k'}+\frac{k'}{2}\hbar, \cdots , 
 \beta_{n-k'}-\frac{k'-2}{2}\hbar). \no 
\ena
\end{lem}

\rem 
As in Lemma \ref{resIIlem}, we see that the residue (\ref{resI}) 
is skew-symmetric with respect to $J_{n-k-s}, \cdots , J_{n}$. 
\newline 

\proof 
Let us consider the case $J_{n-k-1}\not= 0$ and calculate the residue for $s=1$. 
%prove the case of $s=0$. 
{}From Lemma \ref{resIIlem}, it sufficies to calculate the residue at 
$\beta_{n-k+1}=\beta_{n-k}-\frac{k+1}{2}\hbar$ of the following integral: 
\bea
&& 
\left( \prod_{a=1}^{\ell-k+1} \int_{C^{(k)}} d\alpha_{a} \right) 
\phi^{(k-1)}(\{\alpha_{a}\}|\{\beta_{j}\}_{j \le n-k}| \beta_{n-k+1}) \no \\
&& \qquad \qquad \qquad {}\times 
w_{J}^{(k-1)}(\{\alpha_{a}\}|\{\beta_{j}\}_{j \le n-k}|\beta_{n-k+1}) 
W^{(k)}(\{e^{-\alpha_{a}}\}).
\label{pfint11}
\ena 
%In the same way as the proof of Proposition 3 in \cite{NT}, 
%we see that the point $\beta_{n-k+1}=\beta_{n-k}-\frac{k+1}{2}\hbar$ is a s%imple pole 
%of this integral. 
As in the proof of Lemma \ref{resIIlem}, we can replace $W^{(k)}$ by 
\bea
\frac{\prod_{a=1}^{\ell-k}(1-e^{-(\alpha_{a}-\beta_{n-k})})
      \widehat{P}^{(k)}(\{e^{-\alpha_{a}}\})}
{\prod_{a=1}^{\ell-k+1}\left( 
 \prod_{j=1}^{n-k}(1-e^{-(\alpha_{a}-\beta_{j})}) 
 (1-e^{-(\alpha_{a}-\beta_{n-k+1}+\frac{k+1}{2}\hbar)})\right)}. 
\label{cycle11} 
\ena
Then the calculation of the residue is quite similar to 
that in the proof of Lemma \ref{resIIlem}. 
The singularity of the integral at $\beta_{n-k+1}=\beta_{n-k}-\frac{k+1}{2}\hbar$ 
comes from the pinch of the contour by the poles of the integrand 
at $\alpha_{a}=\beta_{n-k}$ and $\alpha_{a}=\beta_{n-k+1}+\frac{k+1}{2}\hbar$. 
Since (\ref{cycle11}) is regular at 
$\alpha_{a}=\beta_{n-k}, \, (1 \le a \le \ell-k)$, 
only the contour for $\alpha_{\ell-k+1}$ may be pinched. 
 
Expand $w_{J}^{(k-1)}$ in (\ref{pfint11}) as follows: 
\bea
&& 
w_{J}^{(k-1)}=
\sum_{\sigma \in S_{\ell-k+1}} 
({\rm sgn}\sigma) 
\prod_{a=1}^{\ell-k}\left( 
\frac{1}{\alpha_{\sigma(a)}-\beta_{m_{a}}} 
\prod_{j=1}^{m_{a}-1}\frac{\alpha_{\sigma(a)}-\beta_{j}-\hbar}{\alpha_{\sigma(a)}-\beta_{j}}
\right)  \label{resrat1} \\
&& \quad {}\times 
\frac{1}{\alpha_{\sigma(\ell-k+1)}-\beta_{n-k+1}+\frac{k-1}{2}\hbar}
\prod_{j=1}^{n-k}\frac{\alpha_{\sigma(a)}-\beta_{j}-\hbar}{\alpha_{\sigma(a)}-\beta_{j}} \no \\
&& \quad {}\times 
G_{\bar{J}_{1}, \cdots , \bar{J}_{\ell-k},
   [\bar{J}_{\ell-k+1}, \cdots , \bar{J}_{\ell}]}
 (\alpha_{\sigma(1)}, \cdots , \alpha_{\sigma(\ell-k+1)}, 
  \beta_{n-k+1}+\frac{k-1}{2}\hbar, \cdots , \beta_{n-k+1}-\frac{k-3}{2}\hbar). \no
\ena
It is easy to see that the pinch of the contour for $\alpha_{\ell-k+1}$ 
occurs only when $\sigma(\ell-k+1)=\ell-k+1$. 
By calculating the residue of (\ref{resrat1}) at $\alpha_{\ell-k+1}=\beta_{n-k}$ 
and taking the limit as $\beta_{n-k+1} \to \beta_{n-k}-\frac{k+1}{2}\hbar$, 
%of the other part of the integrand in (\ref{pfint11}), 
we get (\ref{resI}) for $s=1$. 

Repeating this calculation, we find (\ref{resI}) for $s>1$. 
Note that, if $J_{a}=0$ for some $a, \, (n-k-s \le a \le n-k-1)$, 
the pinch of the contour does not occur in the limit 
$\beta_{a+1} \to \beta_{a}-\frac{n-a+1}{2}\hbar$, 
and hence (\ref{reszero}) holds. 
\qed

Now we set 
\bea
f^{(k)}(\beta_{1}, \cdots , \beta_{n-k+1}):=
{\rm RES}_{k}\circ \cdots \circ {\rm RES}_{1} f_{P_{m}}. 
\ena
{}From Remarks in Lemma \ref{resIIlem} and Lemma \ref{resIlem},  
we see that 

\begin{cor}
\bea
f^{(k)}(\beta_{1}, \cdots , \beta_{n-k+1}) \in (V^{(1)})^{\otimes(n-k)}\otimes V^{(k)}, 
\quad (k=2, \cdots , N-1).
\ena
\end{cor}

Let us calculate the residue of $f^{(N-1)}$ at 
$\beta_{n-N+2}=\beta_{n-N+1}-\frac{N}{2}\hbar$. 
{}From (\ref{Geval1}) and (\ref{Geval3}), we see that 
the point $\beta_{n-N+2}=\beta_{n-N+1}-\frac{N}{2}\hbar$ is a simple pole of $f^{(N-1)}$ 
in the same way as the proof of Proposition 3 in \cite{NT}. 
Hence it sufficies to calculate the residue of 
\bea 
F_{J}^{*}[\widehat{W}^{(N-1)}]:=
\left( \prod_{a=1}^{\ell-N+2} \int_{C^{(N-2)}} d\alpha_{a} \right) 
\phi^{(N-2)}(\{\alpha_{a}\}) w_{J}^{*}(\{\alpha_{a}\}) 
\widehat{W}^{(N-1)}(\{e^{-\alpha_{a}}\}), 
\ena  
where 
\bea
&& 
*=(N-2), \quad {\rm if} \quad J_{n-a} \not=0, \, (a=0, \cdots , N-2), \no \\
&& 
*=(k, N-2), \quad {\rm if} \quad J_{n-k}=0 \,\, {\rm and} \,\,
J_{a} \not=0, \, (1 \le a \le N-2, a \not=k), \no 
\ena 
and 
\bea 
\widehat{W}^{(N-1)}:=
\frac{
\prod_{a=1}^{\ell -N+1}(1-e^{-(\alpha_{a}-\beta_{n-N+1})})(1-e^{-(\alpha_{a}-\beta_{n-N+1}-\hbar)})
\widehat{P}^{(N-1)}}
{\prod_{a=1}^{\ell-N+2}\left( 
 \prod_{j=1}^{n-N+1}(1-e^{-(\alpha_{a}-\beta_{j})}) 
 (1-e^{-(\alpha_{a}-\beta_{n-N+2}+\frac{N-2}{2}\hbar)})\right)}.
\ena

Consider the decomposition 
\bea
\widehat{W}^{(N-1)}=\widehat{W}_{0}+\widehat{W}_{1}, \quad 
\widehat{W}_{\delta}:=
\frac{(-e^{\hbar})^{\delta}(1-e^{-(\alpha_{\ell-N+2}-\beta_{n-N+2}+(\frac{N}{2}+\delta-1)\hbar)})}
     {1-e^{\hbar}} \widehat{W}^{(N-1)}. 
\ena 
Set 
\bea
\widehat{\psi}:=\widehat{\psi}_{0}+\widehat{\psi}_{1}, \quad 
\widehat{\psi}_{\delta}(\beta_{1}, \cdots , \beta_{n-N+1} | \beta_{n-N+2}) 
:=\sum_{J}F_{J}^{*}[\widehat{W}_{\delta}]v_{J}.
\ena

First let us calculate the residue of $\widehat{\psi}_{0}$. 
%The residue of $F_{M}^{*}[\widehat{W}_{0}]$ can be calculated 
%in a similar way to the proof of Lemma \ref{resIIlem} and Lemma \ref{resIlem}. 
%Then we get the following result. 
The result is the following. 

\begin{lem}\label{lastres+}
The residue of $\widehat{\psi}_{0}$ is given by 
\bea 
&& 
2\pi i{\rm res}_{\beta_{n-N+2}=\beta_{n-N+1}-\frac{N}{2}\hbar}\widehat{\psi}_{0} \no \\
&& {}=
(-1)^{\frac{(N-1)(N-2)}{2}m}(-2\pi i)^{\ell-N+1}a_{m, N-1} \no \\
&& {}\times \prod_{j=1}^{n-N}
\Bigl( e^{\frac{1}{2}(\beta_{n-N+1}-\beta_{j})} 
 \Gamma( \frac{\beta_{n-N+1}-\beta_{j}-\hbar}{-2\pi i}+1) 
 \Gamma( \frac{\beta_{n-N+1}-\beta_{j}}{2\pi i}) 
\Bigr) \no \\
&& 
{}\times d_{m}^{-1}e^{-\frac{(N-1)(2n-N)}{2}\beta_{n-N+1}}
\Psi_{P_{m-1}}(\beta_{1}, \cdots , \beta_{n-N}) \otimes v_{[0, 1, \cdots , N-1]}. 
\label{lastres+eq}
\ena
\end{lem}
Here $\Psi_{P_{m-1}}$ is the solution of the qKZ equation defined by 
\bea 
\Psi_{P_{m-1}}:=\psi_{W_{m-1}}, \quad {\rm where} \quad 
W_{m-1}:=\frac{P(\{e^{-\alpha_{a}}\})} 
         {\prod_{a=1}^{\ell-N+1}\prod_{j=1}^{n-N}(1-e^{-(\alpha_{a}-\beta_{j})})}. 
\ena 

\proof
The singularity of $F_{J}^{*}[\widehat{W}_{0}]$ at the point 
$\beta_{n-N+2}=\beta_{n-N+1}-\frac{N}{2}\hbar$ comes from the pinch of the contour $C$
by poles 
%of the integrand $\phi^{(N-2)}w_{J}^{*}\widehat{W}_{0}$ 
at $\alpha_{a}=\beta_{n-N+1}-(N-1)\hbar, \beta_{n-N+1}, \beta_{n-N+1}+\hbar$ 
and 
$\alpha_{a}=\beta_{n-N+2}-(\frac{N}{2}-1)\hbar, \beta_{n-N+2}+\frac{N}{2}\hbar, 
 \beta_{n-N+2}+(\frac{N}{2}+1)\hbar$, respectively. 
On the other hand 
we can see that the integrand $\phi^{(N-2)}w_{J}^{*}\widehat{W}_{0}$ is regular at 
$\alpha_{a}=\beta_{n-N+1}-(N-1)\hbar, \beta_{n-N+1}, 
 \beta_{n-N+1}+\hbar, \, (1 \le a \le \ell-N+1)$, 
hence only the contour for $\alpha_{\ell-N+2}$ may be pinched. 
Moreover, the integrand is regular at 
$\alpha_{\ell-N+2}=\beta_{n-N+2}-(\frac{N}{2}-1)\hbar, \beta_{n-N+2}+(\frac{N}{2}+1)\hbar$. 
Therefore the contour for $\alpha_{\ell-N+2}$ may be pinched only by the poles 
at $\beta_{n-N+1}$ and $\beta_{n-N+2}+\frac{N}{2}\hbar$. 
In order to avoid this pinch, 
we deform the contour $C$ by taking the residue at $\alpha_{\ell-N+2}=\beta_{n-N+1}$ 
in the same way as the proof of Lemma \ref{resIIlem}. 

Then, after the similar calculation to 
that in the proof of Lemma \ref{resIIlem}, 
we get the following integral:
\bea
2\pi i{\rm res}F_{J}^{*}[\widehat{W}_{0}]&=& \!\!\!\!
({\rm a \, \, certain \,\, function \,\, of} \,\,  \beta_{1}, \cdots , \beta_{n-N+1}) \no \\
&\times& \!\!\!\!
\left( \prod_{a=1}^{\ell -N+1}\int_{C} d\alpha_{a} \right) 
\phi(\{\alpha_{a}\}|\{\beta_{j}\}_{1 \le j \le n-N}) 
\prod_{a=1}^{\ell-N+1}\frac{1}{\alpha_{a}-\beta_{n-N+1}-\hbar} \\
&& {}\times 
w_{J}^{(k, N-1)}(\{\alpha_{a}\}|\{\beta_{j}\}_{j \le n-N}|\beta_{n-N+1}) 
\frac{P_{m-1}(\{e^{-\alpha_{a}}\})}
{\prod_{a=1}^{\ell-N+1}\prod_{j=1}^{n-N}(1-e^{-(\alpha_{a}-\beta_{j})})}. \no
\ena
By using (\ref{Geval1}) and (\ref{Geval3}), we get (\ref{lastres+eq}). 
\qed

Next we write down the formula for the residue of $\widehat{\psi}_{1}$. 

\begin{lem}\label{lastres-}
The residue of $\widehat{\psi}_{1}$ is given by 
\bea 
&& 
2\pi i{\rm res}_{\beta_{n-N+2}=\beta_{n-N+1}-\frac{N}{2}\hbar}\widehat{\psi}_{1} \no \\
&& {}=
(-1)^{\frac{(N-1)(N-2)}{2}m}(-2\pi i)^{\ell-N+1}a_{m, N-1} \no \\
&& {}\times \prod_{j=1}^{n-N}
\Bigl( e^{\frac{1}{2}(\beta_{n-N+1}-\beta_{j}-(N-1)\hbar)}
 \Gamma( \frac{\beta_{n-N+1}-\beta_{j}-N\hbar}{-2\pi i}+1) 
 \Gamma( \frac{\beta_{n-N+1}-\beta_{j}-(N-1)\hbar}{2\pi i}) 
\Bigr) \no \\
&& {}\times 
R_{n-N+1, n-N}(\beta_{n-N+1}-\beta_{n-N}) \cdots R_{n-N+1, 1}(\beta_{n-N+1}-\beta_{1}) \no \\
&& {}\times e^{(1-\frac{N}{2})\hbar}d_{m}^{-1}e^{-\frac{(N-1)(2n-N)}{2}(\beta_{n-N+1}-(N-1)\hbar)} 
\Psi_{P_{m-1}}(\beta_{1}, \cdots , \beta_{n-N})\otimes v_{[0, 1, \cdots , N-1]}. 
\label{lastres-eq}
\ena
\end{lem} 

\proof 
Note that $\widehat{\psi}_{1}$ satisfies 
\bea
\widehat{\psi}_{1}( \cdots , \beta_{i+1}, \beta_{i}, \cdots | \beta_{n-N+2})=
P_{i, i+1}R_{i, i+1}(\beta_{i}-\beta_{i+1})
\widehat{\psi}_{1}(\cdots , \beta_{i}, \beta_{i+1}, \cdots | \beta_{n-N+2}). 
\ena
Hence we have 
\bea
&& 
\widehat{\psi}_{1}(\beta_{1}, \cdots , \beta_{n-N+1}| \beta_{n-N+2})=
R_{n-N+1, n-N}(\beta_{n-N+1}-\beta_{n-N}) \cdots R_{n-N+1, 1}(\beta_{n-N+1}-\beta_{1}) \no \\
&& \qquad \qquad \qquad \qquad  {}\times 
P_{n-N+1, n-N} \cdots P_{2, 1} 
\widehat{\psi}_{1}(\beta_{n-N+1}, \beta_{1}, \cdots , \beta_{n-N}| \beta_{n-N+2}). 
\ena
The second line of the rhs above is given by 
\bea
&& 
P_{n-N+1, n-N} \cdots P_{2, 1} 
\widehat{\psi}_{1}(\beta_{n-N+1}, \beta_{1}, \cdots , \beta_{n-N}| \beta_{n-N+2})  \\
&& {}=
\sum_{M}F_{J_{n-N+1}, J_{1}, \cdots , J_{n-N}, J_{n-N+2}, \cdots , J_{n}}^{*}
[\widehat{W}_{1}](\beta_{n-N+1}, \beta_{1}, \cdots , \beta_{n-N} | \beta_{n-N+2})
v_{J_{1}} \otimes \cdots \otimes v_{J_{n}}. \no
\ena 

Now set $\beta_{n-N+1}':=\beta_{n-N+1}+2\pi i$ and consider the singularlity 
of $F_{J_{n-N+1}, J_{1}, \cdots , J_{n-N}, J_{n-N+2}, \cdots , J_{n}}^{*}$ 
at $\beta_{n-N+2}=\beta_{n-N+1}'+\frac{N}{2}\hbar$. 
In the same way as the proof of Lemma \ref{lastres+}, 
this singularity comes from the pinch of the contour, 
and we see that only the contour for $\alpha_{\ell-N+2}$ may be pinched. 
%by the poles at $\alpha_{\ell-N+2}=\beta_{n-N+1}'+\hbar$ and 
%$\alpha_{\ell-N+2}=\beta_{n-N+2}-(\frac{N}{2}-1)\hbar$. 
%In order to avoid this pinch, we deform the contour by taking the residue at 
%$\alpha_{\ell-N+2}=\beta_{n-N+2}-(\frac{N}{2}-1)\hbar$ in the following way. 
Now we rewrite the integrand in 
$F_{J_{n-N+1}, J_{1}, \cdots , J_{n-N}, J_{n-N+2}, \cdots , J_{n}}^{*}$ 
in terms of $\beta_{1}, \cdots , \beta_{n-N}, \beta_{n-N+1}'$ and $\beta_{n-N+2}$ 
by using 
\bea 
\phi(\alpha | \beta_{1}, \cdots , \beta_{n-N}, \beta_{n-N+1})= 
\phi(\alpha | \beta_{1}, \cdots , \beta_{n-N}, \beta_{n-N+1}')
\frac{\alpha-\beta_{n-N+1}'-N\hbar}{\alpha-\beta_{n-N+1}'-(N+1)\hbar}. 
\ena 
Then the integrand $\phi^{(N-2)}w_{J}^{*}\widehat{W}_{1}$ is given by 
\bea 
&& 
\phi^{(N-2)}(\{\alpha_{a}\}|\{\beta_{j}\}_{j \le n-N}, \beta_{n-N+1}' | \beta_{n-N+2}) 
\widehat{W}_{1}(\{e^{-\alpha_{a}}\}) 
\prod_{a=1}^{\ell-N+2} 
\frac{\alpha_{a}-\beta_{n-N+1}'-N\hbar}{\alpha_{a}-\beta_{n-N+1}'-(N+1)\hbar} 
\no  \\ 
&& {}\times                                                                             
\sum_{\sigma \in S_{\ell-N+2}}({\rm sgn}\sigma) 
g_{J}^{*} 
(\{\alpha_{\sigma(a)}\} | \beta_{n-N+1}'+N\hbar, \beta_{1}, \cdots , \beta_{n-N} | \beta_{n-N+2}). 
\label{rewrite1} 
\ena 
In the case of $J_{n-N+2}>0$, we also change the integration variable 
$\alpha_{\sigma(1)} \to \alpha_{\sigma(1)}-2\pi i$ and 
set $\tau :=\sigma  \cdot (1, 2, \cdots , \ell-N+2) \in S_{\ell-N+2}$. 
Then we get the following integral: 
\bea 
&& 
\phi^{(N-2)}(\{\alpha_{a}\}|\{\beta_{j}\}_{j \le n-N}, \beta_{n-N+1}' | \beta_{n-N+2}) 
\widehat{W}_{1}(\{e^{-\alpha_{a}}\}) \label{rewrite2}  \\ 
&& {}\times 
\sum_{\tau \in S_{\ell-N+2}}(-1)^{\ell-N+1}({\rm sgn}\tau) 
\prod_{a=1}^{\ell-N+1}
\frac{\alpha_{\tau(a)}-\beta_{n-N+1}'-N\hbar}{\alpha_{\tau(a)}-\beta_{n-N+1}'-(N+1)\hbar} 
\prod_{j=1}^{n-N}\frac{\alpha_{\tau(\ell-N+2)}-\beta_{j}-\hbar}
                      {\alpha_{\tau(\ell-N+2)}-\beta_{j}} \no  \\ 
&& \quad {}\times 
\frac{\alpha_{\tau(\ell-N+2)}-\beta_{n-N+2}-\frac{N}{2}\hbar} 
     {\alpha_{\tau(\ell-N+2)}-\beta_{n-N+2}+(\frac{N}{2}-1)\hbar} 
g_{J}^{*}
(\{\alpha_{\tau(a)}\} | \beta_{n-N+1}'+N\hbar , \beta_{1}, \cdots , \beta_{n-N} | \beta_{n-N+2}). 
\no 
\ena 
We see that the contour for $\alpha_{\ell-N+2}$ may be pinched 
by poles at $\alpha_{\ell-N+2}=\beta_{n-N+1}'+\hbar$ and 
$\alpha_{\ell-N+2}=\beta_{n-N+2}-(\frac{N}{2}-1)\hbar$. 
This pinch occurs only when $\sigma(\ell-N+2)=\ell-N+2$ in (\ref{rewrite1}), 
and $\tau(\ell-N+1)=\ell-N+2$ or $\tau(\ell-N+2)=\ell-N+2$ in (\ref{rewrite2}). 
Hence it suffices to calculate the residue at 
$\alpha_{\ell-N+2}=\beta_{n-N+2}-(\frac{N}{2}-1)\hbar$ for such terms. 
%we can calculate the residue at $\beta_{n-N+2}=\beta_{n-N+1}'+\frac{N}{2}\hbar$ 
%in the same way as in the proof of Lemma \ref{resIIlem} and Lemma \ref{lastres+}. 
In this calculation, we use the following formula
\bea
&& \!\!\!\!
\frac{\alpha-\beta}{\alpha-\beta+N\hbar} 
G_{\epsilon_{1}, \cdots , \epsilon_{\ell-N}, 
   [\epsilon_{\ell-N+1}, \cdots , \epsilon_{\ell-1}], \epsilon_{\ell}} 
 (\alpha_{1}, \cdots , \alpha_{\ell-N}, 
  \beta-(N-1)\hbar, \beta-\hbar, \cdots , \beta-(N-2)\hbar, \alpha) \no \\
&& {}+(N-1) 
G_{\epsilon_{1}, \cdots , \epsilon_{\ell-N}, 
   [\epsilon_{\ell-N+1}, \cdots , \epsilon_{\ell-1}], \epsilon_{\ell}} 
 (\alpha_{1}, \cdots , \alpha_{\ell-N}, 
  \alpha, \beta-\hbar, \cdots , \beta-(N-1)\hbar) \\
&&  \!\!\!\! {}=
\frac{\alpha-\beta+(N-1)\hbar}{\alpha-\beta} 
G_{\epsilon_{1}, \cdots , \epsilon_{\ell-N}, 
   [\epsilon_{\ell-N+1}, \cdots , \epsilon_{\ell-1}, \epsilon_{\ell}]} 
 (\alpha_{1}, \cdots , \alpha_{\ell-N}, 
  \alpha, \beta-\hbar, \cdots , \beta-(N-1)\hbar) \no
\ena
instead of (\ref{keyformula}). 
This formula can be obtained from (\ref{Grel1}), (\ref{Geval1}) and (\ref{Geval2}). 

After this calculation, we get the following integral:
\bea
&& 
\left( \prod_{a=1}^{\ell -N+1}\int_{C} d\alpha_{a} \right) 
\phi(\{\alpha_{a}\}|\{\beta_{j}\}_{1 \le j \le n-N}) 
\prod_{a=1}^{\ell-N+1}\frac{1}{\alpha_{a}-\beta_{n-N+1}}  \\
&& \qquad \qquad \qquad {}\times 
w_{J}^{(k, N-1)}(\{\alpha_{a}\}|\{\beta_{j}\}_{j \le n-N}|\beta_{n-N+1}-\hbar) 
\frac{P_{m-1}(\{e^{-\alpha_{a}}\})}
{\prod_{a=1}^{\ell-N+1}\prod_{j=1}^{n-N}(1-e^{-(\alpha_{a}-\beta_{j})})}. \no
\ena
{}From (\ref{Geval1}) and (\ref{Geval3}), we get (\ref{lastres-eq}). 
\qed 

Note that 
\bea
&&
\prod_{j=1}^{n-N} 
\frac{\Gamma( \frac{\beta_{n-N+1}-\beta_{j}-N\hbar}{-2\pi i}+1) 
      \Gamma( \frac{\beta_{n-N+1}-\beta_{j}-(N-1)\hbar}{2\pi i})}
     {\Gamma( \frac{\beta_{n-N+1}-\beta_{j}-\hbar}{-2\pi i}+1) 
      \Gamma( \frac{\beta_{n-N+1}-\beta_{j}}{2\pi i})}
=\prod_{j=1}^{n-N}S_{0}(\beta_{n-N+1}-\beta_{j}). 
\ena
Therefore we get 
\bea
&& 
2\pi i{\rm res}_{\beta_{n-N+2}=\beta_{n-N+1}-\frac{N}{2}\hbar}\widehat{\psi} \no \\
&& {}=
(-1)^{\frac{(N-1)(N-2)}{2}m}a_{m, N-1}(-2\pi i)^{\ell-N+1} \no \\
&& {}\times \prod_{j=1}^{n-N}
\Bigl( e^{\frac{1}{2}(\beta_{n-N+1}-\beta_{j})} 
 \Gamma( \frac{\beta_{n-N+1}-\beta_{j}-\hbar}{-2\pi i}+1) 
 \Gamma( \frac{\beta_{n-N+1}-\beta_{j}}{2\pi i}) 
\Bigr) \no \\
&& 
{}\times d_{m}^{-1}e^{-\frac{(N-1)(2n-N)}{2}\beta_{n-N+1}} \no \\
&& {}\times 
\left(
I+e^{-\frac{2\pi i}{N}+\frac{N-1}{N}(n-N)\pi i}
 S_{n-N+1, n-N}(\beta_{n-N+1}-\beta_{n-N})\cdots S_{n-N+1, 1}(\beta_{n-N+1}-\beta_{1}) 
\right) \no \\
&& \quad {}\times
\Psi_{P_{m-1}}(\beta_{1}, \cdots , \beta_{n-N})\otimes v_{[0, 1, \cdots , N-1]}.
\ena

At last we write down the formula for ${\rm res}f^{(N-1)}$. 
Note that, for any regular function $F(\beta_{1}, \cdots , \beta_{n})$, 
we have 
\bea 
&& 
2\pi i{\rm res}_{\beta_{n-N+2}=\beta_{n-N+1}-\frac{N}{2}\hbar} \circ 
{\rm RES}_{N-2} \circ \cdots \circ {\rm RES}_{1} (F\psi_{W}) \no \\
&& {}= 
F(\beta_{1}, \cdots , \beta_{n-N}, \beta_{n-N+1}, \beta_{n-N+1}-\hbar, \cdots , 
   \beta_{n-N+1}-(N-1)\hbar) \no \\
&& \quad {}\times 
2\pi i{\rm  res}_{\beta_{n-N+2}=\beta_{n-N+1}-\frac{N}{2}\hbar} \circ 
{\rm RES}_{N-2} \circ \cdots \circ {\rm RES}_{1} (\psi_{W}). 
\ena 
By using 
\bea
&& 
\prod_{k=0}^{N-1}\zeta(\beta+k\hbar)=
\prod_{s=0}^{N-2}
\left\{
\Gamma_{1}(\frac{-i\beta+\frac{2(s+1)}{N}\pi}{2\pi})
\Gamma_{1}(\frac{\beta+\frac{2s}{N}\pi}{2\pi}) \right\}^{-1} \no \\
&& \quad {}=
(2\pi)^{\frac{(N-1)(N+1)}{N}}
\prod_{s=0}^{N-2}
\left\{ 
\Gamma(\frac{\beta+(N-s-1)\hbar}{2\pi i}+1)
\Gamma(\frac{\beta+s\hbar}{-2\pi i}) \right\}^{-1},
\ena
we get 
\bea
&& 
2\pi i{\rm res}_{\beta_{n-N+2}=\beta_{n-N+1}-\frac{N}{2}\hbar}f^{(N-1)}={} \no \\
&& {}=
d_{m}^{-1}
(-i)^{\frac{(N-1)(N-2)}{2}m}e^{\frac{(N-1)m}{2}\pi i}
(-2\pi i)^{(N-1)(m-1)}(2\pi )^{(N-1)(N+1)(m-1)}
\prod_{s=1}^{N-1}\zeta(s\hbar)^{N-s}\prod_{k=1}^{N-1}a_{m, k} \no \\
&& {}\times 
\left(
I+e^{-\frac{2\pi i}{N}+\frac{N-1}{N}(n-N)\pi i}
 S_{n-N+1, n-N}(\beta_{n-N+1}-\beta_{n-N})\cdots S_{n-N+1, 1}(\beta_{n-N+1}-\beta_{1}) 
\right) \no \\
&& \quad {}\times
f_{P_{m-1}}(\beta_{1}, \cdots , \beta_{n-N})\otimes v_{[0, 1, \cdots , N-1]}.
\ena
Hence, if $d_{m}$ is given by 
\bea
&& 
d_{m}=(-i)^{\frac{(N-1)(N-2)}{2}m}e^{\frac{(N-1)m}{2}\pi i}
(-2\pi i)^{(N-1)(m-1)}(2\pi )^{(N-1)(N+1)(m-1)} \no \\
&& \qquad \quad {}\times 
\prod_{s=1}^{N-1}\zeta(s\hbar)^{N-s}\prod_{k=1}^{N-1}a_{m, k},
\label{defdm}
\ena
then (\ref{res2}) holds. This completes the proof of Proposition \ref{rescond}.

\subsection{Deformed cycles associated with energy momentum tensor}

Hereafter we use the following notation: 
\bea
A_{a}:=e^{-\alpha_{a}}, \quad B_{j}:=e^{-\beta_{j}}, \quad {\rm and } 
\quad \omega:=e^{\hbar}=e^{-\frac{2\pi i}{N}}. 
\ena

The $n$ rank-1 particle form factor $f_{\mu\nu}$ of 
the energy momentum tensor $T_{\mu\nu}$ was determined in \cite{smirbook}. 
In terms of our formula, it is given by 
\bea
f_{\mu\nu}(\beta_{1}, \cdots , \beta_{n})= 
C_{0}f_{P_{\mu\nu}}(\beta_{1}, \cdots , \beta_{n}), 
\ena
where $C_{0}$ is a constant independent of $n, \mu, \nu$, and  
$P_{\mu\nu}$ is given by 
\bea 
&& 
P_{\mu\nu}(A_{1}, \cdots , A_{\ell}):=c_{m}
\left( \sum_{j=1}^{n}B_{j}^{-1}-(-1)^{\mu}\sum_{j=1}^{n}B_{j} \right) \label{emtcycle} \\ 
&& {}\times 
\left( (-1)^{\nu+\frac{(N-1)(N-2)}{2}m}\omega^{-\frac{m(m-1)}{2}} 
       (\prod_{a=1}^{\ell}A_{a})^{n} \, w(A_{2}^{-1}, \cdots , A_{\ell}^{-1}) 
       +w(A_{2}, \cdots , A_{\ell}) \right). \no
\ena 
Note that $n=Nm$ and $\ell=(N-1)m$. 
Here  
\bea 
c_{m}:=\omega^{-\frac{N(N-1)(N-2)}{3}m}\prod_{j=1}^{m}d_{j}^{-1}, \qquad
w(A_{2}, \cdots , A_{\ell}):=\prod_{a=2}^{\ell}A_{a}^{a+[\frac{a-1}{N-1}]}, 
\ena 
where $[ \, \cdot \, ]$ is Gauss' symbol. 

In this subsection, we prove that 
$f_{\mu\nu}$ satisfies (\ref{res0}), (\ref{res1}) and (\ref{res2}).  

First we consider the case of $m>1$. 
Fix $m$ such that $m>1$,  
and set   
\bea 
&& 
P^{-}:= c_{m}(-1)^{\frac{(N-1)(N-2)}{2}m}\omega^{-\frac{m(m-1)}{2}} 
       (\prod_{a=1}^{\ell}A_{a})^{n} \, 
       w^{-}(A_{1}^{-1}, \cdots , A_{\ell}^{-1}), \quad {\rm and} \\ 
&& 
P^{+}:=c_{m}w^{+}(A_{1}, \cdots , A_{\ell}), 
\ena 
where 
\bea 
w^{\pm}(A_{1}, \cdots , A_{\ell}):= 
\prod_{j=0}^{N-1}(1-\omega^{j}B_{n-1}^{\mp 1}A_{1}) 
w(A_{2}, \cdots , A_{\ell}). 
\ena 
{}From Corollary \ref{zerocycle}, we have $f_{w^{+}(\{A_{a}\})}=f_{w(\{A_{a}\})}$ and 
$f_{(\prod_{a}A_{a})^{n}w^{-}(\{A_{a}^{-1}\})}=f_{(\prod_{a}A_{a})^{n}w(\{A_{a}^{-1}\})}$. 
Hence we have 
\bea
f_{P_{\mu\nu}}=f_{P'_{\mu\nu}}, \quad 
P'_{\mu\nu}:= 
\left( \sum_{j=1}^{n}B_{j}^{-1}-(-1)^{\mu}\sum_{j=1}^{n}B_{j} \right)
\left( (-1)^{\nu}P^{-}+P^{+} \right). 
\ena

\begin{prop}\label{emtrec}
Set $P_{m}:=P^{\pm}$ and 
\bea 
&& 
P_{m-1}:=c_{m-1}w(A_{2}, \cdots , A_{\ell-N+1}), \quad {\rm for} \quad P_{m}=P^{+}, 
\label{cyclesum1} \\
&& 
P_{m-1}:=c_{m-1}(-1)^{\frac{(N-1)(N-2)}{2}(m-1)}\omega^{-\frac{(m-1)(m-2)}{2}} \no \\
&& \qquad \qquad {}\times 
       (\prod_{a=1}^{\ell-N+1}A_{a})^{n-N} \, w(A_{2}^{-1}, \cdots , A_{\ell-N+1}^{-1}), 
\quad {\rm for} \quad P_{m}=P^{-}. 
\label{cyclesum2} \ena 
Then $P_{m}$ and $P_{m-1}$ satisfy the assumption in Proposition \ref{rescond} 
for certain polynomials $\widehat{P}^{(k)}$ and $P^{(k)}, \, (k=1, \cdots , N-1)$.  
\end{prop} 

Note that 
\bea 
\sum_{j=0}^{N-1}(\omega^{j}B_{n-N+1})^{\pm 1}=0. 
\label{omegasum} 
\ena 
Therefore Proposition \ref{emtrec} implies that $f_{P_{\mu\nu}}$ satisfies 
(\ref{res0}), (\ref{res1}) and (\ref{res2}) for $m>1$. 

In the proof of Proposition \ref{emtrec}, 
we use the following lemmas: 

\begin{lem}\label{skew1}
Set 
\bea 
P_{k}(A_{1}, \cdots , A_{N} |B):= 
\sum_{j=0}^{k}\omega^{-\frac{j(j-1)}{2}}{ \, k \,  \atopwithdelims[] \, j\,  } 
B^{-j}\prod_{a=1}^{N-j-1}A_{a}^{a}\prod_{a=N-j}^{N-1}A_{a}^{a+1}\cdot A_{N}^{N+1}. 
\ena 
for $k=0, \cdots , N-2$. 
Here ${ \, k\,  \atopwithdelims[] \, j\,  }$ is defined by 
\bea 
{ \, k\,  \atopwithdelims[] \, j\,  }:=
\frac{(1-\omega^{-k})(1-\omega^{-(k-1)}) \cdots (1-\omega^{-(k-j+1)})}
     {(1-\omega^{-j})(1-\omega^{-(j-1)}) \cdots (1-\omega^{-1})},
\ena 
that is, the $q$-binomial coefficient with $q=\omega^{-1}$. 
Then 
\bea 
{\rm Skew}\left( P_{k}|_{B \to \omega B} \right)= 
{\rm Skew}\left( \prod_{a=1}^{N-1}(1-B^{-1}A_{a}) P_{k+1} \right), 
\label{skew1eq} 
\ena 
where ${\rm Skew}$ is the skew-symmetrization with respect to $A_{1}, \cdots , A_{N}$. 
\end{lem}

We can prove this lemma easily by using 
\bea
{ \, k\,  \atopwithdelims[] \, j\,  }=
{ \, k-1\,  \atopwithdelims[] \, j-1\,  }+
\omega^{-j}{ \, k-1\,  \atopwithdelims[] \, j\,  }. 
\ena 

It is also easy to see that the following lemma holds: 

\begin{lem}\label{skew2}
Suppose that 
\bea 
c_{1}B^{-2}\omega+c_{2}B^{-1}(1+\omega)+c_{3}=0 
\ena 
for three constants $c_{1}, c_{2}$ and $c_{3}$. 
Then 
\bea 
&& 
{\rm Skew} \left( 
c_{1}A_{2}A_{3}^{3}+c_{2}A_{2}^{2}A_{3}^{3}+c_{3}A_{1}A_{2}^{2}A_{3}^{3} \right) \no \\
&& {}=
{\rm Skew} \left( 
(1-B^{-1}A_{1})(1-\omega B^{-1}A_{1})(c_{1}A_{2}A_{3}^{3}+c_{2}A_{2}^{2}A_{3}^{3}) \right) , 
\ena 
where ${\rm Skew}$ is the skew-symmetrization with respect to $A_{1}, A_{2}$ and $A_{3}$. 
\end{lem} 

From Lemma \ref{skew2} we find the following formula: 

\begin{lem}\label{skew3} 
\bea 
&& 
{\rm Skew}\left( 
P_{N-2}(A_{1}, \cdots , A_{N} | \omega B) \right) \no \\
&& {}= 
{\rm Skew}\left( 
\prod_{a=1}^{N-1}(1-B^{-1}A_{a})(1-\omega B^{-1}A_{a}) 
P_{0}(A_{1}, \cdots , A_{N} |B) \right). \label{skew3eq}
\ena
\end{lem} 

\proof 
The lhs is given by 
\bea 
P_{N-2}(A_{1}, \cdots , A_{N} | \omega B)= 
\sum_{j=0}^{N-2}\omega^{-\frac{j(j+1)}{2}}{ \, N-2 \,  \atopwithdelims[] \, j\,  } 
B^{-j}\prod_{a=1}^{N-j-1}A_{a}^{a}\prod_{a=N-j}^{N-1}A_{a}^{a+1}\cdot A_{N}^{N+1}. 
\ena 
It is easy to see that 
\bea 
{ \, N-2 \,  \atopwithdelims[] \, j\,  }\omega+
\omega^{-(j+1)}{ \, N-2 \,  \atopwithdelims[] \, j+1\,  }(1+\omega)+ 
\omega^{-(j+1)-(j+2)}{ \, N-2 \,  \atopwithdelims[] \, j+2\,  }=0, 
\ena 
for $j=0, \cdots , N-4$. 
Hence, from Lemma \ref{skew2}, we get 
\bea 
&& 
P_{N-2}(A_{1}, \cdots , A_{N} | \omega B) \\ 
&& {}\sim 
\prod_{a=1}^{N-2}(1-B^{-1}A_{a})(1-\omega B^{-1}A_{a}) 
\prod_{a=1}^{N-1}A_{a}^{a}
\left( 
1+\omega^{-1}{ \, N-2 \,  \atopwithdelims[] \, 1\,  }B^{-1}A_{N-1} \right) 
A_{N}^{N+1}. \no 
\ena 
Here we write $f \sim g$ if ${\rm Skew}(f-g)=0$. 

Note that 
\bea 
1+\omega^{-1}{ \, N-2 \,  \atopwithdelims[] \, 1\,  }B^{-1}A_{N-1}=
(1-B^{-1}A_{N-1})(1-\omega B^{-1}A_{N-1})-\omega B^{-2}A_{N-1}^{2}. 
\ena 
Therefore (\ref{skew3eq}) holds. 
\qed 

Set 
\bea 
P'_{k}(A_{1}, \cdots , A_{N}| B):=
\prod_{j=0}^{N-k-1}(1-\omega^{j}B^{-1}A_{1})P_{k}(1, A_{2}, \cdots , A_{N} | B)
\ena 
for $k=0, \cdots , N-2$. 
Then we can also see that 
\bea 
&& 
{\rm Skew}\left( P'_{k}|_{B \to \omega B} \right)= 
{\rm Skew}\left( \prod_{a=1}^{N-1}(1-B^{-1}A_{a}) P'_{k+1} \right), \quad {\rm and} \\ 
&& 
{\rm Skew}\left( 
P'_{N-2}(A_{1}, \cdots , A_{N} | \omega B) \right) \no \\
&& {}= 
{\rm Skew}\left( 
\prod_{a=1}^{N-1}(1-B^{-1}A_{a})(1-\omega B^{-1}A_{a}) 
P_{0}(1, A_{2}, \cdots , A_{N} |B) \right). 
\ena 

\noindent{\it Proof of Proposition \ref{emtrec}} 

First let us prove for $P_{m}=P^{+}$. 
Note that 
\bea 
&& 
P^{+}=
c_{m}P'_{0}(A_{1}, \cdots , A_{N-1}, 1) \\
&& \qquad {}\times 
\prod_{s=1}^{m-2}\left( \prod_{a=s(N-1)+1}^{(s+1)(N-1)}A_{a} \right)^{sN} 
\!\! P_{0}(A_{s(N-1)+1}, \cdots , A_{(s+1)(N-1)}, 1) 
\prod_{a=\ell-N+2}^{\ell}A_{a}^{n-\ell-1+a}. \no
\ena 

We set 
\bea 
&& 
\widehat{P}^{(k)}:= 
c_{m}\prod_{j=0}^{n-k+1+j}(\omega^{j}B_{n-k})^{n-k+1+j} 
P'_{k}(A_{1}, \cdots , A_{N-1}, 1 | B_{n-k}) \\
&& \qquad {}\times 
\prod_{s=1}^{m-2}\left( \prod_{a=s(N-1)+1}^{(s+1)(N-1)}A_{a} \right)^{sN} 
\!\! P_{k}(A_{s(N-1)+1}, \cdots , A_{(s+1)(N-1)}, 1 | B_{n-k}) 
\prod_{a=\ell-N+2}^{\ell-k+1}A_{a}^{n-\ell-1+a}, \no
\ena 
for $k=1, \cdots , N-2$, and 
\bea 
&& 
\widehat{P}^{(N-1)}:= 
c_{m}\prod_{j=0}^{N-3}(\omega^{j}B_{n-N+1})^{n-N+2+j} 
P_{0}(A_{1}, \cdots , A_{N-1}, 1)  \\
&& \qquad \quad {}\times 
\prod_{s=1}^{m-2}\left( \prod_{a=s(N-1)+1}^{(s+1)(N-1)}A_{a} \right)^{sN} 
\!\! P_{0}(A_{s(N-1)+1}, \cdots , A_{(s+1)(N-1)}, 1 ) 
A_{\ell-N+2}^{n-N+1}. \no
\ena 
We define $P^{(k)}$ from $\widehat{P}^{(k)}$ by (\ref{cond3}). 
Then we can check that $\widehat{P}^{(k)}$ and 
$P^{(m)}, \, (k=1, \cdots , N-1)$ satisfy the assumption 
in Proposition \ref{rescond} by using Lemma \ref{skew1}, Lemma \ref{skew3} and 
\bea 
P_{k}(A_{1}, \cdots , A_{N} | B)=A_{1}P_{k}(1, A_{2}, \cdots , A_{N} | B) 
=P_{k}(A_{1}, \cdots , A_{N-1}, 1 | B)A_{N}^{N+1}. 
\ena

We can prove the case of $P_{m}=P^{-}$ in a similar way by using 
\bea 
&& 
{\rm Skew}\left( P_{k}(A_{1}^{-1}, \cdots , A_{N}^{-1}| \omega^{-1} B^{-1}) \right) \no \\
&& {}= 
{\rm Skew}\left( \prod_{a=1}^{N-1}(-BA_{a}^{-1})(1-B^{-1}A_{a}) \cdot P_{k+1} \right),  
\ena 
and 
\bea 
&& 
{\rm Skew}\left( 
P_{N-2}(A_{1}^{-1}, \cdots , A_{N}^{-1} | \omega^{-1} B^{-1}) \right)  \\
&& {}= 
{\rm Skew}\left( 
\prod_{a=1}^{N-1}(\omega^{-1}B^{2}A_{a}^{-2})(1-B^{-1}A_{a})(1-\omega B^{-1}A_{a}) \cdot
P_{0}(A_{1}^{-1}, \cdots , A_{N}^{-1} |B^{-1}) \right) \no
\ena
instead of (\ref{skew1eq}) and (\ref{skew3eq}), respectively. 
\qed

At last we show that $f_{\mu\nu}$ satisfies (\ref{res0}), (\ref{res1}) and (\ref{res2}) 
in the case of $m=1$. 
In this case, we set 
\bea 
\widehat{P}^{(k)}:=P_{k}'(A_{1}, \cdots , A_{N-k}, 
                      \omega B_{N-k}, \cdots , \omega^{k-1} B_{N-k}, 1 | B_{N-k})
\ena 
for $k=1, \cdots , N-2$. 
Then the assumption in Proposition \ref{rescond} is satisfied for $P_{1}=P^{+}$ 
except (\ref{cond3.5}) and (\ref{cond4}). 
Similarly, we can see that the assumption except (\ref{cond3.5}) and (\ref{cond4}) holds 
for $P_{1}=P^{-}$. 
Hence in the same way as the proofs of Lemma \ref{resIIlem} and Lemma \ref{resIlem} 
we can calculate the residue 
\bea 
{\rm RES}_{N-2} \circ \cdots \circ {\rm RES}_{1}f_{P^{\pm}_{\mu}}, 
\label{minimality} 
\ena 
where 
\bea 
P^{\pm }_{\mu}:=
\left( \sum_{j=1}^{n}B_{j}^{-1}-(-1)^{\mu}\sum_{j=1}^{n}B_{j} \right)P^{\pm }.
\ena 
Then we see that the residue of (\ref{minimality}) 
at $\beta_{2}=\beta_{1}+\pi i$ equals zero because of (\ref{omegasum}). 

Therefore, the form factor $f_{P_{\mu\nu}}$ satisfies 
(\ref{res0}), (\ref{res1}) and (\ref{res2}) for all $m>0$.

\section{Supplements and proofs}

\subsection{Properties of Smirnov's basis}\label{app1}

First we extend the definition of 
$\omega_{\epsilon_{1}, \cdots , \epsilon_{n}}(\beta_{1}, \cdots , \beta_{n})$ 
in Section \ref{sectiononetime} as follows. 
%
%We denote by ${\cal Z}_{m_{0}, \cdots , m_{N-1}}, \, (\sum_{i}m_{i}=n)$ 
%the set of all $n$-tuples $(\epsilon_{1}, \cdots , \epsilon_{n})$ such that 
%\bea
%\# \{j | \epsilon_{j}=k\}=m_{k}, \quad {\rm for \,\, all} \quad k=0, \cdots , N-1. 
%\ena
For $\epsilon=(\epsilon_{1}, \cdots , \epsilon_{n}) \in {\cal Z}_{\nu_{1}, \cdots , \nu_{N-1}}$, we set 
\bea
v_{\epsilon}:=v_{\epsilon_{1}} \otimes \cdots \otimes v_{\epsilon_{n}}. 
\ena
Define a partial order in ${\cal Z}_{\nu_{1}, \cdots , \nu_{N-1}}$: 
\bea
(\epsilon_{1}, \cdots , \epsilon_{n}) \le (\epsilon_{1}', \cdots , \epsilon_{n}') 
\qquad {\rm if \, \, and \,\, only \,\, if} \qquad 
 \sum_{i=1}^{r}\epsilon_{i} \le \sum_{i=1}^{r}\epsilon_{i}' \quad 
{\rm for \, \, all }\quad r. 
\ena
We define two elements $\epsilon^{{\rm max}}$ and $\epsilon^{{\rm min}}$ of 
${\cal Z}_{\nu_{0}, \cdots , \nu_{N-1}}$ by 
\bea
&& 
\epsilon^{{\rm max}}:=(
\underbrace{N-1, \cdots , N-1}_{\nu_{N-1}}, \cdots , 
\underbrace{1, \cdots , 1}_{\nu_{1}-\nu_{2}}, 
\underbrace{0, \cdots , 0}_{n-\nu_{1}}), \\
&& 
\epsilon^{{\rm min}}:=(
\underbrace{0, \cdots , 0}_{n-\nu_{1}}, 
\underbrace{1, \cdots , 1}_{\nu_{1}-\nu_{2}}, \cdots , 
\underbrace{N-1, \cdots , N-1}_{\nu_{N-1}}). 
\ena 

We define 
$\{\omega_{\epsilon_{1}, \cdots , \epsilon_{n}}(\beta_{1}, \cdots , \beta_{n})\}_
 {(\epsilon_{1}, \cdots , \epsilon_{n}) \in {\cal Z}_{\nu_{1}, \cdots , \nu_{N-1}}}$ 
by the conditions (\ref{omegarel}) and $\omega_{\epsilon^{{\rm min}}}:=v_{\epsilon^{{\rm min}}}$. 
Then we see that 
\bea 
\omega_{\epsilon_{1}, \cdots , \epsilon_{n}}(\beta_{1}, \cdots , \beta_{n})=
\prod_{a<b \atop (\epsilon_{a} > \epsilon_{b})} 
\frac{\beta_{a}-\beta_{b}}{\beta_{a}-\beta_{b}-\hbar} 
v_{\epsilon_{1}, \cdots , \epsilon_{n}} + 
({\rm lower \,\, term}). 
\label{triangular} 
\ena

\begin{lem}\label{omegahw}
For $(\epsilon_{1}, \cdots , \epsilon_{n}) \in {\cal Z}_{\nu_{1}, \cdots , \nu_{N-1}}$, 
the following formula holds: 
\bea 
E_{k}\omega_{\epsilon_{1}, \cdots , \epsilon_{n}}= 
\sum_{a \atop (\epsilon_{a}=k)} 
\prod_{b \atop (\epsilon_{b}=k, b \not= a)} \!\!\!
\frac{\beta_{b}-\beta_{a}-\hbar}{\beta_{b}-\beta_{a}} \,
\omega_{\epsilon_{1}, \cdots , \epsilon_{a}-1, \cdots , \epsilon_{n}}, \quad (k=1, \cdots N-1). 
\label{omegahweq} 
\ena
\end{lem}

\proof 
The proof in the case of $N=2$ is given in \cite{smirbook}. 
Here let us prove the case of $N>2$. 

Note that the action of $E_{k}$ commutes with that of $R_{i, i+1}$ for all $i$. 
Hence we see that both sides satisfy (\ref{omegarel}). 
Moreover, it can be checked that (\ref{omegahweq}) holds for 
$(\epsilon_{1}, \cdots , \epsilon_{\ell})=\epsilon^{{\rm min}}$ 
by using (\ref{omegahweq}) in the case of $N=2$. 
Therefore, (\ref{omegahweq}) holds for all $(\epsilon_{1}, \cdots , \epsilon_{n})$. 
\qed 

In the rest of this subsection, we use the following simple lemma. 

\begin{lem}\label{tec}
Suppose that a $(V_{N})^{\otimes n}$-valued function 
\bea
F(x_{1}, \cdots , x_{n})=
\sum_{(\epsilon_{1}, \cdots , \epsilon_{n}) \in {\cal Z}_{\nu_{1}, \cdots , \nu_{N-1}}}
F_{\epsilon_{1}, \cdots , \epsilon_{n}}(x_{1}, \cdots , x_{n}) 
v_{\epsilon_{1}} \otimes \cdots \otimes v_{\epsilon_{n}} 
\ena
satisfies 
\bea 
F( \cdots , x_{j+1}, x_{j}, \cdots )=
P_{j, j+1}R_{j, j+1}(x_{j}-x_{j+1}) F( \cdots , x_{j}, x_{j+1}, \cdots ). 
\label{Frel} 
\ena 
and $F_{\epsilon_{1}, \cdots , \epsilon_{n}}=0$ 
for some $(\epsilon_{1}, \cdots , \epsilon_{n})$. 
Then $F=0$. 
\end{lem} 

By using $\{\omega_{\epsilon_{1}, \cdots , \epsilon_{n}}\}$, 
we can get another formula for the special solution at level one (\ref{ssol}), 
and prove the highest weight condition as follows. 

\begin{prop}\label{Hanotherformula}
\bea 
&& 
\sum_{(\epsilon_{1}, \cdots , \epsilon_{\ell}) \in {\cal Z}_{(N-2)m, \cdots ,2m, m}}
H_{\epsilon_{1}, \cdots , \epsilon_{\ell}}(\alpha_{1}, \cdots , \alpha_{\ell}) 
v_{\epsilon_{1}} \otimes \cdots \otimes v_{\epsilon_{\ell}} \no \\
&&\qquad  {}= 
\sum_{(\epsilon_{1}, \cdots , \epsilon_{\ell}) \in {\cal Z}_{(N-2)m, \cdots , 2m, m}}
\prod_{a, b \atop (\epsilon_{a}<\epsilon_{b})} 
\frac{1}{\alpha_{a}-\alpha_{b}}
\omega_{\epsilon_{1}, \cdots , \epsilon_{\ell}}(\alpha_{1}, \cdots , \alpha_{\ell}), 
\label{anotherformula} 
\ena 
where $\ell=(N-1)m$. 
The function (\ref{anotherformula}) satisfies the highest weight condition. 
\end{prop} 

\proof 
{}From (\ref{rel1}) and (\ref{omegarel}), 
it is easy to see that both sides satisfy (\ref{Frel}). 

Now we consider the coefficients of $v_{\epsilon^{{\rm max}}}$ of both sides. 
{}From (\ref{triangular}), we can calculate the coefficient of the rhs easily, 
and see that it suffices to prove that 
\bea 
H_{\epsilon^{{\rm max}}}=
\prod_{a, b \atop (\epsilon^{{\rm max}}_{a}<\epsilon_{b}^{{\rm max}})}
\frac{1}{\alpha_{a}-\alpha_{b}+\hbar}. 
\label{extcoeff}
\ena
First consider the case of $N=3$. 
Then we can calculate $H_{\epsilon^{{\rm max}}}$ explicitly 
{}from (\ref{rel2}) and (\ref{rel3}), and get (\ref{extcoeff}). 
In the case of $N>3$, we have the following from (\ref{extcoeff}) for $N=3$: 
\bea
H_{1, \cdots , 1, 0, \cdots , 0, 2, \cdots , 2, \cdots , N-2, \cdots , N-2}=
\prod_{a, b \atop (\epsilon_{a}=0, \epsilon_{b}=1)} 
 \frac{1}{\alpha_{a}-\alpha_{b}+\hbar} 
\prod_{a, b \atop (\epsilon_{a}<\epsilon_{b}, 2m<b)} 
 \frac{1}{\alpha_{a}-\alpha_{b}-\hbar}. 
\ena
Repeating this calculation, we get (\ref{extcoeff}) for $N>3$. 
In this way we find (\ref{anotherformula}) from Lemma \ref{tec}. 

Let us prove the highest weight condition for the rhs of (\ref{anotherformula}). 
{}From (\ref{omegahw}), we have 
\bea
&& 
E_{k}({\rm the \, \, rhs \,\, of} \,\, (\ref{anotherformula}))= 
\sum_{\epsilon \in {\cal Z}_{(N-2)m, \cdots , (N-k-1)m-1, \cdots , m}}
\omega_{\epsilon}
\prod_{a, b \atop (\epsilon_{a}<\epsilon_{b}, (\epsilon_{a}, \epsilon_{b}) \not= (k, k+1))}
 \frac{1}{\alpha_{a}-\alpha_{b}} \no \\
&& \qquad {}\times 
\sum_{a \atop (\epsilon_{a}=k-1)} 
\prod_{j \atop (\epsilon_{j}=k-1, j \not= a)}
 \frac{1}{\alpha_{j}-\alpha_{a}}
\prod_{j, b \atop (\epsilon_{j}=k-1, j \not=a, \epsilon_{b}=k-1)} 
 \frac{1}{\alpha_{j}-\alpha_{b}} 
\prod_{j \atop (\epsilon_{j}=k)}
 \frac{\alpha_{j}-\alpha_{a}-\hbar}{\alpha_{j}-\alpha_{a}}. 
\ena
The second line above eqauls zero from the following lemma.
\qed

\begin{lem}
For $x_{1}, \cdots , x_{r+1}$ and $y_{1}, \cdots , y_{r-1}$, the following equality holds: 
\bea
\sum_{s=1}^{r+1}
\prod_{j \not=s}\left( 
\frac{1}{x_{j}-x_{s}}\prod_{t=1}^{r-1}\frac{1}{x_{j}-y_{t}} \right) 
\prod_{t=1}^{r-1}\frac{y_{t}-x_{s}-\hbar}{y_{t}-x_{s}}=0. 
\ena
\end{lem}

This lemma is easy to prove by induction on $r$. 
\newline

\noindent{\it Proof of Proposition \ref{basechange}}

The case of $N=2$ is proved in \cite{NPT}. 

{}From (\ref{wrel1}) for $w_{M}$ and (\ref{omegarel}), 
we see that both sides of (\ref{basechangeeq}) satisfy (\ref{Frel}). 
Hence it suffices to check the coefficients of $v_{\epsilon^{{\rm max}}}$ 
of both sides are equal, that is 
\bea
w_{\epsilon^{{\rm max}}}=(-1)^{\frac{\ell(\ell-1)}{2}} 
\prod_{r=1}^{N-1}\prod_{a<b \atop (\epsilon_{a}=r=\epsilon_{b})}
 \frac{(\beta_{b}-\beta_{a}-\hbar)(\beta_{a}-\beta_{b}-\hbar)}{\beta_{a}-\beta_{b}} 
\tilde{w}_{\epsilon^{{\rm max}}}. 
\label{coeffprove2}
\ena
This equality can be proved by using (\ref{coeffprove2}) with $N=2$. 
\qed

\subsection{Proofs of equalities of rational functions}\label{app2}

\noindent{\it Proof of Lemma \ref{hw1}} 

Here we set $r_{1, m}:=r_{1, m-1}^{J}, \, (2 \le m \le \nu_{1})$ 
and $r_{2, m}:=r_{2, m}^{J}, \, (1 \le m \le \nu_{2})$. 
We set $r_{1, 1}=0=r_{2, 0}$ and $r_{1, \nu_{1}+1}=n+1=r_{2, \nu_{2}+1}$. 

Define functions $f_{a}, \, (1 \le a \le n)$ as follows.  

For $r_{1, t}<a<r_{1, t+1}$ such that $r_{2, q}<a<r_{2, q+1}$, we set 
\bea
f_{a}&:=&(-1)^{t-1}
g_{M_{1}^{J}\cup \{a\}}
(\alpha_{2}, \cdots , \alpha_{t}, \alpha_{1}, \alpha_{t+1}, \cdots , \alpha_{\nu_{1}}| 
 \{\beta_{j}\}) \no \\
&& {}\times 
g_{M_{2}^{J}}(\{\gamma_{m}\}| \alpha_{2}, \cdots , \alpha_{\nu_{1}}) 
\prod_{k=q+1}^{\nu_{2}}\frac{\gamma_{k}-\alpha_{1}-\hbar}{\gamma_{k}-\alpha_{1}}. 
\ena
Note that ${\rm Skew}f_{a}=w_{J_{1}, \cdots , J_{a}+1, \cdots , J_{n}}^{(3)}$. 

For $a=r_{1, t}$ such that $r_{2, q}<r_{1, t}<r_{2, q+1}$, set 
\bea
&& 
f_{r_{1, t}}:=(-1)^{t} g_{M_{1}^{J}}(\alpha_{2}, \cdots , \alpha_{\nu_{1}} | \{\beta_{j}\})
g_{M_{2}^{J}}(\{\gamma_{m}\}| \alpha_{2}, \cdots , \alpha_{\nu_{1}}) 
\prod_{k=q+1}^{\nu_{2}}\frac{\gamma_{k}-\alpha_{1}-\hbar}{\gamma_{k}-\alpha_{1}}
\no \\
&& \quad {}\times 
\prod_{j < r_{1, t}}\frac{\alpha_{1}-\beta_{j}-\hbar}{\alpha_{1}-\beta_{j}}
\left\{
(\alpha_{1}-\alpha_{t}-\hbar)+
\frac{\alpha_{1}-\beta_{r_{1, t}}-\hbar}{\alpha_{1}-\beta_{r_{1, t}}}
(\alpha_{t}-\alpha_{1}-\hbar) \right\} \no \\
&& \quad {}\times 
\prod_{b=2}^{t-1}(\alpha_{b}-\alpha_{1}-\hbar) 
\prod_{b=t+1}^{\nu_{1}}(\alpha_{1}-\alpha_{b}-\hbar).
\ena
Note that $f_{r_{1, t}}$ is symmetric with respect to $\alpha_{1}$ and $\alpha_{t}$. 

For $a=r_{2, q}=r_{1, t}$, we set 
\bea
&& 
f_{r_{2, q}}:=(-1)^{t}g_{M_{1}^{J}}(\alpha_{2}, \cdots , \alpha_{\nu_{1}}| \{\beta_{j}\})
g_{M_{2}^{J}}(\{\gamma_{m}\}| \alpha_{2}, \cdots , \alpha_{\nu_{1}}) \no \\
&& \quad {}\times
\prod_{k=q+1}^{\nu_{2}}\frac{\gamma_{k}-\alpha_{1}-\hbar}{\gamma_{k}-\alpha_{1}} 
\prod_{j < r_{2, q}}\frac{\alpha_{1}-\beta_{j}-\hbar}{\alpha_{1}-\beta_{j}} 
\prod_{b=2}^{t-1}(\alpha_{b}-\alpha_{1}-\hbar) 
\prod_{b=t+1}^{\nu_{1}}(\alpha_{1}-\alpha_{b}-\hbar) \no \\
&& \quad {}\times 
\left\{ 
(\alpha_{1}-\alpha_{t}-\hbar)\frac{\gamma_{q}-\alpha_{1}-\hbar}{\gamma_{q}-\alpha_{1}}+
\frac{\alpha_{1}-\beta_{r_{2, q}}-\hbar}{\alpha_{1}-\beta_{r_{2, q}}}(\alpha_{t}-\alpha_{1}-\hbar)
\right\}. 
\ena
Note that $f_{r_{2, q}}$ is symmetric with respect to $\alpha_{1}$ and $\alpha_{t}$. 

It is easy to check that 
\bea
&& 
\hbar \sum_{a=1}^{n}f_{a}= 
g_{M_{1}^{J}}(\{\alpha_{a}\}_{2 \le a \le \nu_{1}}|\{\beta_{j}\})
g_{M_{2}^{J}}(\{\gamma_{m}\}| \{\alpha_{a}\}_{2 \le a \le \nu_{1}})  \\
&& \qquad \qquad {}\times 
\left(\prod_{a=2}^{\nu_{1}}(\alpha_{1}-\alpha_{a}-\hbar)
\prod_{m=1}^{\nu_{2}}\frac{\gamma_{m}-\alpha_{1}-\hbar}{\gamma_{m}-\alpha_{1}}-
\prod_{j=1}^{n}\frac{\alpha_{1}-\beta_{j}-\hbar}{\alpha_{1}-\beta_{j}}
\prod_{a=2}^{\nu_{1}}(\alpha_{1}-\alpha_{a}+\hbar)\right). \no
\ena
By skew-symetrizing both sides above, we have (\ref{hw11}). 
\qed

\noindent{\it Proof of Lemma \ref{simplehw}} 

The proof is quite similar to that of Lemma \ref{hw1}. 

We set $M_{1}^{J}=\{m_{2}, \cdots , m_{\ell}\}, \, m_{2}< \cdots m_{\ell}$, 
$m_{1}=0, m_{\ell+1}=n+1$ and 
$\epsilon_{r}:=\bar{J}_{r-1}, \, (2 \le r \le \ell)$. 

Define functions $f_{a}, \, (1 \le a \le n)$ as follows. 

For $m_{r}<a<m_{r+1}$, we set 
\bea
&&
f_{a}:=(-1)^{r-1}g_{M_{1}^{J}\cup\{a\}}(
\alpha_{2}, \cdots , \alpha_{r}, \alpha_{1}, \alpha_{r+1}, \cdots , \alpha_{\ell} | \{\beta_{j}\})
\no \\
&& \qquad \qquad {}\times  
H_{\epsilon_{2}, \cdots , \epsilon_{r}, 0, \epsilon_{r+1}, \cdots , \epsilon_{\ell}}
 (\alpha_{2}, \cdots , \alpha_{r}, \alpha_{1}, \alpha_{r+1}, \cdots , \alpha_{\ell}). 
\ena
Note that ${\rm Skew}f_{a}=w_{J_{1}, \cdots , J_{a}+1, \cdots , J_{n}}$. 

For $a=m_{r}$, we set 
\bea
&& 
f_{m_{r}}:=(-1)^{r-2}g_{M_{1}^{J}}(\alpha_{2}, \cdots , \alpha_{\ell}| \{\beta_{j}\}) 
\prod_{j<m_{r}}\frac{\alpha_{1}-\beta_{j}-\hbar}{\alpha_{1}-\beta_{j}} 
\prod_{b=2}^{r-1}(\alpha_{b}-\alpha_{1}-\hbar) 
\prod_{b=r+1}^{\ell}(\alpha_{1}-\alpha_{b}-\hbar) \no \\
&& \quad {}\times \Bigl\{
(\alpha_{1}-\alpha_{r}-\hbar) 
H_{\epsilon_{2}, \cdots , \epsilon_{r-1}, 0, \epsilon_{r}, \cdots , \epsilon_{\ell}}
 (\alpha_{2}, \cdots , \alpha_{r-1}, \alpha_{1}, \alpha_{r}, \cdots , \alpha_{\ell})  \\
&& \qquad {}+ 
(\alpha_{r}-\alpha_{1}-\hbar)
\frac{\alpha_{1}-\beta_{m_{r}}-\hbar}{\alpha_{1}-\beta_{m_{r}}} 
H_{\epsilon_{2}, \cdots , \epsilon_{r}, 0, \epsilon_{r+1}, \cdots , \epsilon_{\ell}}
 (\alpha_{2}, \cdots , \alpha_{r}, \alpha_{1}, \alpha_{r+1}, \cdots , \alpha_{\ell}) 
\Bigr\}. \no
\ena
{}From (\ref{rel1}), it can be checked that $f_{m_{r}}$ is symmetric 
with respect to $\alpha_{1}$ and $\alpha_{r}$. 

We can see that 
\bea
&& 
\hbar \sum_{a=1}^{n}f_{a}=
g_{M_{1}^{J}}(\{\alpha_{a}\}_{2 \le a \le \ell}|\{\beta_{j}\}) \times{} \no \\
&& \qquad \qquad {}\times 
\Bigl\{ 
\prod_{a=2}^{\ell-1}(\alpha_{1}-\alpha_{a}-\hbar) 
H_{0, \epsilon_{2}, \cdots , \epsilon_{\ell}}(\alpha_{1}, \alpha_{2}, \cdots , \alpha_{\ell}) \no \\
&& \qquad \qquad \quad {}+ 
(-1)^{\ell} \prod_{a=2}^{\ell}(\alpha_{a}-\alpha_{1}-\hbar) 
\prod_{j=1}^{n}\frac{\alpha_{1}-\beta_{j}-\hbar}{\alpha_{1}-\beta_{j}} 
H_{\epsilon_{2}, \cdots , \epsilon_{\ell}, 0} 
(\alpha_{2}, \cdots , \alpha_{\ell}, \alpha_{1}) \Bigr\}. 
\ena
By using (\ref{rel2}) and skew-symmetrizing both sides above, 
we have (\ref{FMhweq}). 
\qed 

\begin{lem}\label{ratclaim1}
Let $I_{s}, (s=0, \cdots , r)$ be sets of indices such that $\# I_{s}=m$ for all $s$. 
Then the following equality holds:
\bea
&&
\hbar \sum_{s=1}^{r}\prod_{j \in I_{s}}(x-y_{j}-\hbar)
\prod_{t=s+1}^{r}\prod_{j \in I_{t}}\frac{x-y_{j}-\hbar}{x-y_{j}}
\sum_{k \in I_{s}}
\frac{\prod_{j \in I_{0}}(y_{k}-y_{j}-s\hbar)}
     {(x-y_{k}-\hbar)(x-y_{k})\prod_{j \in I_{s} \atop j \not= k}(y_{k}-y_{j})} \no \\
&& {}+
\frac{\prod_{j \in I_{0} \cup \cdots \cup I_{r}}(x-y_{j}-\hbar)} 
     {\prod_{j \in I_{1} \cup \cdots \cup I_{r}}(x-y_{j})}=
\prod_{j \in I_{0}}(x-y_{j}-(r+1)\hbar).
\ena
\end{lem}

This lemma can be proved by induction on $r$. \newline

\noindent{\it Proof of Proposition \ref{putD}}.

Note that both sides of (\ref{DLformula}) are rational functions of $\alpha$ 
with at most simple poles at points $\beta_{j}, (j \in K_{r}, r>0)$, 
and have the same growth $O(\alpha^{m-2})$ as $\alpha \to \infty$. 

We can see that both sides have the same residue at points 
$\alpha=\beta_{b}, (b \in K_{r}^{J}, r>0)$ from Lemma \ref{ratclaim1} with 
\bea
x=\beta_{b}, \quad I_{s}=K_{s}^{J} \quad {\rm and} \quad y_{j}=\beta_{j}. 
\ena 
Moreover, it can be checked that both sides have the same value at points 
$\alpha=\beta_{b}+\hbar, (b \in K_{q}, q>0)$ from Lemma \ref{ratclaim1} with
\bea 
&&
x=\beta_{b}+\hbar, \quad I_{0}=K_{0}^{J}, \quad I_{s}=K_{q+s}^{J}, \, (s>0), \quad r=N-1, \no \\
&& 
y_{j}=\beta_{j}+r\hbar, \,\, (j \in K_{0}^{J}) \quad {\rm and} 
\quad y_{j}=\beta_{j}, \, \, (j \in K_{s}^{J}, s>q).
\ena
Hence (\ref{DLformula}) holds. 
\qed

\begin{lem}\label{ratclaim2}
Let $I_{s}, (s=0, \cdots , d)$ be sets of indices such that $\# I_{s}=m$ for all $s$. 
For $a \in I_{0}$, the following equality holds:
\bea
&& 
\sum_{s=0}^{d}
\prod_{j \in I_{s}}(x-y_{j}-(d+1-s)\hbar)
\frac{\prod_{j \in I_{t} \atop t<s}(y_{a}-y_{j}-\hbar)
      \prod_{j \in I_{t} \atop t>s}(y_{a}-y_{j})}
     {(x-y_{a}-(d-s)\hbar)(x-y_{a}-(d+1-s))} \no \\
&& {}=
\prod_{j \in I_{0} \atop j \not= a}(y_{a}-y_{j}-\hbar)
\prod_{j \in I_{t} \atop t>0}(y_{a}-y_{j})
\frac{1}{x-y_{a}}
\prod_{j \in I_{0} \atop j \not= a}\frac{x-y_{j}-\hbar}{y_{a}-y_{j}-\hbar}
\prod_{j \in I_{t} \atop t>0}\frac{x-y_{j}-\hbar}{x-y_{j}} +{} \no \\
&& {}+
\hbar\sum_{q=1}^{d}\sum_{k \in I_{q}}
\prod_{j \in I_{q} \atop j \not= k}\frac{y_{k}-y_{j}-\hbar}{y_{k}-y_{j}}
\frac{1}{x-y_{k}}
\prod_{j \in I_{q} \atop j \not= k}\frac{x-y_{j}-\hbar}{y_{k}-y_{j}-\hbar}
\prod_{j \in I_{t} \atop t>q}\frac{x-y_{j}-\hbar}{x-y_{j}} \no \\
&& \qquad {}\times 
\sum_{s=0}^{q-1}
\prod_{j \in I_{s}}(y_{k}-y_{j}-(q-s)\hbar)
\frac{\prod_{j \in I_{t} \atop t<s}(y_{a}-y_{j}-\hbar)
      \prod_{j \in I_{t} \atop t>s}(y_{a}-y_{j})}
     {(y_{k}-y_{j}-(q-s-1)\hbar)(y_{k}-y_{j}-(q-s))}.
\label{rateq1}
\ena
\end{lem}

\proof
Let us prove (\ref{rateq1}) by induction on $d$.
 
It is easy to see that (\ref{rateq1}) holds in the case of $d=0$. 

Suppose that (\ref{rateq1}) holds for $0, 1, \cdots , d-1$. 
First note that the the singularity of the lhs is only the simple pole at $x=y_{a}$. 
Hence both sides are rational functions of $x$ with simple poles 
at points $y_{a}$ and $y_{j}, \, (j \in I_{u}, u>0)$, 
and have the same growth $O(x^{m-2})$ as $x \to \infty$. 
It is easy to see that residues of both sides at $x=y_{a}$ are equal. 
We can check that both sides have the same residue also at $x=y_{j}, \, (j \in I_{u}, u>0)$ 
{}from (\ref{rateq1}) with $d=u-1$. 
Moreover, both sides have the same value at $x=y_{j}+\hbar, \, (j \in I_{d})$. 
Therefore (\ref{rateq1}) holds also for $d$.
\qed

\noindent{\it Proof of Proposition \ref{toQ}}.

Consider the following function $f(\alpha, y)$:
\bea
f(\alpha, y):=
\sum_{s=0}^{N-1}L_{J}^{(s)}(\alpha+s\hbar) \,
T_{\hbar}^{\alpha}\!\left(
\frac{U_{J}^{(s)}(\alpha)-U_{J}^{(s)}(y-s\hbar)}{\alpha-y+s\hbar} \right),
\label{Qpf0}
\ena
where $T_{\hbar}^{\alpha}$ is the difference operator $T_{\hbar}$ 
with respect to $\alpha$, and
\bea
U_{J}^{(s)}(\alpha):=
\prod_{k=0}^{s-1}L_{J}^{(k)}(\alpha+(s-1)\hbar)\prod_{k=s+1}^{N-1}L_{J}^{(k)}(\alpha+s\hbar).
\ena

For $a \in K_{r}^{J}$, we have 
\bea
f(\alpha, \beta_{a}+N\hbar)\!\!&=&\!\!
D\Bigr( \prod_{j=1 \atop j \not= a}^{n}(\alpha -\beta_{j}-N\hbar) \Bigr)  \label{Qpf1} \\
&& \!\!{}-
\hbar \sum_{s=r}^{N-1}L_{J}^{(s)}(\alpha+s\hbar)
\frac{\prod_{k=0}^{s}L_{J}^{(k)}(\beta_{a}+(N-1)\hbar)
      \prod_{k=s+1}^{N-1}L_{J}^{(k)}(\beta_{a}+N\hbar)}
     {(\alpha-\beta_{a}+(s-N)\hbar)(\alpha-\beta_{a}+(s-N+1)\hbar)}. \no
\ena
We find that the sum in the rhs of (\ref{Qpf1}) equals 
the rhs of (\ref{Q}) 
by using Lemma \ref{ratclaim2} with 
\bea
I_{s}=K_{r+s}^{J}, \quad d=N-r-1, \quad x=\alpha \quad {\rm and} \quad y_{j}=\beta_{j}. 
\ena

On the other hand, we have 
\bea
\frac{U_{J}^{(s)}(\alpha)-U_{J}^{(s)}(y-s\hbar)}{\alpha-y+s\hbar}=
\sum_{k=1}^{(N-1)m}\left[
\frac{U_{J}^{(s)}(\alpha)}{(\alpha+s\hbar)^{k}}\right]_{+} y^{k-1}. 
\ena 
Hence we get 
\bea
f(\alpha, \beta_{a}+N\hbar)=
\sum_{k=1}^{(N-1)m}(\beta_{a}+N\hbar)^{k-1}Q_{J}^{(k)}(\alpha).
\ena 
This completes the proof. 
\qed

\end{document}